\documentclass[acmsmall]{acmart}
\usepackage[T1]{fontenc}
\usepackage[utf8]{inputenc}
\usepackage{textcomp} % provide euro and other symbols
\usepackage{pdfpages}
\usepackage{framed}
\usepackage[inline]{enumitem}
\newcommand{\highlight}[1]{\begin{leftbar}\noindent \emph{#1}\end{leftbar}}

\AtBeginDocument{%
  \providecommand\BibTeX{{%
    \normalfont B\kern-0.5em{\scshape i\kern-0.25em b}\kern-0.8em\TeX}}}

\setcopyright{acmcopyright}
\copyrightyear{2020}
\acmYear{2020}
\acmDOI{10.1145/1122445.1122456}

\acmJournal{TOSEM}
\acmVolume{1}
\acmNumber{1}
\acmArticle{111}
\acmMonth{1}

\begin{document}

\title{Psychometrics in Behavioral Software Engineering: A Methodological Introduction with Guidelines}

\author{Daniel Graziotin}
\affiliation{%
  \institution{University of Stuttgart, Institute of Software Engineering}
  \streetaddress{Universitätsstraße 38, 70569 Stuttgart, Germany}
  \city{Stuttgart}
  \country{Germany}}
\email{daniel.graziotin@iste.uni-stuttgart.de}
\orcid{0000-0002-9107-7681}

\author{Per Lenberg}
\affiliation{%
  \institution{Chalmers and the University of Gothenburg, Dept. of Computer Science and Engineering}
  \streetaddress{SE-412 96}
  \city{Gothenbug}
  \country{Sweden}}
\email{per.lenberg@chalmers.se}
\orcid{0000-0002-3186-3947}

\author{Robert Feldt}
\affiliation{%
  \institution{Chalmers and the University of Gothenburg, Dept. of Computer Science and Engineering}
  \streetaddress{SE-412 96}
  \city{Gothenbug}
  \country{Sweden}}
\email{robert.feldt@chalmers.se}
\orcid{0000-0002-5179-4205}

\author{Stefan Wagner}
\affiliation{%
  \institution{University of Stuttgart, Institute of Software Engineering}
  \streetaddress{Universitätsstraße 38, 70569 Stuttgart, Germany}
  \city{Stuttgart}
  \country{Germany}}
\email{stefan.wagner@iste.uni-stuttgart.de}
\orcid{0000-0002-5256-8429}

\renewcommand{\shortauthors}{Graziotin et al.}

\begin{abstract}
A meaningful and deep understanding of the human aspects of software engineering (SE) requires psychological constructs to be considered. Psychology theory can facilitate the systematic and sound development as well as the adoption of instruments (e.g., psychological tests, questionnaires) to assess these constructs. In particular, to ensure high quality, the psychometric properties of instruments need evaluation.
In this paper, we provide an introduction to psychometric theory for the evaluation of measurement instruments for SE researchers. We present guidelines that enable using existing instruments and developing new ones adequately. 
We conducted a comprehensive review of the psychology literature framed by the Standards for Educational and Psychological Testing. We detail activities used when operationalizing new psychological constructs, such as item pooling, item review, pilot testing, item analysis, factor analysis, statistical property of items, reliability, validity, and fairness in testing and test bias. We provide an openly available example of a psychometric evaluation based on our guideline.
We hope to encourage a culture change in SE research towards the adoption of established methods from psychology. To improve the quality of behavioral research in SE, studies focusing on introducing, validating, and then using psychometric instruments need to be more common.
\end{abstract}



\maketitle

\section{Introduction\label{sec:intro}}
Software is developed for people, by people. For decades we have recognized that, no matter the size and importance of the technical side of software engineering, it is humans that ultimately drive the underlying processes and produce the desired artifacts~\citep{weinberg1971}. Software engineers are knowledge workers and have knowledge as their main capital~\citep{swart2003}. They need to construct, retrieve, model, aggregate, and present knowledge in all their analytic and creative daily activities~\citep{rus2002}. Operations related to knowledge are cognitive in nature, and cognition is influenced by characteristics of human behavior, including personality, affect, and motivation~\citep{hilgard1980}. It is no wonder that industry and academia have explored psychological aspects of software development and the assessment of psychological constructs at the individual, team, and organization level~\citep{feldt2008,lenberg2015,carver2021,capretz2003personality,wang2020,tripp2016,gren2016}.

Psychological assessment is the gathering of psychology-related data towards an evaluation that is accomplished through the use of tools such as tests, interviews, case studies, behavioral observation, and other procedures~\citep{cohen1995}. With the present paper, we are primarily focused on psychological tests, i.e. quantitative research\footnote{We are interested in qualitative research as well, and we value it the same as quantitative research in behavioral software engineering, as we state in section~\ref{sec:limitations}.}, as behavioral software engineering has turned much attention to employing theory and measurement instruments from psychology~\citep{capretz2003personality,feldt2008,lenberg2015,graziotin2015,cruz2015}.

Psychological tests are instruments (e.g., questionnaires) used to measure unobserved constructs~\citep{cohen1995}. We can not assess these constructs directly, like when measuring the source lines of code or specific properties of a UML diagram. Hence, the underlying variables are called latent variables~\citep{nunnally1994psychometric}. Examples of such unobserved constructs include attitude, mood, happiness, job satisfaction, commitment, motivation, intelligence, abilities, cognitive skills, and performance. We need to create good measurement instruments\footnote{In this paper, we use the terms psychological test, measurement instrument, and questionnaire interchangeably.} for the assessment of such constructs.
For ensuring a systematic and sound development of psychological tests and their interpretation, the field of psychometrics was born~\citep{rust2009,nunnally1994psychometric}.

Psychometrics is the development of measurement instruments and the assessment on whether these instruments are reliable and valid forms of measurement~\citep{ginty2013}. Psychometrics is also the branch of psychology and education which is devoted to the construction of valid and reliable psychological tests~\citep{rust2009}.

The origin of psychometrics go back way before a hundred years~\citep{rust2009,nunnally1994psychometric}. The establishment of testing as a mathematically-oriented sub-discipline of psychology are to be found in the 1930s, with three milestone publication events~\citep{jones2006}: 
\begin{enumerate*}
  \item \textit{Psychometrika} volume 1, issue 1, in March 1936, a venue ``devoted to the development of psychology as a quantitative rational science'' (subheading of the journal).
  \item ~\citeauthor{guilford1938}'s first edition of \textit{Psychometric Methods}~\citep{guilford1938}.
  \item ~\citeauthor{thurstone1937}'s article \textit{Psychology as a quantitative rational science} in the journal \textit{Science}~\citep{thurstone1937} to address the establishing of the \textit{Psychometric Society} in 1936.
\end{enumerate*}

~\citeauthor{thurstone1937}'s paper~\citep{thurstone1937} served as a plea to recognize quantitative roots of psychology, by which rationalizing problems in psychology with mathematics language and methods 
from statistics would allow the development of psychological science. Proper development and validation of tests has the potential to result in better decisions on individuals, while, on the opposite end, improper development and validation of tests might result in invalid interpretation of results, economic loss, and even harm of individuals~\citep{apa2014}. The individual level also affects teams and, ultimately, organizations, and even when these are in focus, measuring constructs that concern them is typically built up from measurements on the individual level (e.g., organizational readiness for change, at the organization level, is computed from aggregated responses of individuals who work at the organization~\citep{shea2014}). Personality assessment is a classic example of psychological testing in personnel selection, which has been employed across domains, including companies related to information technology~\citep{darcy2005,wyrich2019}. Another example is assessment of job satisfaction of workers, which is an evaluative judgment one makes about one's job or job situation~\citep{weiss2002}. A company would, arguably, attempt to foster job satisfaction of individuals, for example by employing agile methodologies instead of traditional processes~\citep{tripp2016}. Similarly, IT-related companies are interested in fostering motivation of software engineers, which is related to factors that energize, channel and sustain human behavior over time~\citep{franca2018}. Stress of software developers can also be measured by psychological tests coupled with physiological measurements~\citep{ostberg2017towards,ostberg2017towards} and biometrics~\citep{Fagerholm2020}.
Finally, to bring examples related to cognition and behavior, research in software engineering has recently turned attention to understanding and addressing cognitive biases while developing software~\citep{chattopadhyay2020,wyrich2020} and identifying and reducing gender bias in development teams~\citep{wang2020}. 

Improper development, administration, and handling of psychological tests could harm the company by hiring a non-desirable person, and it could harm the interviewee because of missed opportunities.

We believe that solid theoretical and methodological foundations should be the first step when designing a measurement instrument. The reality, however, is that not all tests are well developed in psychology~\citep{apa2014}. Software engineering research, especially when studying psychological constructs quantitatively, is far from adopting rigorous and validated research artifacts.

\subsection{Misconceptions with psychological tests in software engineering research}

Already in 2007, \citet{mcdonald2007} subtitled their paper ``Examining the use and abuse of personality tests in software engineering''. The authors anticipated the issue that we attempt to address in the present manuscript, that is the ``the lack of progress in this [personality research in software engineering] field is due in part to the inappropriate use of psychological tests, frequently coupled with basic misunderstandings of personality theory by those who use them'' (p. 67). While their focus was on personality, their concerns are hold in a broader way to other psychometric instruments and constructs.

Instances of mishandle\footnote{We believe that direct accusations bring no value to our contribution and are counterproductive to our advancement of knowledge, so we discuss resources that point to specific issues rather than specific papers or authors.} can, for example, be observed in the papers found by a systematic literature review of personality research in software engineering by~\citet{cruz2015}. We noted in the results of~\citet{cruz2015} that 48\% of the personality studies in software engineering have employed the Myers-Brigg Type Indicator (MBTI) questionnaire, which has been shown to possess low to no reliability and validity properties~\citep{pittenger1993} up to the point of being called a ``little more than an elaborate Chinese fortune cookie''~\citep{hogan2017}. \citet{feldt2010} similarly pointed out deficiencies of MBTI and proposed and used an alternative (IPIP) with more empirical support in the psychological literature.

\citet{feldt2008} have argued in favor of systematic studies on human aspects of software engineering. More specifically, to adopt measurement instruments coming  from psychology and related fields. \citet{graziotin2015affect} have echoed the call seven years after but found that research on the affect of software developers had been threatened by a misunderstanding of related constructs and how to assess them. In particular, the authors noted that peers in software engineering tend to confuse affect-related psychological constructs such as emotions and moods with related, yet different, constructs such as motivation, commitment, and well-being.

\citet{lenberg2015} have conducted a systematic literature review of studies on human aspects in software development and engineering that made use of behavioral science, calling the field behavioral software engineering. Among their results, they found that software engineering research is threatened by several knowledge gaps when performing behavioral research, and that there have been very few collaborations between software engineering and behavioral science researchers.

\citet{graziotin2015}, meanwhile, extended their prior observations on affect to a broader view of software engineering research with a psychological perspective. The work offered what we can consider the sentiment for the present article, that is brief guidelines to select a theoretical framework and validated measurement instruments from psychology. ~\citet{graziotin2015} called the field ``psychoempirical software engineering'' but later agreed with~\citet{lenberg2015} to unify the vision under ``behavioral software engineering''. Hence, the present collaboration.

Our previous studies have also reported that, when a validated test from psychology is adopted by software engineering researchers, its items are often modified, causing the destruction of its psychometric reliability and validity properties. This includes a thorough evaluation of the psychometric properties of candidate instruments.~\citet{gren2016} have argued in favor of ``useful statistical methods for human factors research in software engineering'' (paper title), which include underused  methods such as Test-Retest, Cronbach's $\alpha$, and exploratory factor analysis---all of which are covered in this paper.
~\citet{gren2018} has also offered a psychological test theory lens for characterizing validity and reliability in behavioral software engineering research, further enforcing our view that software engineering research that investigates any psychological construct quantitatively should maintain fair psychometric properties. We agree with~\citet{gren2018} that we should ``change the culture in software engineering research from seeing tool-constructing as the holy grail of research and instead value [psychometric] validation studies higher.'' (p. 3).

A \textit{mea culpa} works better than a \textit{j'accuse} in further building our case, so we bring a negative example from one of our previous studies. As reported in a very recent work by~\citet{ralph2020} (which we appreciate in the next paragraph), ``there is no widespread consensus about how to measure developers' productivity or the main antecedents thereof. Many researchers use simple, unvalidated productivity scales'' (p. 6). In one of the earliest works by the first author of the present paper~\citep{graziotin2015you}, we compared the affect triggered by a software development task with the self-assessed productivity of individual programmers. While we were very careful to select a validated measurement instrument of emotions and to highlight how self-assessment of productivity converges to objective assessment of productivity, we used a single Likert item scale to represent productivity. This choice was to reduce as much as possible the items of the measurement instrument, which had to be used every ten minutes, for a total of nine times for each participant. While the results of the study are not invalidated by this choice, the productivity scale itself was not validated, making the results less valuable from a psychometric perspective and, thus, our interpretation of its results. The study was also (successfully) independently replicated twice, and both replications suffer from the same unfortunate choice.

We wish to refrain from being overly negative. The field of software engineering does have positive cases---excluding those from the present authors---that we can showcase here. For example,~\citet{fagerholm2014} developed a questionnaire on lean and agile values and applied psychometric approaches to inspect the structure of value dimensions.~\citet{fagerholm2015} has also embodied psychometric approaches in their PhD dissertation by analyzing the validity of the constructs they studied. A more recent example is by~\citet{ralph2020}, who analyzed through a questionnaire the effects of the COVID-19 pandemic on developers' well-being and productivity. The authors constructed their measurement instrument by incorporating psychometrically validated scales on constructs such as perceived productivity, disaster preparedness, fear and resilience, ergonomics, and organizational support. Furthermore, they employed confirmatory factor analysis (which we touch upon in the present paper) to verify that the included items do indeed cluster and converge into the factors that are claimed to converge to.

While positive cases do exist, we notice that they are fairly recent and that we can do better than that. We want to synthesize knowledge from psychology fields to software engineering research towards better quantitative studies of behavioral aspects.

\highlight{Filling the knowledge gap: introduction and guidelines to psychometric evaluation for behavioral software engineering research.}

Overall, we argue that one thing that is missing is an introduction to the field of psychometrics for behavioral software engineering researchers. Such an introduction can help improve the understanding of the available measurement instruments and, also, the development of new tests, allowing researchers as well as practitioners to explore the human component in the software construction process more accurately.

\subsection{Objective}
Our overall objective is to address the lack of understanding and use of psychometrics in behavioral software engineering research including its limitations.

We also hope to increase software engineering researchers' awareness and respect of theories and tools developed in established fields of the behavioral science, towards stronger methodological foundations of behavioral software engineering research.

\subsection{Contribution}

With this paper, we contribute to the behavioral software engineering body of knowledge with a set of guidelines which enable a better understanding of psychological constructs in research activities when we interpret them through measurement instruments. This improvement in research quality is achieved by either (1) reusing psychometrically validated measurement instruments, as well as understanding why and how they are validated, or, if no such questionnaires exist, (2) developing new psychometrically validated questionnaires that are better suited for the software engineering domain.

Our contribution is enabled by offering one theoretical deliverable and one companion, practice-oriented deliverable.

\begin{enumerate}
\item We offer a review and synthesis of psychometric guidelines in form of several textbooks, review papers, as well as empirical studies, packaged in a style that is familiar to the software engineering researchers, including concrete examples and how to execute each activity. The guidelines enable evaluation of existing measurement instruments as well as developing new ones.
\item We offer a \textit{hands-on} counterpart to our review by providing a fully reproducible implementation of our guidelines as R Markdown.
\end{enumerate}

The Standards for Educational and Psychological Testing (SEPT,~\citet{apa2014}) is a set of gold standards in psychological testing jointly developed by the American Psychological Association (APA), National Council on Measurement in Education (NCME), and the American Educational Research Association (AERA). The book defines areas and standards that should be met when developing, validating, and administering psychological tests. We adopted SEPT as a framework to guide the paper construction, for ensuring that the standards are met and that the various other references are framed in the correct context.

Additionally, we organized the scoping of the paper by comparing related work from the fields of psychology research. 
While the present paper is not a systematic literature review or a mapping study---the discipline is so broad that entire textbooks have been written on it---we systematically framed its construction to ensure that all important topics were covered.

\subsection{Scope}
\begin{figure}
\includegraphics[width=\columnwidth]{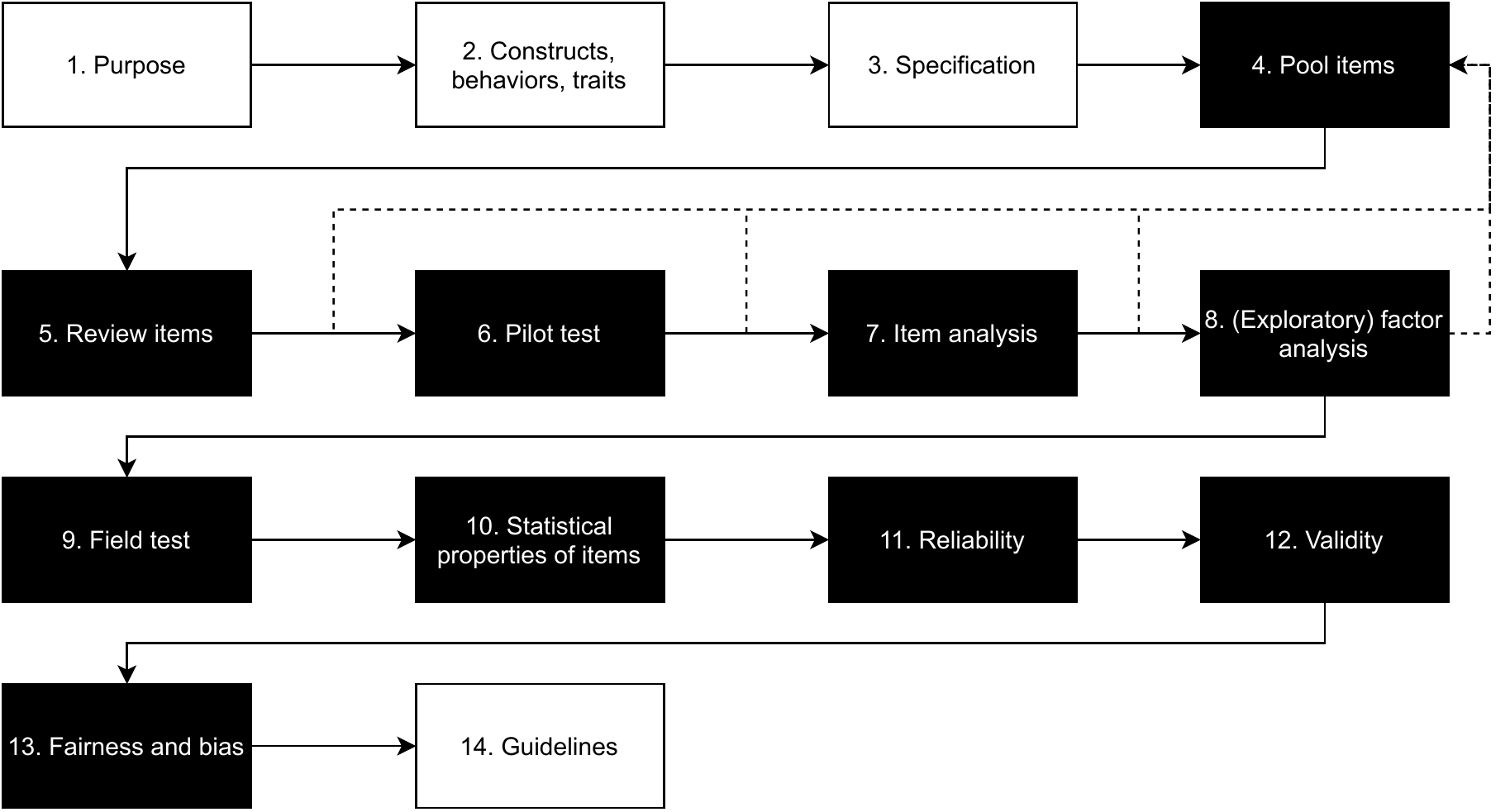}
\caption{Steps for developing a psychological test. Phases with dark background are uncommon in software engineering research and are covered in the present paper.}
\Description{Steps for developing a psychological test. Phases with dark background are uncommon in software engineering research and are covered in the present paper.}
\label{fig:outline}
\end{figure}

Several authors, e.g.,~\citet{crocker2006,singh2016,rust2009}, have proposed different phases for the psychometric development and evaluation of measurement instruments. Through our review, we identified 14 phases that we summarize visually in Figure~\ref{fig:outline} and outline as follows.

\begin{enumerate}
\item Identification of the primary purpose for which the test scores will be employed.
\item Identification of constructs, traits, and behaviors that are reflected by the purpose of the instrument.
\item Development of a test specification, delineation of the items proportion that should focus on each type of constructs, traits, and behaviors of the test.
\item Construction of an initial pool of items.
\item Review of the items.
\item Conduction of a pilot test with the revised items.
\item Execution of an item analysis to possibly reduce the number of items.
\item Evaluation of an exploratory factor analysis to possibly reduce and group items into components or factors.
\item Execution of a field test of the items with a larger, representative sample.
\item Determination of statistical properties of item scores.
\item Design and execution of reliability studies.
\item Design and execution of validity studies.
\item Evaluation of fairness in testing and test bias.
\item Development of guidelines for administering, scoring, and interpreting test scores.
\end{enumerate}

We focus mainly on those with a dark background in Figure~\ref{fig:outline} as they are the most challenging and usually not covered in software engineering research. 

As represented by the dashed lines, the process is linear only idealistically, when everything goes according to plans. Realistically, the construction of tests following a psychometric approach is failure-prone and iterative for correcting issues. 

Furthermore, not all steps have to necessarily be followed when developing a measurement instrument. We are illustrating a wide range of possibilities, some of which are often brought by in future validation studies.

As a final note, the present paper, as well as any psychometric construction of measurement instruments, is not a checklist. A psychometric evaluation does not include all elements reported in this paper, as many facets of psychometrics are influenced by the research questions, study design, and data at hand. Yet, a proper psychometric evaluation requires a consideration of all elements reported in the present paper. 

\subsection{Structure}

After a brief introduction to the key concepts of psychometrics (section~\ref{sec:concepts}), that are required to understand the rest of the paper, we focus on test construction in psychometrics and the phases highlighted in Figure~\ref{fig:outline}, namely pool items (section~\ref{sec:pool-items}), item review (section ~\ref{sec:item-review}), pilot test (section~\ref{sec:pilot-test}), item analysis (section~\ref{sec:item-analysis}), factor analysis (section~\ref{sec:factor-analysis}), statistical properties (section~\ref{sec:standardization}), reliability (section~\ref{sec:reliability}), validity (section~\ref{sec:validity}), and fairness in testing and test bias (section~\ref{sec:fairness}). We close our guidelines with two opposing sources for inspiration: 
\begin{enumerate*}
    \item A comprehensive list for further reading (section~\ref{sec:further-reading}) to deepen what we are able to merely surface in this paper.
    \item A review of limitations of psychometrics and their critique (section~\ref{sec:limitations})
\end{enumerate*}. 

Finally, we provide a hands-on running example (Section~\ref{sec:example}) of a psychometric evaluation. We provide R code and generated datasets openly~\citep{graziotin2020} following open science principles in software engineering~\citep{mendez2020}.

\section{Concepts\label{sec:concepts}}

This section provides an overview of basic terms and concepts from psychometrics that will enable an understanding of all remaining sections. In particular, we clarify on psychometric models and test types (and types of testing) as these will sometimes have an influence on the statistical methods and lens to adopt when designing and evaluating a measurement instrument.

\subsection{Building blocks}
The fundamental idea behind psychological testing is that what is being assessed is not a physical object, such as height and weight. Rather, we are attempting to assess a construct, that is a hypothetical entity~\citep{apa2014} constructed by humans to represent concepts referring to various, concrete entities that are perceived in the moment, such as behaviors, experience, and attitudes~\citep{uher2021}.

If we assess the job satisfaction of a software developer, we are not directly measuring the satisfaction of the individual. Instead, we compare the developer's score with other developers' scores or a set of established norms for job satisfaction. When comparing the satisfaction scores between developers, we are limited to seeing how the scores differentiate between satisfied and unsatisfied developers according to the knowledge and ideas we have about satisfied an dissatisfied individuals.

There are two common models of psychometrics, namely functionalist and trait~\citep{rust2009,wood2015}. Functionalist psychometrics often occurs in educational and occupational tests; it deals with how the design of a test is determined by its application and not about the constructs being measured~\citep{rust2009,green2009}. For functionalist design, a good test is one that is able to distinguish between individuals who perform well and individuals who perform less well on a job or in school activities. This is also called local criterion-based validity (explained in section~\ref{sec:validity}). The functionalist paradigm can be applied to most cases where a performance assessment or an evaluation is required.

Trait psychometrics attempts to address  notions such as human intelligence, personality, and affect scientifically~\citep{costajr1992,morgan1980}. The classic trait approach was based on the notion that, for example, intelligence is related to biological individual differences, and trait psychometric tests aimed to measure traits that would represent biological differences among people~\citep{morgan1980}.

Both schools of thought have several aspects in common, including test construction and validation methods, and they are linked by the theory of true scores~\citep{hambleton1991}. The theory of true scores, or latent trait theory, is governed by formulas of the form:

\begin{equation}\label{eq:true-scores}
X = T + E
\end{equation}

where $X$ is the observed score, $T$ is the true score, and $E$ is the error.
There are three assumptions with the theory of true scores. (1) all errors $E$ are random and normally distributed, (2) true scores $X$ are uncorrelated with the errors, and (3) different measures of the observed score $T$ on the same participants are independent from each other. Besides all issues that come with the three assumptions, the theory has been criticized with the major point being that there is arguably no such thing as a true score, and that all that tests can measure are abstractions of psychological constructs~\citep{loevinger1957}.

Elaborations and re-interpretations of the theory of true scores have been proposed, among which is the statistical true score~\citep{carnap1962}. The statistical true score defines the true score $T$ as the score we would obtain by averaging an infinite number of measures $X$ from the same individual. With an infinite number of measures the random errors $E$ cancel each other, leaving with the true score $T$. The statistical form of the theory of true scores should not be completely new to readers of software engineering, as most quantitative methods that are in use in our field nowadays are based on it. The statistical interpretation of the theory of true scores applies both to trait and functional psychometrics. A difference lies in generalization. Functional tests can only be specific to a certain context while trait tests attempt to generalize to an overall construct present in a group of individuals.

\subsection{Test types and types of testing}

Items on psychological tests can be knowledge-based or person-based~\citep{rust2009,anastasi1997}. Knowledge-based tests assess whether an individual performs well regarding the knowledge of certain information, including possessing skills favoring performance or quality in knowledge-based tasks. For a software engineering example, debugging skills would be assessed by a knowledge-based test. Person-based tests, on the other hand, assess typical performance, or how a person is represented, with respect to a construct. Examples of constructs in person-based tests include personality, mood, and attitudes. The personalities of programmers in pair programming settings would be assessed by a person-based test.

Knowledge-based tests are usually uni-dimensional as they gravitate towards the notion of possessing or not possessing certain knowledge. We can also easily rank individuals on their scores and state who ranks better. Person-based tests are usually multi-dimensional and do not allow direct ranking of individuals without some assumptions. For example, a developer could score high on extroversion. A high score on extroversion does not make a developer with a lower extroversion score a ``worse'' developer in any way, because of the lower score.

A second distinction to be made is between criterion-referenced and norm-referenced testing~\citep{glaser1963}. Criterion-referenced tests are constructed with reference to performance on a-priori defined values for establishing excellence~\citep{glaser1963,berger2013b}. 

Continuing with the example on debugging skills, a criterion-referenced test would assess, with a score from 0 to 10,  whether a developer is able to open a debugging tool and use its ten basic functionalities. A score of 10 out of 10 would mean that the developer is able to debug software.

Norm-referenced tests lack a-priori defined scores. What constitutes a high score is in relation to how everyone else scores. A test for assessing the happiness of software developers will return scores for each participant. The test itself will have a theoretical range, say $-10$ for strong unhappiness and $+10$ for strong happiness. When a developer scores $+4$ on our happiness scale, all we can say is that the developer is  happy rather than unhappy. If we know that software developers score $-3$ on average, with a standard deviation of $1$, then we know that the developer is quite a happy one. The development and evolution of norm-referenced testing attempts, in addition to developing valid and reliable instruments to establish norms, values for populations and sub-populations of individuals. That is, norm-referenced tests allow us to compare scores with respect to what is considered~\textit{normal}~\citep{glaser1963,berger2013a}.

\section{Pool Items\label{sec:pool-items}}
The first steps in developing a measurement instrument is, of course, developing an initial set of items. \citet{coaley2014} summarizes a planning and designing process in a way that reminds us of the Goal-Question-Metrics (GQM) model~\citep{caldiera1994goal}:

\begin{enumerate*}
  \item Set clear aims---on defining the purpose of the measurement instrument, which constructs we are targeting, and what is the intended target population.
  \item Define the attribute(s)---on moving in the empirical lenses from the construct, or the object, to its attributes being measured. In this step, it is advised to perform a comprehensive literature review of the theoretical concepts being assessed.
  \item Write a plan---on drafting a specification of the measurement instrument as if it was completed already, including the test content, target group and population, kinds and number of needed participants, administration instructions, time constraints, and how scores should be interpreted.
  \item Writing items---on designing and constructing a pool of items related to all previously defined steps.
\end{enumerate*}

Sources for generating items could be experts in the domain or fields, interviewing potential respondents, and prior work~\citep{rattray2007} paired with constant comparison or update of research questions~\citep{oppenheim1992}.

We do not want to spend too much space about writing items because of existing good literature: on the questionnaire construction process~\citep{bradburn2004,rattray2007,oppenheim1992}, on question effects and question-wording effects~\citep{kalton1982,sigelman1981,schwarz2016}, and, for our field, on experiences and reviews on conducting surveys in software engineering research~\citep{ciolkowski2003,wagner2020,molleri2016,kitchenham2008,ji2008}. 

We will rather focus on selecting items from the produced pool. While some guidelines suggest to generate a double amount of items than those that are likely required~\citep{kline2015handbook}, we note in section~\ref{sec:factor-analysis} that factors require three to five associated answers to possess meaningful variance properties. Thus, a better strategy would be to develop six to ten items per envisioned factor, or sub-construct.

\section{Item Review\label{sec:item-review}}

When developing a new measurement instrument, we are likely (and encouraged by the previous section) to create more items than are really needed.

\begin{figure}
\includegraphics[width=\columnwidth]{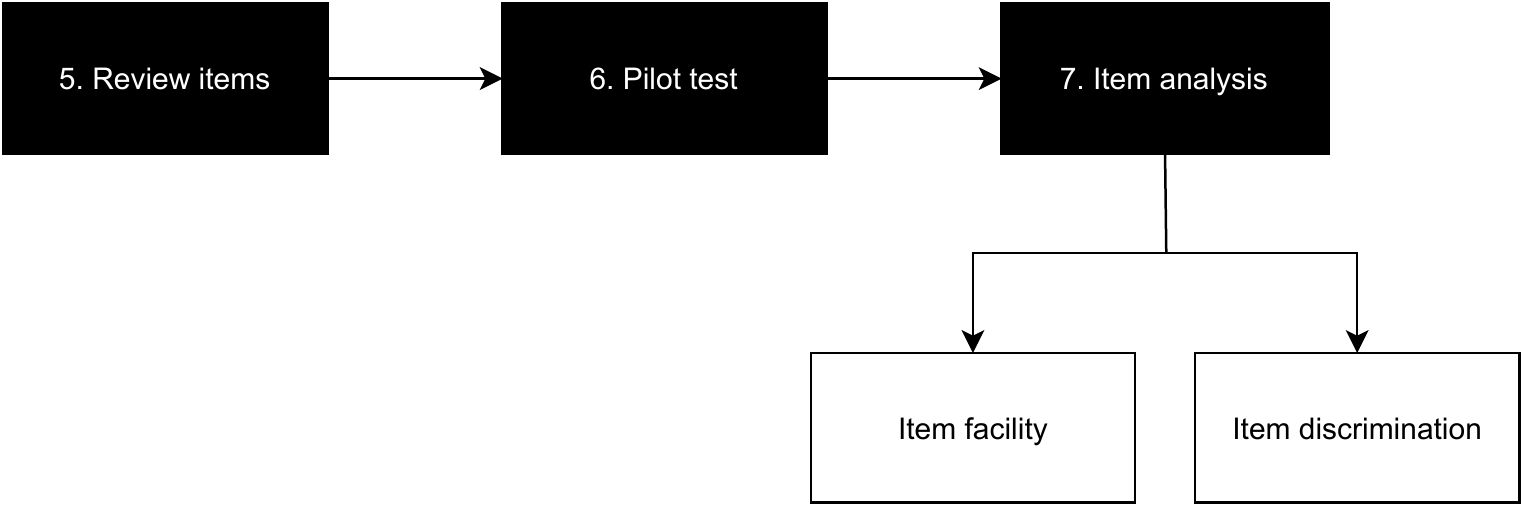}
\caption{Phases for item review and item analysis}\label{fig:item-review-analysis}
\Description{Phases for item review and item analysis}
\end{figure}

Item review and item analysis are a series of methods to reduce the number of items of a measurement instrument and keep the best performing ones~\citep{guilford1954,kline2015handbook}. This is a two-step process, as shown in Figure~\ref{fig:item-review-analysis}. First, it requires a review by experts; then, a pilot study and statistical calculations. During the first step (\textit{item review}), experts in the domain of knowledge evaluate items one by one and argue for their presence in the test~\citep{schoenherr2016,kline2015handbook}. During the second step (\textit{item analysis}), the developers of the measurement instrument calculate item facility and item discrimination based on a pilot study that uses the tentative set of items.

During item review, experts in the domain of knowledge discuss candidate items and argue in favor or against them, as it happens when discussing inclusion and exclusion of publications in systematic literature reviews~\citep{kitchenham2007}. 

Given its usage in the medical domain~\citep{boulkedid2011}, we recommend using the Delphi technique~\citep{dalkey1963} with domain experts to identify best candidates after initial qualitative probes to identify sub-constructs and items. To assess the degree of agreement among raters, we recommend using inter-rater reliability measures such as Cronbach's $\alpha$~\citep{cronbach1951} and Krippendorff's $\alpha$~\citep{de2012calculating}. 

After reaching an agreement on the items to be included, a pilot study is required for an analysis of the items.

\section{Pilot Test\label{sec:pilot-test}}

A pilot test is necessary to probe the effectiveness of the developed items. After item review we are left with a set of items that have been qualitatively evaluated by experts in the field. These items are too many, and we have no idea yet on how they would group together or contribute to the total variance related to the constructs we intend to represent. Hence, a pilot test.

Pilot tests allow to discover early issues related to questionnaire items, for example on wording problems, lack of clarity, confusing steps, or even discrimination of and between respondents~\citep{rattray2007}.

There are no clear guidelines on how big the sample size for pilot tests should be. First criterion is rather on representativeness of the pilot sample with respect to the target group or population~\citep{thabane2010}.~\citet{johanson2010} have identified suggestions from the literature that sees $n=30$ participants as the lower limit to gather meaningful data and to reason for the test construction.~\citet{johanson2010} have conducted a cost-benefit analysis to identify the point at which adding participants to a pilot sample size would incur in notably lesser effects in estimating population parameters. Analysis of sample size changes for a range of Pearson's correlations, range of proportions, and confidence intervals for reliability coefficients. All results confirm the indicated value of $n=30$ as reasonable minimum recommendation for a pilot study for a psychological test. 

A further possible novelty that we would like to report to the software engineering research audience is brought by~\citeauthor{collins2003}'s~\cite{collins2003} overview of cognitive methods to pretest survey instruments. The authors argue, in their review of the literature, that the tradition of survey pre-testing has mostly dealt with quantitative issues, e.g., standardizing data collection procedures such as question wording, worrying about clustering, and variance. All this, while valuable, should be coupled with techniques that go beyond the insofar assumption that respondents all understand the questions being asked in the same way, and that they are willing to respond to the same set of questions the same. Many of these techniques are about qualitative data rather than quantitative data. Cognitive methods enable exploring the process by which respondents think and act with when responding (or not) to questions, and which factors influence these answers. Cognitive  methods include think-aloud and probing in cognitive interviewing, paraphrasing, card sorts, vignettes, confidence ratings and response latency timing. In absence of consensus on the best process~\citep{hilton2017}, we will briefly summarize cognitive interviewing because it is an established method~\citep{miller2014,drennan2003,willis2004} and it brings qualitative methods to psychometrics.

Cognitive interviewing is a qualitative method that studies the question-response process by participants when answering questions on psychological tests~\citep{miller2014}. The aim is to use theories from the cognitive sciences to understand how participants perceive and interpret items, and to identify issues that may arise when distributing the instrument~\citep{drennan2003}. Cognitive interviewing requires a small sample of participants purposely selected for in-depth interviews on how and why they answered the question as they did~\citep{miller2014}.

The cognitive interview process is characterized by two verbal report methods: think-aloud and verbal probing~\citep{willis2004}. In think-aloud, the participant is asked to (preferably) verbally explicating what they are thinking from the point of reading an item to the point of scoring the item. In verbal probing, the interviewer intervenes with specific questions and probes~\citep{collins2003}. The former is respondent-driven, the latter is interviewer-driven. Examples of think-aloud questions are of open-ended nature and inviting for more thoughts, e.g.: ``What are you thinking while reading this item?'', ``What makes you think that?'', or ``I noticed you hesitate for a while there, what made you hesitate?''. Examples of verbal-probing (from ~\citet{collins2003}, Table 2) include comprehension (``What does the term X mean to you?'', retrieval (``How did you calculate your answer?''), confidence judgment (``How well do you remember this?''), and response (``How did you feel about answering this question?'').

\section{Item analysis\label{sec:item-analysis}} 

While qualitative methods are useful for identifying issues, we can apply quantitative techniques to reduce items. Item analysis refers to several statistical methods for the selection of items of a psychological test~\citep{guilford1954,kline2015handbook}. Item analysis is often the only method for item reduction when a single construct should be evaluated, whereas multiple constructs and sub-constructs require further steps to be taken. Two of the most known techniques can be found in item facility and item discrimination. This section will explain both techniques according to the test type.

\subsubsection{Item facility}

Item facility for an item is a measure of tendency in answering one item with the same score. This has different meanings according to the type of test.

\subsubsection*{Knowledge-based test}
Item facility for an item, also known as item difficulty, is defined as the ratio of the number of participants who provided a right answer over the number of all participants to a test~\citep{rust2009,kline2015handbook}. The value of item facility ranges from $0$ (all respondents are wrong) to $1$ (all respondents are right). In other words, item facility is the probability of obtaining the right answer for the item~\citep{kline2015handbook}. Of interest for test construction is the variance of an item. An item variance is the calculated variance of a set of item scores, which is a set of zeros and ones. That is, the item variance for an item with facility $0$ is $0$ (all are wrong), and an item variance for an item with facility $1$ is $1$ (all are right). Both these extremes would render the item rather void, as all individuals scoring the same on an item would not tell us anything interesting on the individuals. What usually happens is that some individuals will get the answer right, some will get the answer wrong. For such cases, we can compute the variance for an item by using the formula in \ref{eq:item-var}:

\begin{equation}\label{eq:item-var}
var_{i} = if_{i} * (1 - if_{i})
\end{equation}

where $i$ is the item and $if_{i}$ is the item facility for $i$. The highest possible value for $var_{i}$ is $0.25$, and this is the case when items are neither very easy or very difficult. When $var_{i}$ has small values, for example $0.047$, it means that most respondents tend to reply the same way for that item, making it either extremely easy or extremely difficult. A value of $var_{i}$ near $0.00$ does not warrant automatic exclusion of item $i$, but the value should solicit a review.

\subsubsection*{Person-based tests}

Facility suggests that participants to all tests can be either right or wrong on an answer. What about trait measurement, where participants are not exactly right or wrong but are assessed in terms of a psychological construct? We can calculate item facility for these cases as well. The issue relies only in the naming, because item facility was developed for knowledge-based tests first. Some scholars prefer to use the term item endorsement or item location~\citep{revelle2009} to better reflect how calculations can be done on traits. 

For trait measurement, it is common to have Likert items~\citep{likert1932}. Item facility for Likert items can be calculated with the mean value of the item. If a Likert item maps to the values 0 (strongly disagree) to 5 (strongly agree), the extreme values for the item will be 0 and 5 instead of 0 and 1. An item with average value of 4.8 with variance of 0.09 is a candidate for deletion, whereas an item with average value of 2.72 with variance of 3.02 is deemed interesting, because we prefer items that reflect on diversity of participants rather than having most participants score the same. Items such as those worded negatively should be reversed in their scoring prior to item analysis, so that all items have comparable values.

\subsubsection{Item discrimination} 

Item discrimination is a technique to discover items that behave oddly with respect to what we expect participants to score the item. The meaning of item discrimination differs according to the type of test.

\subsubsection*{Knowledge-based test}

Item discrimination reflects items that behave oddly, in the sense that individuals that tend to score very high (or very low) on a test as a whole tend to be wrong (right) on the same item~\citep{masters1988}. Such an item would possess a negative (positive) discrimination. Ideally, a test should have items with zero discrimination~\citep{kline2015handbook}. 

From a statistical point of view, if an item is uncorrelated with the overall test score, then it is almost certainly uncorrelated with the other items and making very little contribution to the overall variance of the test~\citep{rust2009,singh2016}. Therefore, we calculate item discrimination by comparing the correlation coefficient of an item score and the overall test score. If the computed correlation coefficient is 0 or below, we should consider removing the item.

\subsubsection*{Person-based tests}

What holds for knowledge-based tests holds to a wide extent with person-based tests. Instead of assessing how well an item behaves with respect to the test score, we instead assess how an item is in fact measuring the overall trait in question. By calculating the correlation coefficient of an item and the overall test score for a specific trait, we will have an initial estimate of how well an item represents the trait in question. If the computed correlation coefficient is 0 or below, we should consider removing the item.

\subsubsection*{The case of norm-referenced tests}

The variance of an item, here calculated using the classic definition from statistics, is interesting also in the context of norm-referenced testing. Item facility also applies with norm-referenced testing, as the purpose of the test is to spread out individuals' scores as much as possible on a continuum. A larger spread is due to a larger variance, and we are interested in including items that make a contribution to the variance~\citep{rust2009,kline2015handbook}. Furthermore, if an item has a high correlation to other items and has a large variance, it derives that the item makes high contribution to the total variance of a test and it will be kept in the pool of items.

\subsubsection*{The case of criterion-referenced tests}

Item analysis is often seen as applicable to norm-referenced testing exclusively~\citep{rust2009}. With criterion-referenced testing, it is still possible to calculate item facility and item discrimination, and these can be conducted, for example, before and after teaching and formative activities (and this might include workshops of, say, Scrum methods, at IT companies). A difference in item facility before and after the teaching activities would indicate that the item is a valid measure of the skill taught. This would turn the measure of item facility into a measure of item discrimination as well.

\subsubsection{Limitations of item analysis}

Item analysis, while valuable and still in use today, is part of the classical test theory (CTT), which assumes that an individual’s observed score is the same as a true score plus an error score\citep{traub2005} (see formula~\ref{eq:true-scores}). Modern replacements for CTT have been proposed, and the most prominent one is item response theory (IRT)~\citep{embretson2013}. IRT models build upon a function (called item response function, IRF, or item characteristics curve, ICC) that defines the probability of being right or wrong on an item~\citep{vanalphen1994}. IRT is outside the scope of the present paper as CTT is still in place to this day~\citep{petrillo2015} and explaining IRT requires a publication on its own.

Item analysis, as presented in this section, assumes that there is a single test score, meaning that a single construct is being measured. Whenever multiple constructs or a construct of multiple factors are being measured, item analysis requires to be accompanied by factor analysis~\citep{singh2016}.

\section{Factor Analysis\label{sec:factor-analysis}}

Factor analysis is one of the most widely employed psychometric tools~\citep{kline2015handbook,singh2016,kootstra2006} and it can be applied to any dataset where the number of participants is higher than the number of item scores under observation. Factor analysis is for understanding which test items ``go together'' to form factors in the data that ideally should correspond to the constructs that we are aiming to assess~\citep{rust2009}. At the same time, factor analysis allows to reduce the dimensionality of the problem space (i.e., reducing factors and/or associated items) and explaining the variance in the observed variables compared to underlying latent factors~\citep{kootstra2006}. In case we intend to assess a single construct, factor analysis helps in identifying those items that (best) represent the construct we are interested in, so that we can exclude the other items.

Factor analysis techniques are based on the notion that those constructs that we observe through our measurement instruments can be reduced to fewer latent variables which are unobservable but share a common variance~\citep{yong2013} (see Section~\ref{sec:concepts}). 

Factor analysis starts with computed correlation coefficients as its first building block. A way to summarize correlation coefficients is through a correlation matrix, which is a $n x n$ matrix of $n$ items that displays the correlation coefficient between all $n$ items of the tentative test. Table~\ref{tbl:corr-matrix} provides an example correlation matrix for five items $a, b, c, d,$ and $e$. Given that two items ($a$ and $b$) correlate with each other the same, no matter their order ($cor(a, b) = cor(b, a)$), and that a single item $a$ correlates with itself with a perfect correlation coefficient ($cor(a, a) = 1.00$), the matrix displays only its lower triangle, omitting the repeated upper part, with the value of $1.00$ as its diagonal~\citep{russell2016,yong2013}.

\begin{table}[ht]
\Description{Example correlation matrix for items a, b, c, d, e}
\caption{Example correlation matrix for items a, b, c, d, e}
\label{tbl:corr-matrix}
\begin{tabular}{|l|l|l|l|l|l|}
\hline
\textbf{items} & \textbf{a} & \textbf{b} & \textbf{c} & \textbf{d} & \textbf{e} \\ \hline
\textbf{a}            & 1.00       &            &            &            &            \\ \hline
\textbf{b}     & 0.73       & 1.00       &            &            &            \\ \hline
\textbf{c}     & 0.03       & 0.20       & 1.00       &            &            \\ \hline
\textbf{d}     & 0.89       & 0.84       & -0.18      & 1.00       &            \\ \hline
\textbf{e}     & -0.04      & 0.46       & 0.12       & 0.04       & 1.00       \\ \hline
\end{tabular}
\end{table}

A high correlation among certain items, in our case $a$ with $b$ and $d$, and $b$ with $d$, indicate these items might belong to the same factor. This approach, however, lacks part of the story. Questions such as ``how do our candidate factor explain the total variance of the measurement instrument'', ``to which candidate factor does an item belong \textit{more}?'' and ``how are factors related to each other'' are better answered by further analysis. 

There are two main factor analysis techniques, namely Exploratory Factor Analysis (EFA) and Confirmatory Factor Analysis (CFA)~\citep{singh2016, revelle2009, kootstra2006}. EFA attempts to uncover patterns and clusters of items by exploring predictions, while CFA attempts to confirm hypotheses on existing factors~\citep{yong2013}.

\subsection{Exploratory Factor Analysis}

Exploratory factor analysis (EFA) is a family of analysis techniques aimed to reduce the number of items by retaining the items that are most relevant to certain factors~\citep{kootstra2006}. Strictly speaking, when developing a measurement instrument, after item analysis, it is desirable to observe whether the measurements for the items tend to \textit{cluster}. These clusters are likely to represent different factors that might or might not pertain to the construct being measured~\citep{rust2009,kline2015handbook}. EFA provides tools to group and select items from a correlation matrix.

EFA operates on the equation in~\ref{eq:factor-analysis} for a measure $x_1$~\citep{yong2013,singh2016}:

\begin{equation}\label{eq:factor-analysis}
X_{1} = w_{11}F_{1} + w_{21}F_{2} + .. + w_{n1}F{n} + w_{1}U_{1} + e_{1}
\end{equation}

where $F_{s}$ are those factors grouping the items being analyzed, $U_{s}$ are factors that are unique to each measure, $w_{s}$ are loading of each item on respective factors, and 
$e_{s}$ are random measurement errors. Factor loadings are, in practice, weights that provide us with an idea of how much an item contributes to a factor~\citep{yong2013}.

From the equation we derive that the variance of the constructs being measured is explained by three parts: (1) the common factors, also known as \textit{communality} of a variable~\citep{yong2013,singh2016, fabrigar1999} (2), the influence of factors that are \textit{unique} to that measure, and (3) \textit{random error}, or $V = V_{h} + V_{c} + V_+{e}$.

Estimates for \textit{communalities} of an item are often referred to as $h^{2}$. $h^{2}$ is the calculated proportion of variance that is free of error and is thus shared with other variables in a correlation matrix~\citep{yong2013,singh2016, fabrigar1999}. Several techniques calculate the communality of a variable by summing the squared loadings of each variable associated with a variable.

Estimates for the \textit{unique variance}, denoted as $u^{2}$, is the proportion of variance that is associated with communalities, that is $u^{2} = 1 - h^{2}$~\citep{yong2013,singh2016, fabrigar1999}. Determining a value for $u^{2}$ for an item allows us to find how much specific variance can be attributed to that variable.

Lastly, the \textit{random error} that is associated with an item is the last component of the total variance. Random error is also often called the unreliability of variance~\citep{yong2013, fabrigar1999}.

Unique factors are never correlated with common factors, but common factors may or may not be correlated with each other~\citep{yong2013}.

\begin{figure}
\includegraphics[width=\columnwidth]{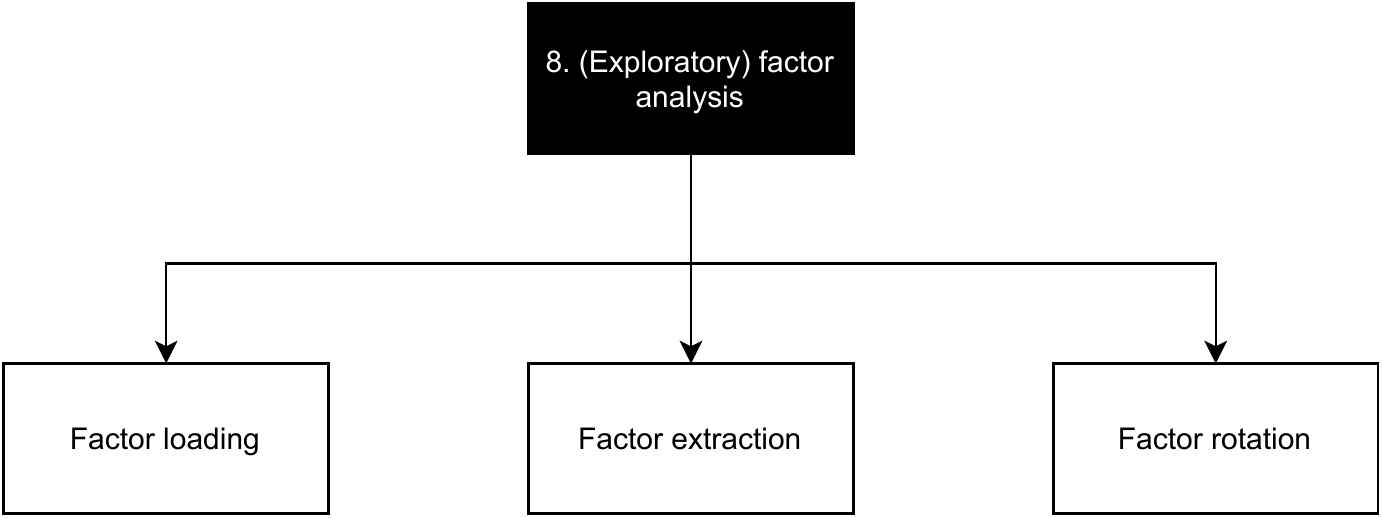}
\caption{Phases for exploratory factor analysis}\label{fig:exploratory-factor-analysis}
\Description{Phases for exploratory factor analysis}\
\end{figure}

EFA encompasses three phases~\citep{rust2009,singh2016, revelle2009, kootstra2006}, described in Figure~\ref{fig:exploratory-factor-analysis}. First, we have to select a fitting procedure to estimate the factor loadings and unique variances of the model. Then, we need to define and extract a number of factors. Finally, we need to rotate the factors to be able to properly interpret the produced factor matrices.

Many statistical programs allow to either perform all these phases separately or to perform more than one at the same time. It is not an easy task to assign a methodology to one of the three categories below. The reader is advised that some textbooks avoid our classification of phases and simply revert to a more practical set of questions, e.g., ``how to calculate factors'' and ``how many factors should we retain''. We also note that recent studies have formulated Bayesian versions of these classical exploratory factor analysis techniques and claimed several benefits~\cite{conti2014bayesian}.

\subsubsection{Factor loading}

The most common technique for estimating the loadings and variance is to use the standard statistical technique of principal component analysis (PCA)~\citep{pearson1901}. PCA assumes that the communalities for the measures are equal to 1.0. That is, all the variance for a measure is explained only by the factors that are derived by the methodology, and hence there is no error variance.
PCA operates on the correlation matrix, mostly on its eigenvalues, to extract factors that correlate with each other. The eigenvalue of a factor represents the variance of the variables accounted for by that factor. The lower the eigenvalue, the least the factor contributes to the variance explanation in the variables~\citep{Norris2010}. 
Factor weights are computed to extract the maximum possible variance, with successive factoring continuing until there is no further meaningful variance left. PCA is not a factor analysis method \textit{strictu sensu}, as factor analysis  assumes a presence of error variance rather than being able to explain all variance. Some advocates prefer to state that its output should be referred to as a series of components rather than factors. While more simplistic than other techniques to estimate factor loading, performing a PCA is still encouraged as a first step in EFA before performing the actual factor analysis~\citep{russell2016,yong2013,rust2009}.

Among the proper factor analysis techniques that exist, we are interested in a widely recommended technique for estimating loadings and variance named principal axis factoring (PAF)~\citep{russell2016, widaman1993,kline2015handbook}. PAF does not operate under the assumption that the communalities are equal to 1.0, so the diagonal of the correlation matrix (e.g., the one in Table~\ref{tbl:corr-matrix}) is substituted with estimations of communalities, $h^{2}$. PAF estimates the communalities using different techniques (e.g., the squared multiple correlation between a measure and the rest of measures) and a covariance matrix of the items. Factors are estimated one at a time up until there is a large enough variance accounted for in the correlation matrix. Under PAF, the ordering of the factors determines their importance in terms of fitting, e.g., the first factor accounts for as much variance as possible, followed by the second factor, and so on. ~\citet{russell2016} provides a detailed description of the underlying statistical operations of PAF, that we will omit for the sake of brevity. Most statistical software provides functions to implement PAF.

\subsubsection{Factor extraction\label{ssec:factor-extraction}}

Both PCA and PAF result in values assigned to candidate factors. Therefore, there has to be a strategy to extract meaningful factors. The bad news here is that there is no unique way, let alone a single proper way, to extract factors~\citep{russell2016, singh2016, revelle2018}. More than one strategy should be adopted in parallel to allow a comparison of results, ending with a sense-making analysis to take an ultimate decision~\citep{fabrigar1999}. Several factor extraction techniques exist, which are mentioned in the cited references in the present section. We provide here those that are used most widely as well as those that are easier to apply and understand. Our running example (Section
~\ref{sec:example}) enables an easier visualization of the concepts below.

Perhaps the simplest strategy to extract factors is~\citeauthor{kaiser1960}'s eigenvalue-greater-than-one (K1) rule~\citep{kaiser1960}. The rule simply states that factors with eigenvalue \textit{higher} than 1.0 should be retained. Kaiser's rule is quite easy to apply but it is highly controversial~\citep{russell2016,revelle2018, fabrigar1999, courtney2013}. First, the rule was originally designed for PCA and not for PAF or other factor analysis methods, which might make it unsuitable for methodologies that provide estimation for commonalities as diagonals for correlation matrices~\citep{courtney2013}. Second, the cut-off value for 1.0 might discriminate for factors that are just above or just below 1.0~\citep{courtney2013}. Third, computer simulations found that K1 tends to overestimate the number of factors~\citep{fabrigar1999}. Yet, K1 is still the default option for some statistical software suites, making it an unfortunate de-facto main method for factor extraction~\citep{courtney2013}.

~\citet{cattel1966} scree test, also based on eigenvalues, foresees the plot of the eigenvalues extracted from either the correlation matrix or the reduced correlation matrix (thus making it suitable for both PCA and PAF) against the factor they are associated with, in descending order. One then inspects the curved line for a break in the values (or an elbow) up to when a substantial drop in the eigenvalues cannot be observed anymore. The break is a point at which the shape of the curve becomes horizontal. The strategy is then to keep all factors before the breaking point. The three major criticisms of this approach is that it is subjective~\citep{courtney2013}, that more than one scree might exist~\citep{tinsley1987}, and that data often does not offer a discernible scree and a conceptual analysis of the candidate factor is thus always required~\citep{rust2009}.

\citet{revelle1979} proposed the Very Simple Structure (VSS) method for factor extraction that is based on assessing how the original correlation matrix can be reproduced by a simplified pattern matrix, for which only the highest loading of each item is retained (everything else set to zero)~\citep{courtney2013}. The VSS criterion to assess how well the pattern matrix performs is a number from 0 to 1, making it a goodness-of-fit measure almost of a confirmatory nature rather than an exploratory one~\citep{revelle2020}. The VSS criterion is gathered for solutions involving a number of factors that goes from 1 to a user-specified value. The strategy ends with selecting the number of factors that provide the highest VSS criterion.

Finally, the method of parallel analysis (PA), introduced by~\citet{horn1965}, was found to be very robust for factor extraction~\citep{courtney2013, fabrigar1999}. PA starts with the K1 concept that only factors with eigenvalue larger than 1.0 should be kept.~\citet{horn1965} has argued that the K1 rule was developed with population statistics and was thus unsuitable when sampling data. Sampling errors would then cause some components from uncorrelated variables to have eigenvalues higher than one in the population~\citep{courtney2013}. PA takes into account the proportion of variance that results from sampling rather than being able to access to the population. The way it achieves this is a constant comparison of the solution with randomly generated data~\citep{revelle2020}. PA generates a large number of matrices from random data in parallel with the real data. Factors are retained as long as they are greater than the mean eigenvalue generated from the random matrices.

\subsubsection{Factor Rotation}

The last step is to rotate the factors in the dimensional space for improving our interpretation of the results~\citep{darton1980,kootstra2006,kline2015handbook}. 

An unrotated output, that is the one that often results after factor extraction, maximizes the variance accounted for by the first factor, followed by the second factor, the third factor, and so on. That is, most items would load on the first factors and many of them would load on more than one factor in a substantial way.

Rotating factors build on the concept that there are a number of ``factor solutions'' that are mathematically equivalent to the solution found after factor extraction. By performing a rotation of the factors, we retain our solution but allow an easier interpretation. We rotate the factors to seek a \textit{simple structure}. A simple structure is a loadings pattern such that each item loads strongly on one factor only and weakly on other factors. If the reader is interested in mathematical foundations of factor rotation, two deep overviews of factor rotation are offered by \citet{darton1980, browne2001}.

There are two families of rotations, namely orthogonal and oblique~\citep{russell2016}. Orthogonal rotations force the assumption of independency between the factors, whereas oblique rotations allows the factors to correlate with each others. Which methodology to use is influenced by the statistics software; for example, R psych package~\citep{revelle2017} provides ``varimax'', ``quartimax'', ``bentlerT'', ``equamax'', ``varimin'', ``geominT'', and``bifactor'' for orthogonal rotations and ``promax'', ``oblimin'', ``sim- plimax'', ``bentlerQ, ``geominQ'', ``biquartimin'', and``cluster'' for oblique rotations. Several rotation methodologies are summarized by~\citet{browne2001, herve2003, russell2016}.

Perhaps the most known and employed~\citep{darton1980} orthogonal rotation method is the Varimax rotation~\citep{kaiser1958}. Varimax maximizes the variance (hence the name) of the squared loadings of a factor on all variables. Each factor will tend to have either large or small loadings of any particular variable. While this solution makes it rather easy to identify factors and their loading on items, the independency condition of orthogonal rotation techniques is hard to achieve. The assumption of independency of factors, especially in the context of behavioral research, belittles the value of orthogonal rotation techniques, to the point that ``we see little justification for using orthogonal rotation as a general approach to achieving solutions with simple structure''~\citep[p.~283]{fabrigar1999}. 

Oblique rotation is preferred for behavioral software engineering studies, as it is sensible to assume that behavioral, cognitive, and affective factors are separated by soft walls of independence (e.g., motivation and job satisfaction)~\citep{rust2009, fabrigar1999, russell2016}. If any, one would have to first conduct an investigation using oblique rotation and observe if the solution shows little to no correlation between factors and, in that case, switch to orthogonal rotation~\citep{fabrigar1999}. The two most employed and recommended oblique rotation techniques are Direct Oblimin (and its slight variation Direct Quartimin) and Promax, both of which perform well~\citep{fabrigar1999}.

~\citet{fabrigar1999,russell2016} recommended to use a Promax rotation because it provides the best of both approaches. A Promax rotation first performs an orthogonal rotation (a Varimax rotation) to maximize the variance of the loadings of a factor on all the variables~\citep{russell2016}. Then, Promax relaxes the constraint that the factors are independent from each other, turning the rotation to oblique. The advantage of this technique is that it will reveal whether factors really are uncorrelated with each other~\citep{russell2016}.

\subsubsection{Further recommendations}

There have been several recommendations regarding the required sample size, number of measures per factor, number of factors to retain, and interpretation of loadings~\citep{russell2016, singh2016, yong2013}. 

The recommended overall sample size as reported by~\citet{yong2013} is at least 300 participants, with each variable that is subject to factor analysis with at least 5 to 10 observations. This recommendation has, however, low empirical validation. As reported by~\citet{russell2016}, a Monte Carlo study by~\citet{maccallum1999} analyzed how different sample sizes and communalities of the variables were able to reproduce the population factor loadings. They found that with item communalities higher or equal to 0.60, results were very consistent with sample sizes as low as 60 cases. Communality levels around 0.50 required samples of 100 to 200 cases. In their review,~\citet{russell2016} also found that 39\% of EFA studies involved samples of 100 or fewer cases.

On the number of measures (items) per factor,~\citet{yong2013} report that for something to be labeled as a factor it should have at least 3 variables, especially in cases when factors receive a rotation treatment, where only a high correlation (coefficient higher than 0.70) with each other and mostly uncorrelation with other items would make them worthy of consideration. Generally speaking, the correlation coefficient for an item to belong to a factor should be 0.30 or higher~\citep{tabachnick2007}.~\citet{russell2016} identifies that prior work has requested at least three items per factor; however, four or more items per factor was found to be a better holistic way to ensure an adequate identification of the factors. In his review he identified 25\% studies with three or fewer measures per factor.

We reported on the number of factors to retain in section~\ref{ssec:factor-extraction}, so we will not repeat ourselves here. There is not a recommended number and one would follow (possibly more than) one extraction method to identify the best number of factors according to the case. ~\citet{tabachnick2007} add that cases with missing values should be deleted to prevent overestimation of factors.~\citet{russell2016} wrote something that is worthy of mentioning for the uninitiated behavioral software engineering researchers, that is, even when constructing a new measurement instrument there is already an expectation of possible factors in the mind of the researcher. The reason is that items are developed following an investigation of prior work and/or empirical data (see section~\ref{sec:item-analysis}). That number is a good starting point to base ourselves when conducting EFA. 

~\citet{yong2013} spends some further explanations on interpretation of loadings when they are produced by a statistical software. There should be few item crossloadings (i.e., split loadings, when an item loads at 0.32 or higher with two or more factors) so that each factor defines a distinct cluster of interrelated variables. There are exceptions to this that require an analysis of the items. Sometimes it is useful to retain an item that crossloads, with the assumption that it is the latent nature of the variable. Furthermore,~\citet{tabachnick2007} report that, with an alpha level of 0.01, a rotated factor loading with a meaningful sample size would need to have a value of at least 0.32 for loadings as this would correspond to approximately 10\% of the overlapping variance.

As a final note, the reader might now be left wondering, why conducting item analysis to reduce items if factor analysis is available?~\citet{kline2015handbook} argues that the so many phases of factor analysis foresee assumptions of data distribution and decisions on techniques, that much could go wrong with it. Furthermore, items selected by item analysis are highly correlated to items selected by factor analysis~\cite{nunnally1978}, making item analysis a cheaper, effective \textit{initial} method for item reduction.
 
\subsection{Confirmatory Factor Analysis}

Contrary to exploratory factor analysis, confirmatory factor analysis (CFA) is for confirming a-priori hypotheses and assessing how well an hypothesized factor structure fits the obtained data~\citep{russell2016}. A hypothesized factor could be derived from existing literature as well as data from a previous study to explore the factor structure.

Once the data is obtained to compare to the hypothesized factor structure, a goodness-of-fit test should be conducted. CFA requires statistical modeling that is outside the scope of this paper and the estimation of the goodness-of-fit in CFA is a long lasting debate as ``there are perhaps as many fit statistics as there are psychometricians'' (~\citep[p. 31]{revelle2018a}). ~\citet{russell2016, singh2016, rust2009} provide several techniques for estimating the goodness-of-fit in CFA, e.g., Chi-squared test, root mean square residual (RMSR), root mean square error of approximation (RMSEA), and the  standardized RMSR. Statistical software implement these techniques, including R psych package~\citep{revelle2020}. A widely employed technique for CFA is to be found in structural equation modeling (SEM), which is a family of models to fit networks of constructs~\citep{kaplan2008structural}.~\citet{maccallum2000} provided a comprehensive review of SEM techniques in the psychological science including their applications and pitfalls.

Conducting both EFA and CFA is very expensive. When designing and validating a measurement instrument, and when obtaining a large enough sample of participants, it is common to split the sample for conducting EFA on a part of it and CFA on the remaining part~\citep{singh2016}. Most authors, however, prefer conducting EFA only~\citep{russell2016} and rely on future independent studies towards a better psychometrics evaluation of a tool. This is also why statistical tools, e.g., R psych package~\citep{revelle2017} provide estimates of fits for EFA as well as convenience tools to adapt the data to CFA packages, e.g., R sem~\citep{fox2017}.

We refer the reader to a prior work of ours in the behavioral software engineering domain~\citep{lenberg2017b} where we conducted a CFA and describe its application. We also note that, like for EFA, there have also been Bayesian methods for CFA proposed~\cite{lee1981,ansari2002heterogeneous}.

\section{Statistical properties of items\label{sec:standardization}}

Assessing characteristics and performance of individuals poses several challenges when interpreting the resulting scores.
One of them is that a raw score is not meaningful without understanding the test standardization characteristics~\citep{kline2015handbook}.
For example, a score of 38 on a debugging performance test is meaningless without knowing that 38 means to be able to open a debugger only. Furthermore, the interpretation of the results vary wildly when knowing that on average, developers score 400 on the test itself compared to if they score 42. The former issue is related to criterion-referenced standardization, the latter to norm-referenced standardization~\citep{rust2009,kline2015handbook}.

Criterion-referenced tests assess what an individual with any score is expected to be able to do or know. Norm-referenced standardization enables to compare an individual's score to the ordered ranking of a population (also see section~\ref{sec:concepts}). 
We concentrate on norm-referenced standardization as criterion-referenced standardization is unique to a test criteria.

A first step to norm-reference a test is to order the results of all participants and rank an individual's score.
Measures such as median and percentiles are useful for achieving the ranking and compare. 
When we can treat our data as interval scales and have it approximately following a normal distribution, 
we can also use the mean and the standard deviation. 
The standard deviation is useful for telling us how much an individual's score is above or below the mean score.
Instead of reporting that an individual's score is, e.g., 13 above the mean score, it is more interesting to know that the score is 1.7 standard deviations above the mean score.
Hence, we norm-standardize scores using different approaches. The remaining of this section is modeled after~\citet{rust2009} text and augmented by further explanations from other sources.

\subsection{Standardization to Z scores}

Whenever a sample approximates a normal distribution, we know that a score above average is in the upper 50\% and by following the three sigma empirical rule~\citep{pukelsheim1994}, 
we know that a score greater than one or two standard deviations from the mean is in the top 68\% and 95\% respectively.
For expressing an individual's score in terms of how distant it is from the mean score, we transform the value to its Z score (also called standard score) using the formula in~\ref{eq:z-score}:

\begin{equation}\label{eq:z-score}
Z_{score}(x_{pc})=\frac{x_{pc}-\bar{x}_{pc}}{{s}_{pc}}
\end{equation}

where $x_{pc}$ is a participant's score,
$\bar{x}_{pc}$ is the mean of all participants' scores, and $s_{pc}$ is
the standard deviation for all participants' scores.

The ideal case would be to use the population mean and standard deviation. 
In software engineering research we lack studies estimating population characteristics (an example of norm studies was provided by ~\citep{graziotin2017unhappy}), so we should either aggregate the results of some studies or gather more samples.

An important note is that transforming scores into a Z scores does not make the scores normally distributed. This would require a normalization procedure, explained below. 

\subsection{Standardization to T scores}

Z scores typically range between -3.00 and +3.00. The range is not always suitable for its application. A software developer could, for example, object to a Z score of -0.89 which, at first glance, might suggest to be low (value) or negative (sign).

A T score, not to be confused with t-statistics of the Student's t-test, is a standard Z score that is scaled and shifted so that it has a mean of 50 and a standard deviation of 10. T scores thus typically range between 20 and 80.

\begin{equation}\label{eq:t-scores}
T_{score}(Z_{pc})= Z_{pc} * 10 + 50
\end{equation}

For transforming a Z score into a T score, we use the formula in \ref{eq:t-scores}. 

The software developer in the previous example would have a T score of $41.1$ from a Z score of $-0.89$.

\subsection{Standardization to stanine and sten scores}

Stanine and sten scores respond to the need of transforming a score to a scale from 1 to 9 (stanine) or 10 (sten) with a mean of 5 (stanine) or 5.5 (sten) and a standard deviation of 2. These scores purposely lose precision by keeping only decimal values.

\begin{table}[ht]
\Description{Conversion from Z scores to Stanine and Sten scores}
\caption {Conversion from Z scores to Stanine and Sten scores} \label{tbl:stanine-sten} 
\begin{tabular}{|l|l|l|l|}
\hline
Stanine score & Z score        & Sten score & Z score        \\ \hline
1             & \textless -1.75        & 1          & \textless -2.00        \\ \hline
2             & -1.75 to -1.25 & 2          & -2.00 to -1.50 \\ \hline
3             & -1.25 to -0.75  & 3          & -1.50 to -1.00 \\ \hline
4             & -0.75 to -0.25 & 4          & -1.00 to -0.50 \\ \hline
5             & -0.25 to +0.25 & 5          & -0.50 to -0.00 \\ \hline
6             & +0.25 to +0.75 & 6          & +0.00 to +0.50 \\ \hline
7             & +0.75 to +1.25 & 7          & +0.50 to +1.00 \\ \hline
8             & +1.25 to +1.75 & 8          & +1.00 to +1.50 \\ \hline
9             & \textgreater +1.75        & 9          & +1.50 to +2.00 \\ \hline
              &                & 10         & \textgreater +2.00        \\ \hline
\end{tabular}
\end{table}
The conversion to stanine and sten scores follows the rules in Table~\ref{tbl:stanine-sten}.

The advantage of stanine and sten scores lies in their imprecision. If our non performing developer with a Z score of -0.89 was compared with other two developers having scores of -0.72 and -0.94, how meaningful would be such a tiny difference in scores? Their stanine scores are 3, 4, and 4, respectively. Their sten score would be 4. Stanine and sten scores provide clear cut-off points for easier comparisons.

There is an important difference between stanine and sten scores, besides their range. A stanine score of 5 represents an average score in a sample. An average sten sore can not be obtained, because the value of 5.5 does not belong to its possible values. A score of 4 represents the low average band (which ranges from 4.5 to 5.5, that is one standard deviation below the man), and a score of 6 represents the high average band (which ranges from 5.5 to 6.5, that is one standard deviation above the mean).

\subsection{Normalization}

The standardization techniques that we presented in the previous section carry the assumption that the sample and population approximate the normal distribution. For all other cases, it is possible to normalize the data. Examples include algebraic transformation, e.g., square-rooting or log transformation, as well as graphical transformation. See introductory statistical texts for more detailed explanations of and a broad set of such transformations.

\section{Reliability\label{sec:reliability}}

Reliability can be seen either in terms of precision, that is the consistency of test scores across replications of the testing procedure (reliability/precision), or as a coefficient, that is the correlation between scores on two equivalent forms of the test (reliability/coefficients)~\citep{apa2014}).
For evaluating the precision of a measurement instrument, it would be ideal to have as many independent replications of the testing procedure as possible on the same very large sample. Scores are expected to generalize over alternative tests forms, contexts, raters, and over time. The reliability/precision of a measurement instrument is then assessed through the range of differences of the obtained scores. The reliability/precision of an instrument should be assessed with as many sub-groups of a population as possible.

The reliability/coefficients of a measurement instrument, which we will simply call reliability from this point on, is the most common way to refer to the reliability of a test~\citep{apa2014}. There are three categories of reliability coefficients, namely alternate-form (derived by administering alternative forms of test), test-retest (same test on different times), and internal-consistency (relationship among scores derived from individual test items during a single session).

\begin{figure}
\includegraphics[width=\columnwidth]{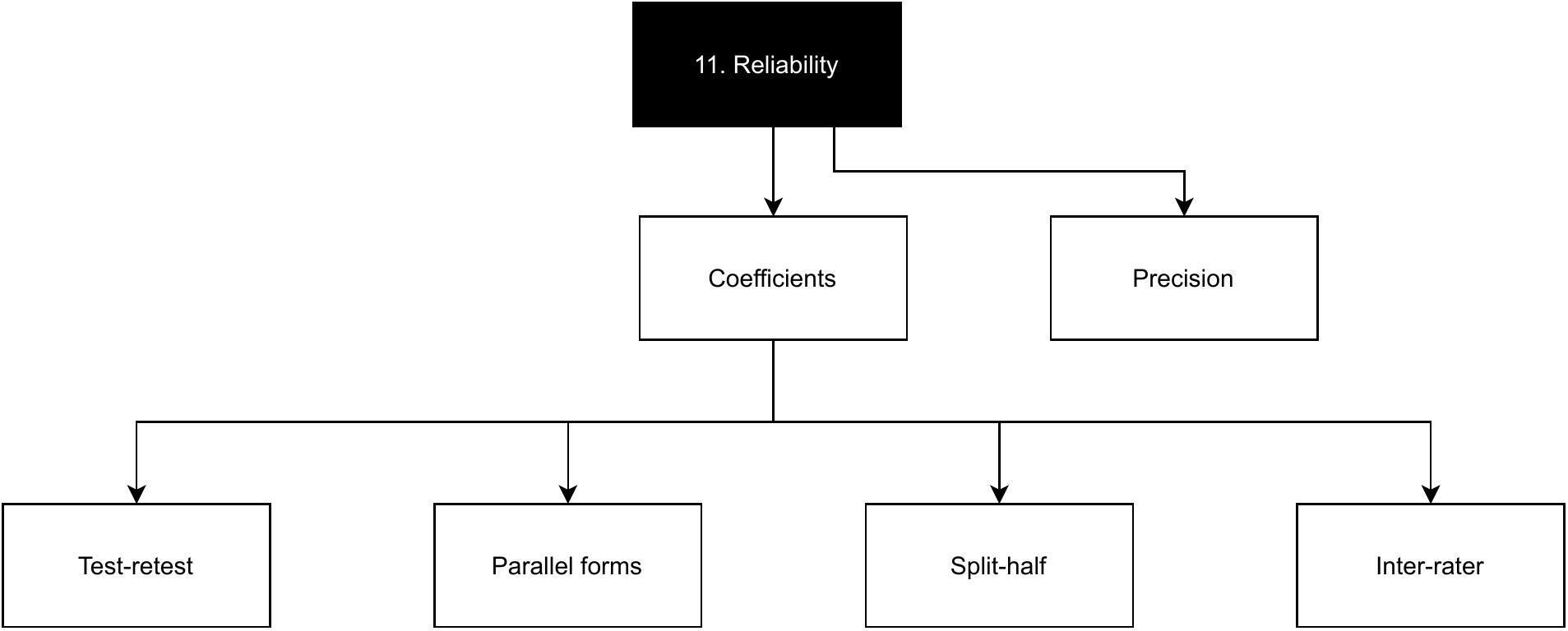}
\Description{Reliability in psychometric theory}
\caption{Reliability in psychometric theory}\label{fig:reliability}
\end{figure}

We adhere to~\citet{rust2009,nunnally1994psychometric} classification of reliability and provide a brief overview of reliability facets in psychometric theory in Figure~\ref{fig:reliability}.

Several factors, as defined in the  SEPT~\citep{apa2014}, affect the reliability of a measurement instrument, especially adding or removing items, changing wording or intervals of items, causing variations in the constructs to be measured (e.g., using a measurement instrument for happiness to assess job satisfaction of developers), and ignoring variations in the intended populations for which the test was originally developed.

We now introduce the most widely employed techniques for establishing the reliability of a test.

\subsection{Test-retest reliability}

Test-retest reliability, also known as test stability, is assessed when administering the measurement instrument twice to the same sample within a short interval of time. The paired set of scores for each participant is then compared with a correlation coefficient such as Pearson product-moment correlation coefficient or Spearman's rank-order correlation. A correlation coefficient of 1.00, while rare, would indicate a perfect test-retest reliability, whereas a correlation coefficient of 0.00 would indicate no test-retest reliability at all. A negative score is no good news either, and it is automatically considered as a value of 0.00. 

\subsection{Parallel forms reliability}

Test-retest reliability is not suitable for certain tests, such as those assessing knowledge or performance in general. Participants either face a learning or motivation effect from the first test session or simply improve (or worsen) their skills between sessions. Fur such cases, the parallel forms method is more suitable. The technique requires a systematic development of two versions of the same measurement instrument, namely two parallel tests, that are assessing the same construct but using different wording or items. Parallel tests for assessing debugging skills would feature the same sections and amount of items, e.g., arithmetic, logic, and syntax errors. The two tests would need different source code snippets that are, however, very similar. A trivial example would be to test for unwanted assignments inside conditions in different places and with different syntax (e.g., using \textit{if (n = foo())} in version one and \textit{if (x = y + 2)} in version two). As with test-retest reliability, each participant faces both tests and a correlation coefficient can be computed.

\subsection{Split-half reliability}

Split-half reliability is a widely adopted and more convenient alternative to parallel forms reliability. Under this technique, a measurement instrument is split into two half-size versions. The split should be as random as possible, e.g., splitting by taking odd and even numbered items. Participants face both halves of the test and, again, a correlation coefficient can be computed. The obtained coefficient, however, is not a measure of reliability yet. The reliability of the whole measurement instrument is computed with the Spearman-Brown formula in \ref{eq:spearman-brown}.

\begin{equation}\label{eq:spearman-brown}
R_{instrument}=2 * \frac{r_{half}}{1 + {r}_{half}}
\end{equation}

where $r_{half}$ is the correlation of the split tests. 
This formula shows that the more discriminating items a test has, the higher will be its reliability.

\subsection{Inter-rater reliability}

Inter-rater reliability is perhaps the most common reliability that is found in software engineering studies. Qualitative studies or systematic literature reviews and mapping studies often have different raters for evaluating the same items. The sets of rates can be assessed using a correlation coefficient. Cohen's $k$ is widely used in the literature for inter-rater coefficient of two entities together with Fleiss' $k$ for the inter-rater coefficient of more than two entities. Cases have also been made for using Krippendorff's $\alpha$~\cite{de2012calculating}.

\subsection{Standard error of measurement}

The standard error of measurement is used for generating confidence intervals around reported scores. The score is strictly related to the reliability coefficient~\citep{rust2009} as shown in formula~\ref{eq:stderr-measurement}

\begin{equation}\label{eq:stderr-measurement}
E_{measurement}= var_{test} * (1 - r_{test})
\end{equation}

where $var_{test}$ is the variance of the test scores and $r_{test}$ is the calculated reliability coefficient of the test. The standard error of measurement also provides an idea of how errors are distributed around observed scores. The standard error of measurement is maximized---and becomes equal to the standard deviation of the observed scores---when a test is completely unreliable. The standard error of measurement is minimized to zero when a test is perfectly reliable.

If the assumption that errors are distributed normally is met, one can calculate the 95\% confidence interval by using the z curve value of 1.96 to construct the interval $x_{obs} \pm E_{measurement} * 1.96$. Confidence intervals could also be used to compare participants' scores. Should one participant score fall below or above the interval, their results would differ significantly from the normality of scores.

\section{Validity\label{sec:validity}}
Validity in psychometrics is defined as ``The degree to which evidence and theory support the interpretation of test scores for proposed uses of tests.''~\citep{apa2014}.
Psychometric validity is therefore a different (but related) concept than the one of study validity that software engineers are used to deal with~\citep{wohlin2012,feldt2010,siegmund2015,petersen2013}.
Validation in psychometric research is related to the interpretation of the test scores. 
For validating a test, 
we should gather relevant evidence for providing a sound scientific basis for the interpretation of the proposed scores.

\begin{figure}
\includegraphics[width=\columnwidth]{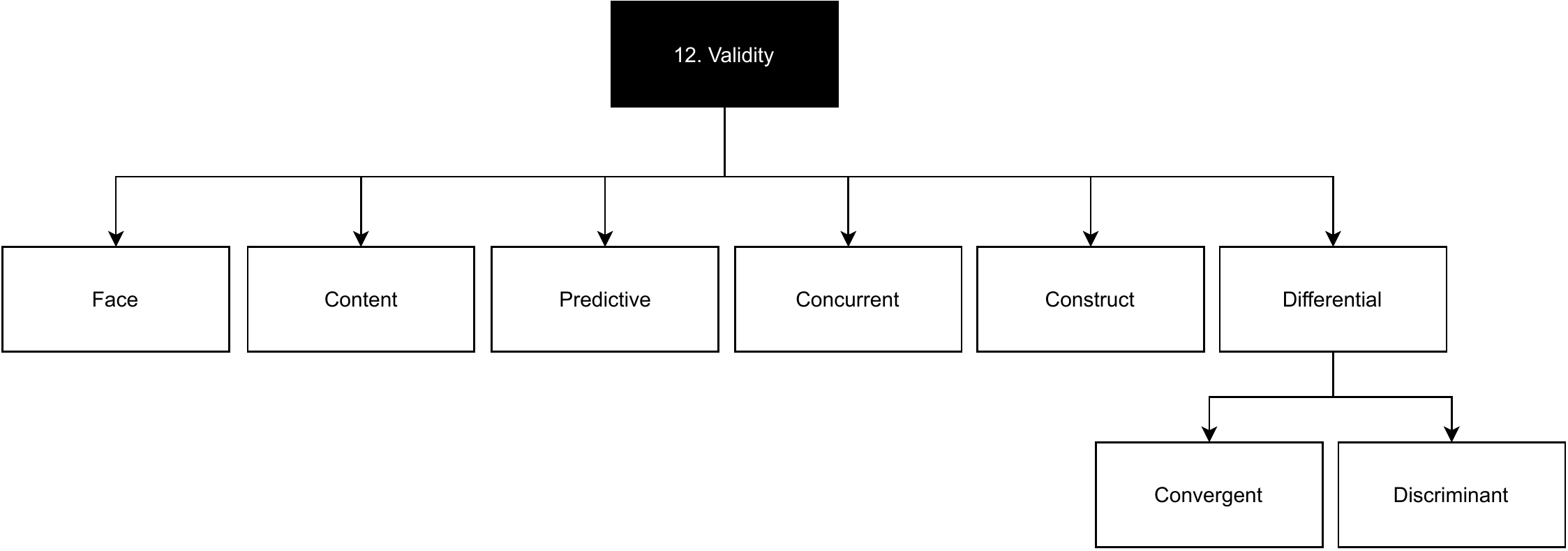}
\Description{Validity in psychometric theory}
\caption{Validity in psychometric theory}\label{fig:validity}
\end{figure}

\citet{rust2009,kline2015handbook} have summarized six major facets of validity in the context of psychometric tests, which we summarize and represent in Figure~\ref{fig:validity} and describe below, augmented with references to material that offers additional explanations.

\subsection{Face validity}
Face validity concerns how the items of a measurement instrument are accepted by respondents.
For example, software developers expect the wording of certain items to be targeted to them instead of say, children.
Similarly, if a test presents itself to be about a certain construct, such as debugging expertise, it could cause face
validity issues if it contained a personality assessment.

\subsection{Content validity}\label{ssec:content-validity}
Content validity (sometimes called criterion validity or domain-referenced validity) 
concerns the extent to which a measurement instrument reflects the purpose for which the instrument is being developed.
If a test was developed under the specifications of job satisfaction, but measured developers' motivation instead, 
it would present issues of content validity.
Content validity is evaluated qualitatively most of the times because the form of deviation matters more
than the degree of deviation, but there are proposals for its quantitative estimation~\cite{wynd2003}.

\subsection{Predictive validity}
Predictive validity is a statistical validity defined as the correlation between the score of a measurement instrument 
and a score of the degree of success of the selected field. 
For example, the degree of success of debugging performance capability is expected to be higher with a higher programming experience.
Computing a score for predictive validity is as simple as calculating a correlation value (such as Pearson or Spearman).
According to the acceptance criterion for predictive validity, a score higher than 0.5 could be considered an acceptable
predictive validity for the items.
We would then feel justified in including programming experience as an item to represent the construct of debugging 
performance capability.

\subsection{Concurrent validity}
Concurrent validity is a statistical validity that is defined as the correlation of a new measurement instrument
and existing measurement instruments for the same construct. 
A measurement instrument tailored to the personality of software developers ought to correlate with existing personality
measurement instruments.
While concurrent validity is a common measure for test validity in psychology, it is a weak criterion as the old measurement
instrument itself could have a low validity.
Nevertheless, concurrent validity is important for detecting low validity issues in measurement instruments.

\subsection{Construct validity}
Construct validity is a major validity criterion in psychometric tests. 
As constructs are not directly measurable, we observe the relationship between the test and the phenomena that the test
attempts to represent.
For example, a test that identifies highly communicative team members should have a high correlation with\ldots 
observations of highly communicative people who have been labelled as such. 
The nature of construct validity is that it is cumulative over the number of available studies~\citep{schmidt1992}.

\subsection{Differential validity}
Differential validity assesses how a measurement instrument correlates with measures from which it should differ, and how a measurement instrument correlates with measures from which it should not differ. In particular, 
\citet{campbell1959} have differentiated between two aspects of differential validity, namely convergent and discriminant validity. ~\citet{rust2009} mentions a straightforward example of both. A test of mathematics reasoning should correlate positively with a test of numerical reasoning (convergent validity). However, the mathematics test should not strongly correlate positively with a test of reading comprehension, because the two constructs are supposed to be different (discriminant validity). In case of a low discriminant validity, there should be an investigation of whether the correlation is a result of a wider underlying trait, say, general intelligence. Differential validity is overall empirically demonstrated by a discrepancy between convergent validity and discriminant validity.

\section{Fairness in testing and test bias\label{sec:fairness}}

Fairness is ``the quality of treating people equally or in a way that is right or reasonable'' (\citet{caefairness2018}, online.). A test is fair when it reflects the same constructs for all participants, and its scores have the same meaning for all individuals of the population~\citep{apa2014}. A fair test does neither advantage or disadvantage any participant through characteristics that are not related to the constructs under observation. From a participant point of view, an unfair test brings a wrong decision based on the test results. An example of a test that requires fairness is an attitude or skills assessment when interviewing candidates for hire in an information technology company.

The SEPT~\citep{apa2014} reports on several facets of fairness. Individuals should have the opportunity to maximize how they perform with respect to the constructs being assessed. Similarly, for a measurement instrument that assesses traits of participants, the test should maximize how it assesses that the constructs being measured are present among individuals. This fairness comes from how the test is administered, which should be as standardized as possible. Research articles should describe the environment for the experimental settings, how the participants were instructed, which time limits were given, and so on. Fairness also comes, on the other hand, from participants themselves. Participants should be able to access the constructs as being measured without being advantaged or disadvantaged by individual characteristics. This is an issue of accessibility to a test and is also part of limiting item, test, and measurement bias.

\begin{figure}
\includegraphics[width=\columnwidth]{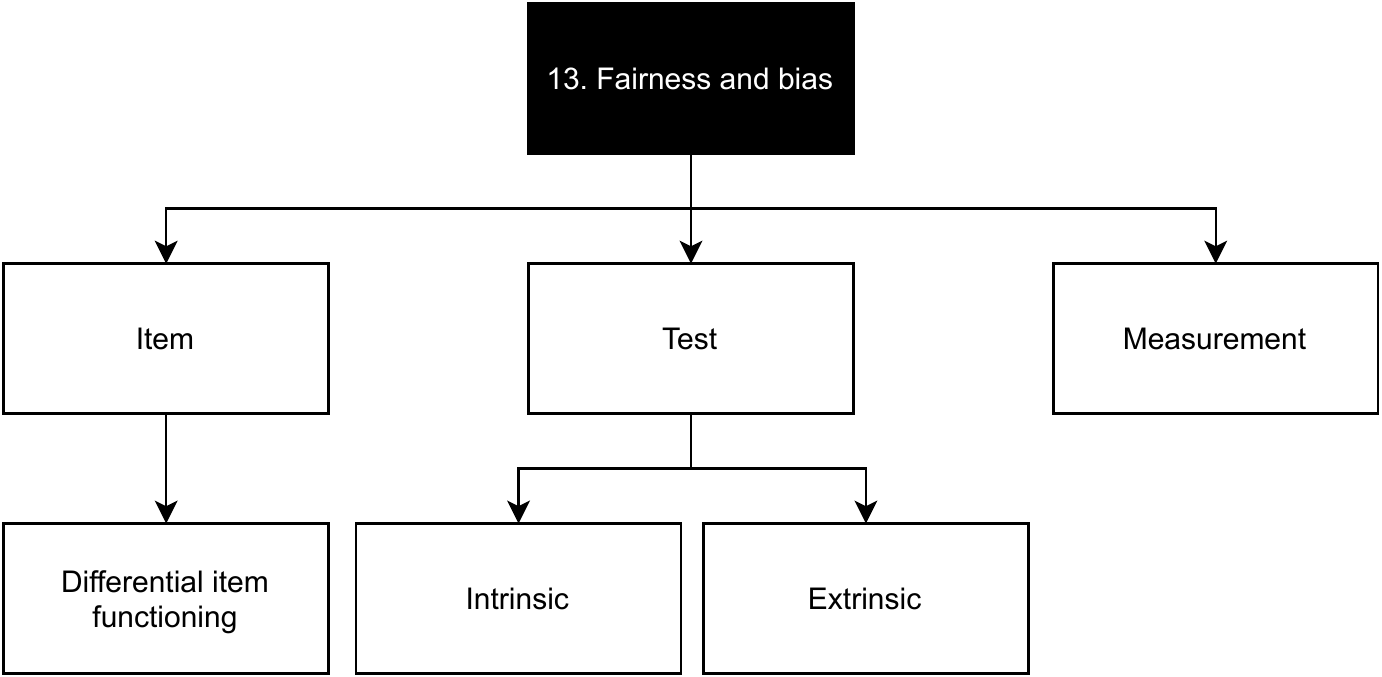}
\Description{Bias in psychometric theory}
\caption{Bias in psychometric theory}\label{fig:bias}
\end{figure}

We provide an overview or bias in psychometric theory in Figure~\ref{fig:bias}.

~\citet{rust2009} provides an overview of item, test, and measurement bias, which we supplement with related work. It almost feels unnecessary to state that a measurement instrument should be free from bias regarding age, sex, gender, and race. These cases are indeed covered by legislation to ensure fairness. In general, there are three forms of bias in tests, namely item bias, intrinsic test bias, and extrinsic test bias~\citep{rust2009}.

\subsection{Item bias}

Item bias, also known as differential item functioning, refers to bias born out of individual items of the measurement instrument. A straightforward example would be to test a (non-UK) European developer about coding snippets dealing with imperial system units. A more common item bias is about the wording of items. Even among native speakers, the use of idioms such as double negatives can cause confusion. Asking a developer to mark a coding snippet that \textit{is free from logic and syntax errors} is clearer than asking to mark code that \textit{does not possess neither logic nor syntax errors}.

A systematic identification of item bias that goes beyond carefully checking an instrument is to carry out an item analysis with all possible groups of potential participants, for example men and women, or speakers of English on different levels. A comparison of the facility values (the proportion of correct answers) of each item can reveal potential item bias. For instruments that assess traits and characteristics of a group instead of function or skills, a strategy is to follow a checklist of questions that researchers and pilot participants can answer~\citep{hambleton1995item}.

Differential item functioning (DIF) is a statistical characteristic of an item that shows potential unfairness of the item among different groups that should provide same test results otherwise~\citep{perrone2006}. A presence of DIF does not necessarily indicate bias but unexpected behavior on an item~\citep{apa2014}. This is why, after the detection of DIF, it is important to review the root causes of the differences. Whenever DIF happens for many items of a test, a test construct or final score is potentially unfair among different groups that should provide same test results otherwise. This situation is called differential test functioning (DTF)~\citep{runnels2013}.
There are three main techniques for identifying DIF, namely Mantel-Haenszel approach, item response theory (IRT) methods, and logistic regression~\citep{zumbo2007}.

\subsection{Intrinsic test bias} 

Intrinsic test bias occurs when there are differences in the mean scores of two groups that are due to the characteristics of the test itself rather than difference between the groups in the constructs being measured. Measurement invariance is the desired property upon lack of which intrinsic test bias occurs. If a test for assessing the knowledge of software quality is developed in English and then administered to individuals who are not fluent in English, the measure for the construct of software quality knowledge would be contaminated by a measure of English proficiency. Differential content validity (see section~\ref{ssec:content-validity}) is the most severe form of intrinsic test bias as it causes lower test scores in different groups. If a measurement instrument for debugging skills has been designed by keeping in mind American software testers, any participant that is not an American software tester will likely perform worse on the test to different degrees.~\citet{rust2009} reports various statistical model proposals over the last 50 years to detect intrinsic test bias which, however, present various issues including the introduction of more unfairness near cut-off points or for certain groups of individuals. There is not a recommended way to detect intrinsic test bias other than perform item bias analysis paired with sensitivity analysis.

\subsection{Extrinsic test bias} Extrinsic test bias occurs whenever unfair decision happens based on a non-biased test. These issues usually belong to tests about demographics dealing with social, economical, and political issues, so they are unlikely to belong to measurement instruments developed for the software engineering domain.

\section{Further reading\label{sec:further-reading}}

The present paper only scratches the surface of psychometric theory and practice, and it is its aim to be broad rather than deep. We collect, in this section, what we consider to be good next steps for a better understanding and expansion of the concepts that we have presented.

The books written by \citet{rust2009, kline2015handbook, nunnally1994psychometric,coaley2014} provide an overall overview of psychometric theory, cover all topics mentioned in the present paper, and more. In particular, we invite to compare how they present measurement theory and their views and classifications of validity and reliability. A natural followup is The SEPT,~\citep{apa2014}, which proposes standards that should be met in psychological testing.

While our summary breaks down fundamental concepts and presents them in an introductory way for researchers of behavioral software engineering, our writing can not honor enough the guidelines and recommendations for factor analysis offered by~\citet{yong2013, russell2016, singh2016, fabrigar1999}. To those we add the work of~\citet{zumbo2007}, who have explored, through data simulations, the conditions that yield reliable exploratory factor analysis with samples below 50, which is unfortunately a condition we often live with in software engineering research. Furthermore, we wish to point the reader to alternatives to factor analysis, especially for confirmatory factor analysis (CFA).~\citet{flora2004} analyzed benefits when using Robust Weighted Least Squares (Robust WLS) regression. With a Monte Carlo simulation, they have shown that robust WLS provided accurate test statistics, parameter estimates and standard errors even when the assumption of CFA were not met. Bayesian alternatives for CFA have been proposed in the early 80s already~\citep{lee1981} and later expanded to cover the exploratory phase as well, see, for example, the works by~\citet{conti2014,muthen2012,lu2016}.

In the above sections we have pointed to several papers that can provide a modern, Bayesian statistical view of many psychometric analysis procedures. We also note that a more general treatment and overview can be found in~\citet{levy2016bayesian}. While it is important for a software engineering researcher who wants to use and develop psychometric instruments to know the key concepts and techniques of the more classical, typically frequentist, psychometric techniques one can then switch to a Bayesian view either on philosophical or for practical reasons (a simpler more unified treatment, for one).

Within the software engineering domain,~\citet{gren2018} has offered an alternative lens on validity and reliability of software engineering studies, also based on psychology, that we advise to read.~\citet{ralph2018} have offered a deep overview of construct validity in software engineering through a psychological lens.

We do want to note that one aspect of the method we have used, that can be seen as a limitation, is that there is much current discussion about the statistical methods that are and\slash or should be applied in behavioral and social science, including psychology~\cite{wagenmakers2018bayesian,schad2019toward}, as well as in applied sciences in general~\cite{wasserstein2019moving}. This has also affected software engineering and, for example, a recent paper argued for transitioning to Bayesian statistical analysis in empirical software engineering
~\cite{furia2019bayesian}. However, it is too early to base guidelines on proposals in this ongoing, scholarly discussion since there is not yet a clear consensus. Thus, since we base our review on the current and more established literature, it is likely that future work will need to consider more powerful and up-to-date statistical methods for the creation and assessment of psychometric instruments. Thus, we foresee future updates to this paper that extends it by using such, more recent analysis methods.

\section{Limitations of psychometrics\label{sec:limitations}}

Our article has insofar showed appreciation of psychometrics and how they could be adopted in the software engineering research fields. We should balance this with reasons for not adopting psychometrics, which we also recommend as a further reading~\footnote{Following principles of open science, we deposited revisions of the present article on arXiv as we wrote it, submitted for peer review, and revised during peer review. One of the good aspects of depositing manuscripts on arXiv is to attract feedback. A previous version of the present paper, deposited on arXiv with identifier \href{https://arxiv.org/abs/2005.09959v2}{2005.09959v2}, was used as a basis by~\citet{lewis2020personality} to offer a critique of our stance on psychometrics as well as a broader warning on relying on psychometrics in absolute. The present paper takes~\citeauthor{lewis2020personality}'s feedback into account, especially on the wording and on the limitations of psychometrics. We recommend reading~\citeauthor{lewis2020personality}'s paper as it offers a view on psychometrics that, on several points, is opposed to ours.}.

Psychometrics are not universally accepted as a perfect tool for assessing psychological constructs. Fronts of discontents of psychometrics have emerged, especially in recent times and in the medical, social, and education fields~\cite{norman2002,schoenherr2016,uher2021}. 

The widest critique to psychometric-based testing is that it might not be the best tool for assessing individuals. On the technical side, there has been some evidence that alternative evaluation systems which matched individuals to a standardized set of holistic and realistic vignettes improved discrimination of the individual's performance and facilitated the identification of severe, medical issues of some participants~\cite{regehr2007}. The technical side of critique is, however, limited, as psychometric theory has a long history of robust statistical methods. What causes the widest concern with psychometrics is the central argument, summarized by~\citet{schoenherr2016}, that following a psychometric approach might cause to focus too narrowly on characteristics of individuals in terms of dimensions, features, and competences, while at the same time missing on context, uniqueness of individuals, and team-perspective. In other words, psychometric-based assessments potentially neglects a richness of information, much of which comes from direct experience with individuals and, in particular, qualitative data~\citet{schoenherr2016}. This missed opportunity may be essential for taking decisions such as related to allocation of resources, promotions, and development or prioritization of skills.

~\citet{uher2018} has argued that discussions for and against psychometrics so far have swayed on the ``qualitative vs. quantitative data'' debate, which links to the issue of clashing worldviews in research.
~\citet{lewis2020personality} has expressed skepticism that psychometric methods can be applied in software engineering in a successful way, one of the reasons being that an ``empirical study that is purely empirical cannot succeed.'' (p. 40). This pushback can go to an opposite extreme, with~\citet{michell2000} defining psychometrics as \textit{pathological science} because of a lack of testing of the hypothesis that some psychological attributes are quantitative at all and psychometric science is based on accepting this hypothesis into its core. We hope to meet the reader's mercy if we do not cover these topics in a concluding section~\footnote{We refer the reader to~\citet{creswell2018} textbook for an introduction to ontology and epistemology, the work of~\citet{koltkorivera2004} for the psychology of worldviews, and the studies by~\citet{petersen2013,siegmund2015} for worldviews and their influence in software engineering research.}, especially because we do not see any winning idea regarding epistemological stances. In short, as concisely summarized by~\citet{schoenherr2016}, adopting psychometric approaches can be met by a clash between postpositivism and constructivism, or, the debate concerning the adequacy of mapping constructs to numbers and the assumptions that such a process implies. We agree with~\citet{lewis2020personality} that the idea of psychometrics, as well as the one to add numbers to people, is seductive. We do not see this issue confined to psychometrics, though. Our stance is that, should one decide to embark in a fully empirical and quantitative approach to assess psychological constructs in software engineering, the psychometric approach is an effective tool. There is little debate on the effectiveness of psychometric tests as tools.~\citet{schoenherr2016}, indeed, argue that ``we should not seek to establish a post-psychometric era'' (p. 720) and that we should rather focus on the lack of understanding of the theoretical context in which measurement instruments tools have been developed as well as the several, sometimes unresolved, debates within the psychometrics discourse. Our paper is a start with this.

On the issue of assigning numbers to individuals, we agree that quantitative studies are not all there is. We want to emphasize the importance of qualitative studies the same way as psychology is rediscovering them.~\citet{thurstone1937}, in the plea to render psychological science quantitative, mathematical, and robust, was clear in specifying that ``the mathematical and rigorous treatment of a psychological problem is in no sense a substitute for the other forms of exploratory and descriptive types of experimentation'' (p. 228). This quote should remind us that quantitative methods, while important, are all but one side of a multifaceted coin that we should supply with a mixed-method mindset that includes qualitative studies and other disciplines. As anticipated in the opening of this paper, we are interested in qualitative research as well. We have offered our proposals for qualitative behavioral software engineering~\citep{lenberg2017a}. In~\citet{lenberg2017a}, we suggest future research in software engineering to benefit from a broader set of methods from qualitative psychology such as interpretive phenomenological analysis, narrative analysis, and discourse analysis. We invite researchers in qualitative studies of software engineering to emphasize reflexivity on how our thinking came to be and how pre-existing understanding is constantly revised in the light of new insights. Finally, we encourage the adoption of qualitative guidelines and criteria to enhance the quality of qualitative studies. 

To sum up, we see discontent with psychometrics about clashing worldviews, ontological and epistemological issues, and on how psychometric tests are used in decision-making processes as razor-blade cut-off methods for taking complex decisions.

There are also limitations brought by following psychometric approaches that are related to practical challenges. The field is enormous---even our introductory paper has passed 150 referenced sources---with many theoretical and practical contributions that require a wide understanding of the concepts at hand, several of which are taken as implicit.~\citet{uher2021}, indeed, argues that much issues with psychometrics arise from psychological jargon and underlying, codified conceptual fallacies that bring misconceptions. We believe that basic introductory contributions such as ours help coping with this issue, and we welcome more.~\citet{uher2021} provides a comprehensive introduction on terms such as psyche, behavior, constructs, operationalization, variables, attributes, and more, which we recommend reading. Furthermore, we highlight the considerable amount of effort it takes for developing a psychometrically validate test.~\citet{gren2018} has argued that ``Spending an entire Ph.D. candidacy on the validation of one single measurement of a construct should be, not only approved, but encouraged.''. We agree on that enabling a Ph.D. student to develop a measurement instrument for software engineering constructs following a psychometric approach should be allowed, but we want to emphasize here the ``entire Ph.D.'' part. It might truly take years to developing and validating a measurement instrument. We highly recommend searching the literature for existing, psychometrically validated tests before developing one.

All in all, we emphasize our view that we should be systematic and rigorous regardless of worldview, ontological and epistemological stance, and preferred methods. Psychometrics is one of several ways of quantitative research, which we embrace but not blindly believe in. Qualitative research is as important as quantitative research and they complement each other. Qualitative research is essential to discover rich data that captures the uniqueness of individuals as such, rather than pooling people as unique data points. Psychometric methods are an excellent tool to enhance our way to select, develop, understand, and test metrics. Psychometrics should, however, not be justification to build a metric for the sake of it. Proper application of psychometrics does also not promote the validity of results in absolutism. We remind that validity in psychometrics is defined as ``The degree to which evidence and theory support the interpretation of test scores for proposed uses of tests.''~\citep{apa2014} for which we now highlight the word \textbf{interpretation}.

\section{Running example of psychometric evaluation\label{sec:example}}

We believe that a methodology description is best complemented by a concrete example of its application. In Appendix~\ref{sec:appendix}, we thus provide a complete scenario of the development of a fictitious measurement instrument and the establishment of its psychometric properties with the R programming language. The evaluation follows the same structure as the present paper for ease of understanding. In the spirit of open science in software engineering~\citep{mendez2020}, we provide the running example as a replication package as well~\citep{graziotin2020}. We wrote the example using R Markdown, making it fully repeatable, as well as the generated dataset, and instructions for replication with newly generated data.

\section{Conclusion\label{sec:conclusion}}
The adoption and development of validated and reliable measurement instruments in software engineering research, whenever humans are to be evaluated, should benefit from psychology and statistics theory; we need not and should not `reinvent the wheel'. This paper provides a brief introduction to the evaluation of psychometric tests. Our guidelines will contribute to a better development of new tests as well as a justified decision-making process when selecting  existing tests. 

After providing basic building blocks and concepts of psychometric theory, we introduced item pooling, item review, pilot testing, item analysis, factor analysis, statistical property of items, reliability, validity, and fairness in testing and test bias. In an appendix, we also provided a running example of an implementation of a psychometric evaluation and shared both its data and source code (scripts) openly to promote self-study and a basis for further exploration. We followed textbooks, method papers, and society standards for ensuring a coverage of all important steps, but we could only offer a brief introduction and invite the reader to explore our referenced material further. Each of these steps is a universe of its own, with dozens of published artifacts related to them.

A proper psychometric evaluation requires a consideration of all elements reported in the present paper. The development of a measurement instrument, however, does not have to execute every single step we summarize, and not even necessarily in the same order. We are illustrating a wide range of possibilities, some of which are often brought by in future validation studies. Critical thinking is required throughout the whole process, which involves choices and trade-offs.

Adding the steps described in this paper will increase the time required for developing measurement instruments. However, the return on investment will be considerable. Psychometric analysis and refinement of measurement instruments can improve their reliability and validity. The software engineering community must value psychometric studies more. This, however, requires a cultural change that we hope to champion with this paper.

\begin{acks}
We acknowledge the support of Swedish Armed Forces, Swedish Defense Materiel Administration and Swedish Governmental Agency for Innovation Systems (VINNOVA) in the project number 2013-01199, as well as the Alexander von Humboldt (AvH) Foundation. We are thankful for the comments of 3 anonymous reviewers and the associate editor, who helped the manuscript mature and grow in value during the peer review process. We are thankful to Marvin Wyrich and Kai Mindermann for offering feedback to early versions of this paper. We are grateful for~\citeauthor{lewis2020personality}'s comments in the open, which we treasured when revising the manuscript. Our thanks to Katharina Plett for proofreading the paper.
\end{acks}

\bibliographystyle{ACM-Reference-Format}
\bibliography{references}


\begin{thebibliography}{153}

%%% the Web address.

\ifx \showCODEN    \undefined \def \showCODEN     #1{\unskip}     \fi
\ifx \showDOI      \undefined \def \showDOI       #1{#1}\fi
\ifx \showISBNx    \undefined \def \showISBNx     #1{\unskip}     \fi
\ifx \showISBNxiii \undefined \def \showISBNxiii  #1{\unskip}     \fi
\ifx \showISSN     \undefined \def \showISSN      #1{\unskip}     \fi
\ifx \showLCCN     \undefined \def \showLCCN      #1{\unskip}     \fi
\ifx \shownote     \undefined \def \shownote      #1{#1}          \fi
\ifx \showarticletitle \undefined \def \showarticletitle #1{#1}   \fi
\ifx \showURL      \undefined \def \showURL       {\relax}        \fi
% invisible to TeX
\providecommand\bibfield[2]{#2}
\providecommand\bibinfo[2]{#2}
\providecommand\natexlab[1]{#1}
\providecommand\showeprint[2][]{arXiv:#2}

\bibitem[\protect\citeauthoryear{Abdi}{Abdi}{2003}]%
        {herve2003}
\bibfield{author}{\bibinfo{person}{Hervé Abdi}.}
  \bibinfo{year}{2003}\natexlab{}.
\newblock \showarticletitle{Factor Rotations in Factor Analyses}.
\newblock In \bibinfo{booktitle}{\emph{Encyclopedia of Social Sciences Research
  Methods}}, \bibfield{editor}{\bibinfo{person}{A.~Lewis-Beck~M., Bryman} {and}
  \bibinfo{person}{Futing T}} (Eds.). \bibinfo{publisher}{SAGE},
  \bibinfo{address}{Thousand Oaks (CA)}, \bibinfo{pages}{792--795}.
\newblock


\bibitem[\protect\citeauthoryear{Alphen, Halfens, Hasman, and Imbos}{Alphen
  et~al\mbox{.}}{1994}]%
        {vanalphen1994}
\bibfield{author}{\bibinfo{person}{Arnold Alphen}, \bibinfo{person}{Ruud
  Halfens}, \bibinfo{person}{Arie Hasman}, {and} \bibinfo{person}{Tjaart
  Imbos}.} \bibinfo{year}{1994}\natexlab{}.
\newblock \showarticletitle{Likert or Rasch? Nothing is more applicable than
  good theory}.
\newblock \bibinfo{journal}{\emph{Journal of Advanced Nursing}}
  \bibinfo{volume}{20}, \bibinfo{number}{1} (\bibinfo{year}{1994}),
  \bibinfo{pages}{196–201}.
\newblock


\bibitem[\protect\citeauthoryear{{American Educational Research Association},
  {American Psychological Association}, {National Council on Measurement in
  Education}, and {Joint Committee on Standards for Educational and
  Psychological Testing (U.S.)}}{{American Educational Research Association}
  et~al\mbox{.}}{2014}]%
        {apa2014}
\bibfield{author}{\bibinfo{person}{{American Educational Research
  Association}}, \bibinfo{person}{{American Psychological Association}},
  \bibinfo{person}{{National Council on Measurement in Education}}, {and}
  \bibinfo{person}{{Joint Committee on Standards for Educational and
  Psychological Testing (U.S.)}}.} \bibinfo{year}{2014}\natexlab{}.
\newblock \bibinfo{booktitle}{\emph{Standards for educational and psychological
  testing}}.
\newblock \bibinfo{publisher}{American Educational Research Association},
  \bibinfo{address}{Washington, DC}.
\newblock
\showISBNx{978-0-935302-35-6}


\bibitem[\protect\citeauthoryear{Anastasi and Urbina}{Anastasi and
  Urbina}{1997}]%
        {anastasi1997}
\bibfield{author}{\bibinfo{person}{Anne Anastasi} {and} \bibinfo{person}{Susana
  Urbina}.} \bibinfo{year}{1997}\natexlab{}.
\newblock \bibinfo{booktitle}{\emph{Psychological testing}}.
\newblock \bibinfo{publisher}{Prentice Hall}, \bibinfo{address}{Upper Saddle
  River, N.J}.
\newblock
\showISBNx{9780023030857}


\bibitem[\protect\citeauthoryear{Ansari, Jedidi, and Dube}{Ansari
  et~al\mbox{.}}{2002}]%
        {ansari2002heterogeneous}
\bibfield{author}{\bibinfo{person}{Asim Ansari}, \bibinfo{person}{Kamel
  Jedidi}, {and} \bibinfo{person}{Laurette Dube}.}
  \bibinfo{year}{2002}\natexlab{}.
\newblock \showarticletitle{Heterogeneous factor analysis models: A Bayesian
  approach}.
\newblock \bibinfo{journal}{\emph{Psychometrika}} \bibinfo{volume}{67},
  \bibinfo{number}{1} (\bibinfo{year}{2002}), \bibinfo{pages}{49--77}.
\newblock


\bibitem[\protect\citeauthoryear{Berger}{Berger}{2013a}]%
        {berger2013b}
\bibfield{author}{\bibinfo{person}{Michael Berger}.}
  \bibinfo{year}{2013}\natexlab{a}.
\newblock \showarticletitle{Criterion-Referenced Testing}.
\newblock In \bibinfo{booktitle}{\emph{Encyclopedia of Autism Spectrum
  Disorders}}. \bibinfo{publisher}{Springer New York},
  \bibinfo{address}{Springer New York}, \bibinfo{pages}{823--823}.
\newblock
\urldef\tempurl%
\url{https://doi.org/10.1007/978-1-4419-1698-3_146}
\showDOI{\tempurl}


\bibitem[\protect\citeauthoryear{Berger}{Berger}{2013b}]%
        {berger2013a}
\bibfield{author}{\bibinfo{person}{Michael Berger}.}
  \bibinfo{year}{2013}\natexlab{b}.
\newblock \showarticletitle{Norm-Referenced Testing}.
\newblock In \bibinfo{booktitle}{\emph{Encyclopedia of Autism Spectrum
  Disorders}}. \bibinfo{publisher}{Springer New York},
  \bibinfo{address}{Springer New York}, \bibinfo{pages}{2063--2064}.
\newblock
\urldef\tempurl%
\url{https://doi.org/10.1007/978-1-4419-1698-3_451}
\showDOI{\tempurl}


\bibitem[\protect\citeauthoryear{Boulkedid, Abdoul, Loustau, Sibony, and
  Alberti}{Boulkedid et~al\mbox{.}}{2011}]%
        {boulkedid2011}
\bibfield{author}{\bibinfo{person}{R Boulkedid}, \bibinfo{person}{H Abdoul},
  \bibinfo{person}{M Loustau}, \bibinfo{person}{O Sibony}, {and}
  \bibinfo{person}{C Alberti}.} \bibinfo{year}{2011}\natexlab{}.
\newblock \showarticletitle{Using and reporting the Delphi method for selecting
  healthcare quality indicators: a systematic review.}
\newblock \bibinfo{journal}{\emph{PLoS One}}  \bibinfo{volume}{6}
  (\bibinfo{year}{2011}), \bibinfo{pages}{e20476}.
\newblock
\urldef\tempurl%
\url{https://doi.org/10.1371/journal.pone.0020476}
\showDOI{\tempurl}


\bibitem[\protect\citeauthoryear{Bradburn, Sudman, and Wansink}{Bradburn
  et~al\mbox{.}}{2004}]%
        {bradburn2004}
\bibfield{author}{\bibinfo{person}{Norman~M. Bradburn},
  \bibinfo{person}{Seymour Sudman}, {and} \bibinfo{person}{Brian Wansink}.}
  \bibinfo{year}{2004}\natexlab{}.
\newblock \bibinfo{booktitle}{\emph{Asking Questions}}.
\newblock \bibinfo{publisher}{John Wiley \& Sons}, \bibinfo{address}{Hoboken,
  New Jersey}. 448 pages.
\newblock


\bibitem[\protect\citeauthoryear{Browne}{Browne}{2001}]%
        {browne2001}
\bibfield{author}{\bibinfo{person}{Michael~W. Browne}.}
  \bibinfo{year}{2001}\natexlab{}.
\newblock \showarticletitle{An Overview of Analytic Rotation in Exploratory
  Factor Analysis}.
\newblock \bibinfo{journal}{\emph{Multivariate Behavioral Research}}
  \bibinfo{volume}{36}, \bibinfo{number}{1} (\bibinfo{year}{2001}),
  \bibinfo{pages}{111–150}.
\newblock


\bibitem[\protect\citeauthoryear{Caldiera and Rombach}{Caldiera and
  Rombach}{1994}]%
        {caldiera1994goal}
\bibfield{author}{\bibinfo{person}{Victor R Basili1~Gianluigi Caldiera} {and}
  \bibinfo{person}{H~Dieter Rombach}.} \bibinfo{year}{1994}\natexlab{}.
\newblock \showarticletitle{The goal question metric approach}.
\newblock \bibinfo{journal}{\emph{Encyclopedia of software engineering}}
  \bibinfo{volume}{1}, \bibinfo{number}{1} (\bibinfo{year}{1994}),
  \bibinfo{pages}{528--532}.
\newblock


\bibitem[\protect\citeauthoryear{Cambridge}{Cambridge}{2018}]%
        {caefairness2018}
\bibfield{author}{\bibinfo{person}{English~Dictionary Cambridge}.}
  \bibinfo{year}{2018}\natexlab{}.
\newblock \showarticletitle{Fairness}.
\newblock \bibinfo{journal}{\emph{Cambridge English Dictionary}}
  \bibinfo{volume}{1}, \bibinfo{number}{1} (\bibinfo{year}{2018}),
  \bibinfo{pages}{1}.
\newblock
\newblock
\shownote{Available:
  https://dictionary.cambridge.org/dictionary/english/fairness.}


\bibitem[\protect\citeauthoryear{Campbell and Fiske}{Campbell and
  Fiske}{1959}]%
        {campbell1959}
\bibfield{author}{\bibinfo{person}{Donald~T. Campbell} {and}
  \bibinfo{person}{Donald~W. Fiske}.} \bibinfo{year}{1959}\natexlab{}.
\newblock \showarticletitle{Convergent and discriminant validation by the
  multitrait-multimethod matrix.}
\newblock \bibinfo{journal}{\emph{Psychological Bulletin}}
  \bibinfo{volume}{56}, \bibinfo{number}{2} (\bibinfo{year}{1959}),
  \bibinfo{pages}{81–105}.
\newblock


\bibitem[\protect\citeauthoryear{Capretz}{Capretz}{2003}]%
        {capretz2003personality}
\bibfield{author}{\bibinfo{person}{Luiz~Fernando Capretz}.}
  \bibinfo{year}{2003}\natexlab{}.
\newblock \showarticletitle{Personality types in software engineering}.
\newblock \bibinfo{journal}{\emph{International Journal of Human-Computer
  Studies}} \bibinfo{volume}{58}, \bibinfo{number}{2} (\bibinfo{year}{2003}),
  \bibinfo{pages}{207--214}.
\newblock


\bibitem[\protect\citeauthoryear{Carnap}{Carnap}{1962}]%
        {carnap1962}
\bibfield{author}{\bibinfo{person}{Rudolf Carnap}.}
  \bibinfo{year}{1962}\natexlab{}.
\newblock \bibinfo{booktitle}{\emph{Logical foundations of probability}}.
\newblock \bibinfo{publisher}{University of Chicago Press},
  \bibinfo{address}{Chicago}.
\newblock
\showISBNx{978-0226093437}


\bibitem[\protect\citeauthoryear{Carver, Muccini, Penzenstadler, Prikladnicki,
  Serebrenik, and Zimmermann}{Carver et~al\mbox{.}}{2021}]%
        {carver2021}
\bibfield{author}{\bibinfo{person}{Jeffrey~C. Carver}, \bibinfo{person}{Henry
  Muccini}, \bibinfo{person}{Birgit Penzenstadler}, \bibinfo{person}{Rafael
  Prikladnicki}, \bibinfo{person}{Alexander Serebrenik}, {and}
  \bibinfo{person}{Thomas Zimmermann}.} \bibinfo{year}{2021}\natexlab{}.
\newblock \showarticletitle{Behavioral Science and Diversity in Software
  Engineering}.
\newblock \bibinfo{journal}{\emph{IEEE Software}} \bibinfo{volume}{38},
  \bibinfo{number}{2} (\bibinfo{year}{2021}), \bibinfo{pages}{107–112}.
\newblock
\urldef\tempurl%
\url{https://doi.org/10.1109/ms.2020.3042683}
\showDOI{\tempurl}


\bibitem[\protect\citeauthoryear{Cattell}{Cattell}{1966}]%
        {cattel1966}
\bibfield{author}{\bibinfo{person}{Raymond~B. Cattell}.}
  \bibinfo{year}{1966}\natexlab{}.
\newblock \showarticletitle{The Scree Test For The Number Of Factors}.
\newblock \bibinfo{journal}{\emph{Multivariate Behavioral Research}}
  \bibinfo{volume}{1}, \bibinfo{number}{2} (\bibinfo{year}{1966}),
  \bibinfo{pages}{245–276}.
\newblock


\bibitem[\protect\citeauthoryear{Chattopadhyay, Nelson, Au, Morales, Sanchez,
  Pandita, and Sarma}{Chattopadhyay et~al\mbox{.}}{2020}]%
        {chattopadhyay2020}
\bibfield{author}{\bibinfo{person}{Souti Chattopadhyay},
  \bibinfo{person}{Nicholas Nelson}, \bibinfo{person}{Audrey Au},
  \bibinfo{person}{Natalia Morales}, \bibinfo{person}{Christopher Sanchez},
  \bibinfo{person}{Rahul Pandita}, {and} \bibinfo{person}{Anita Sarma}.}
  \bibinfo{year}{2020}\natexlab{}.
\newblock \showarticletitle{A tale from the trenches},
  Vol.~\bibinfo{volume}{Proceedings of the ACM/IEEE 42nd International
  Conference on Software Engineering}. \bibinfo{publisher}{ACM},
  \bibinfo{address}{New York, NY, USA}, \bibinfo{pages}{654--665}.
\newblock
\urldef\tempurl%
\url{https://doi.org/10.1145/3377811.3380330}
\showDOI{\tempurl}


\bibitem[\protect\citeauthoryear{Ciolkowski, Laitenberger, Vegas, and
  Biffl}{Ciolkowski et~al\mbox{.}}{2003}]%
        {ciolkowski2003}
\bibfield{author}{\bibinfo{person}{Marcus Ciolkowski}, \bibinfo{person}{Oliver
  Laitenberger}, \bibinfo{person}{Sira Vegas}, {and} \bibinfo{person}{Stefan
  Biffl}.} \bibinfo{year}{2003}\natexlab{}.
\newblock \showarticletitle{Practical Experiences in the Design and Conduct of
  Surveys in Empirical Software Engineering}.
\newblock In \bibinfo{booktitle}{\emph{Empirical methods and studies in
  software engineering}}, \bibfield{editor}{\bibinfo{person}{Gerhard Goos},
  \bibinfo{person}{Juris Hartmanis}, \bibinfo{person}{Jan van Leeuwen},
  \bibinfo{person}{Reidar Conradi}, {and} \bibinfo{person}{Alf~Inge Wang}}
  (Eds.). \bibinfo{publisher}{Springer Berlin Heidelberg},
  \bibinfo{address}{Berlin, Heidelberg}, \bibinfo{pages}{104--128}.
\newblock
\urldef\tempurl%
\url{https://doi.org/10.1007/978-3-540-45143-3_7}
\showDOI{\tempurl}


\bibitem[\protect\citeauthoryear{Coaley}{Coaley}{2014}]%
        {coaley2014}
\bibfield{author}{\bibinfo{person}{Keith Coaley}.}
  \bibinfo{year}{2014}\natexlab{}.
\newblock \bibinfo{booktitle}{\emph{An Introduction to Psychological Assessment
  \& Psychometrics}}.
\newblock \bibinfo{publisher}{SAGE}, \bibinfo{address}{Los Angeles}.
\newblock
\showISBNx{978-1446267158}


\bibitem[\protect\citeauthoryear{Cohen, Swerdlik, and Phillips}{Cohen
  et~al\mbox{.}}{1995}]%
        {cohen1995}
\bibfield{author}{\bibinfo{person}{Ronald~Jay Cohen}, \bibinfo{person}{Mark~E.
  Swerdlik}, {and} \bibinfo{person}{Suzanne~M. Phillips}.}
  \bibinfo{year}{1995}\natexlab{}.
\newblock \bibinfo{booktitle}{\emph{Psychological Testing and Assessment: An
  Introduction to Tests and Measurement}}.
\newblock \bibinfo{publisher}{Mayfield Pub Co}, \bibinfo{address}{California
  City, California, United States}.
\newblock


\bibitem[\protect\citeauthoryear{Collins}{Collins}{2003}]%
        {collins2003}
\bibfield{author}{\bibinfo{person}{D Collins}.}
  \bibinfo{year}{2003}\natexlab{}.
\newblock \showarticletitle{Pretesting survey instruments: an overview of
  cognitive methods.}
\newblock \bibinfo{journal}{\emph{Qual Life Res}} \bibinfo{volume}{12},
  \bibinfo{number}{3} (\bibinfo{year}{2003}), \bibinfo{pages}{229–238}.
\newblock


\bibitem[\protect\citeauthoryear{Conti, Fr{\"u}hwirth-Schnatter, Heckman, and
  Piatek}{Conti et~al\mbox{.}}{2014a}]%
        {conti2014bayesian}
\bibfield{author}{\bibinfo{person}{Gabriella Conti}, \bibinfo{person}{Sylvia
  Fr{\"u}hwirth-Schnatter}, \bibinfo{person}{James~J Heckman}, {and}
  \bibinfo{person}{R{\'e}mi Piatek}.} \bibinfo{year}{2014}\natexlab{a}.
\newblock \showarticletitle{Bayesian exploratory factor analysis}.
\newblock \bibinfo{journal}{\emph{Journal of econometrics}}
  \bibinfo{volume}{183}, \bibinfo{number}{1} (\bibinfo{year}{2014}),
  \bibinfo{pages}{31--57}.
\newblock


\bibitem[\protect\citeauthoryear{Conti, Frühwirth-Schnatter, Heckman, and
  Piatek}{Conti et~al\mbox{.}}{2014b}]%
        {conti2014}
\bibfield{author}{\bibinfo{person}{G Conti}, \bibinfo{person}{S
  Frühwirth-Schnatter}, \bibinfo{person}{JJ Heckman}, {and} \bibinfo{person}{R
  Piatek}.} \bibinfo{year}{2014}\natexlab{b}.
\newblock \showarticletitle{Bayesian Exploratory Factor Analysis.}
\newblock \bibinfo{journal}{\emph{J Econom}} \bibinfo{volume}{183},
  \bibinfo{number}{1} (\bibinfo{year}{2014}), \bibinfo{pages}{31–57}.
\newblock


\bibitem[\protect\citeauthoryear{Costa~Jr and McCrae}{Costa~Jr and
  McCrae}{1992}]%
        {costajr1992}
\bibfield{author}{\bibinfo{person}{Paul~T Costa~Jr} {and}
  \bibinfo{person}{Robert~R McCrae}.} \bibinfo{year}{1992}\natexlab{}.
\newblock \showarticletitle{Comes of Age}, Vol.~\bibinfo{volume}{Psychology and
  aging: Nebraska Symposium on Motivation, 1991 39}.
  \bibinfo{publisher}{University of Nebraska Press}, \bibinfo{address}{Lincoln,
  Nebraska}, \bibinfo{pages}{169}.
\newblock


\bibitem[\protect\citeauthoryear{Courtney}{Courtney}{2013}]%
        {courtney2013}
\bibfield{author}{\bibinfo{person}{Matthew Gordon~Rau Courtney}.}
  \bibinfo{year}{2013}\natexlab{}.
\newblock \showarticletitle{Determining the Number of Factors to Retain in EFA:
  Using the SPSS R-Menu v2.0 to Make More Judicious Estimations}.
\newblock \bibinfo{journal}{\emph{Practical Assessment, Research \&
  Evaluation}} \bibinfo{volume}{18}, \bibinfo{number}{8}
  (\bibinfo{year}{2013}), \bibinfo{pages}{1–14}.
\newblock


\bibitem[\protect\citeauthoryear{Creswell and Creswell}{Creswell and
  Creswell}{2018}]%
        {creswell2018}
\bibfield{author}{\bibinfo{person}{John~W. Creswell} {and}
  \bibinfo{person}{J.~David Creswell}.} \bibinfo{year}{2018}\natexlab{}.
\newblock \bibinfo{booktitle}{\emph{Research Design}}.
  Vol.~\bibinfo{volume}{2nd}.
\newblock \bibinfo{publisher}{SAGE Publications, Incorporated},
  \bibinfo{address}{Thousand Oaks, California}. 304 pages.
\newblock


\bibitem[\protect\citeauthoryear{Crocker}{Crocker}{2008}]%
        {crocker2006}
\bibfield{author}{\bibinfo{person}{Linda Crocker}.}
  \bibinfo{year}{2008}\natexlab{}.
\newblock \bibinfo{booktitle}{\emph{Introduction to classical and modern test
  theory}}.
\newblock \bibinfo{publisher}{Cengage Learning}, \bibinfo{address}{Mason,
  Ohio}.
\newblock
\showISBNx{978-0495395911}


\bibitem[\protect\citeauthoryear{Cronbach}{Cronbach}{1951}]%
        {cronbach1951}
\bibfield{author}{\bibinfo{person}{Lee~J Cronbach}.}
  \bibinfo{year}{1951}\natexlab{}.
\newblock \showarticletitle{Coefficient alpha and the internal structure of
  tests}.
\newblock \bibinfo{journal}{\emph{Psychometrika}} \bibinfo{volume}{16},
  \bibinfo{number}{3} (\bibinfo{year}{1951}), \bibinfo{pages}{297--334}.
\newblock


\bibitem[\protect\citeauthoryear{Cruz, da~Silva, and Capretz}{Cruz
  et~al\mbox{.}}{2015}]%
        {cruz2015}
\bibfield{author}{\bibinfo{person}{Shirley Cruz}, \bibinfo{person}{Fabio~Q.B.
  da Silva}, {and} \bibinfo{person}{Luiz~Fernando Capretz}.}
  \bibinfo{year}{2015}\natexlab{}.
\newblock \showarticletitle{Forty years of research on personality in software
  engineering: A mapping study}.
\newblock \bibinfo{journal}{\emph{Computers in Human Behavior}}
  \bibinfo{volume}{46} (\bibinfo{year}{2015}), \bibinfo{pages}{94–113}.
\newblock
\urldef\tempurl%
\url{https://doi.org/10.1016/j.chb.2014.12.008}
\showDOI{\tempurl}


\bibitem[\protect\citeauthoryear{Dalkey and Helmer}{Dalkey and Helmer}{1963}]%
        {dalkey1963}
\bibfield{author}{\bibinfo{person}{Norman Dalkey} {and} \bibinfo{person}{Olaf
  Helmer}.} \bibinfo{year}{1963}\natexlab{}.
\newblock \showarticletitle{An Experimental Application of the DELPHI Method to
  the Use of Experts}.
\newblock \bibinfo{journal}{\emph{Management Science}} \bibinfo{volume}{9},
  \bibinfo{number}{3} (\bibinfo{year}{1963}), \bibinfo{pages}{458–467}.
\newblock
\urldef\tempurl%
\url{https://doi.org/10.1287/mnsc.9.3.458}
\showDOI{\tempurl}


\bibitem[\protect\citeauthoryear{Darcy and Ma}{Darcy and Ma}{2005}]%
        {darcy2005}
\bibfield{author}{\bibinfo{person}{David~P Darcy} {and} \bibinfo{person}{Meng
  Ma}.} \bibinfo{year}{2005}\natexlab{}.
\newblock \showarticletitle{Exploring individual characteristics and
  programming performance: Implications for programmer selection},
  Vol.~\bibinfo{volume}{Proceedings of the 38th Annual Hawaii International
  Conference on System Sciences}. \bibinfo{publisher}{IEEE},
  \bibinfo{address}{Piscataway, NJ}, \bibinfo{pages}{314a--314a}.
\newblock


\bibitem[\protect\citeauthoryear{Darton}{Darton}{1980}]%
        {darton1980}
\bibfield{author}{\bibinfo{person}{R.~A. Darton}.}
  \bibinfo{year}{1980}\natexlab{}.
\newblock \showarticletitle{Rotation in Factor Analysis}.
\newblock \bibinfo{journal}{\emph{The Statistician}} \bibinfo{volume}{29},
  \bibinfo{number}{3} (\bibinfo{year}{1980}), \bibinfo{pages}{167}.
\newblock


\bibitem[\protect\citeauthoryear{De~Swert}{De~Swert}{2012}]%
        {de2012calculating}
\bibfield{author}{\bibinfo{person}{Knut De~Swert}.}
  \bibinfo{year}{2012}\natexlab{}.
\newblock \bibinfo{booktitle}{\emph{Calculating inter-coder reliability in
  media content analysis using Krippendorff's Alpha}}.
\newblock \bibinfo{type}{{T}echnical {R}eport}. \bibinfo{institution}{Center
  for Politics and Communication, University of Amsterdam}.
  \bibinfo{pages}{1--15} pages.
\newblock


\bibitem[\protect\citeauthoryear{Drennan}{Drennan}{2003}]%
        {drennan2003}
\bibfield{author}{\bibinfo{person}{J Drennan}.}
  \bibinfo{year}{2003}\natexlab{}.
\newblock \showarticletitle{Cognitive interviewing: verbal data in the design
  and pretesting of questionnaires.}
\newblock \bibinfo{journal}{\emph{J Adv Nurs}} \bibinfo{volume}{42},
  \bibinfo{number}{1} (\bibinfo{year}{2003}), \bibinfo{pages}{57–63}.
\newblock
\urldef\tempurl%
\url{https://doi.org/10.1046/j.1365-2648.2003.02579.x}
\showDOI{\tempurl}


\bibitem[\protect\citeauthoryear{Embretson and Reise}{Embretson and
  Reise}{2013}]%
        {embretson2013}
\bibfield{author}{\bibinfo{person}{Susan~E Embretson} {and}
  \bibinfo{person}{Steven~P Reise}.} \bibinfo{year}{2013}\natexlab{}.
\newblock \bibinfo{booktitle}{\emph{Item response theory}}.
\newblock \bibinfo{publisher}{Psychology Press}, \bibinfo{address}{Hove, East
  Sussex, United Kingdom}.
\newblock


\bibitem[\protect\citeauthoryear{Fabrigar, Wegener, MacCallum, and
  Strahan}{Fabrigar et~al\mbox{.}}{1999}]%
        {fabrigar1999}
\bibfield{author}{\bibinfo{person}{Leandre~R Fabrigar},
  \bibinfo{person}{Duane~T Wegener}, \bibinfo{person}{Robert~C MacCallum},
  {and} \bibinfo{person}{Erin~J Strahan}.} \bibinfo{year}{1999}\natexlab{}.
\newblock \showarticletitle{Evaluating the use of exploratory factor analysis
  in psychological research.}
\newblock \bibinfo{journal}{\emph{Psychological methods}} \bibinfo{volume}{4},
  \bibinfo{number}{3} (\bibinfo{year}{1999}), \bibinfo{pages}{272}.
\newblock


\bibitem[\protect\citeauthoryear{Fagerholm}{Fagerholm}{2015}]%
        {fagerholm2015}
\bibfield{author}{\bibinfo{person}{Fabian Fagerholm}.}
  \bibinfo{year}{2015}\natexlab{}.
\newblock \emph{\bibinfo{title}{Software Developer Experience: Case Studies in
  Lean-Agile and Open Source Environments}}.
\newblock \bibinfo{thesistype}{Ph.D. Dissertation}. \bibinfo{school}{Ph. D.
  Dissertation. Department of Computer Science, University of Helsinki},
  \bibinfo{address}{Helsinki}.
\newblock


\bibitem[\protect\citeauthoryear{Fagerholm and Fritz}{Fagerholm and
  Fritz}{2020}]%
        {Fagerholm2020}
\bibfield{author}{\bibinfo{person}{Fabian Fagerholm} {and}
  \bibinfo{person}{Thomas Fritz}.} \bibinfo{year}{2020}\natexlab{}.
\newblock \showarticletitle{Biometric Measurement in Software Engineering}.
\newblock In \bibinfo{booktitle}{\emph{Contemporary Empirical Methods in
  Software Engineering}}, \bibfield{editor}{\bibinfo{person}{Michael Felderer}
  {and} \bibinfo{person}{Guilherme~Horta Travassos}} (Eds.).
  \bibinfo{publisher}{Springer International Publishing},
  \bibinfo{address}{Cham, Switzerland}, \bibinfo{pages}{151--172}.
\newblock
\showISBNx{978-3-030-32489-6}
\urldef\tempurl%
\url{https://doi.org/10.1007/978-3-030-32489-6_6}
\showDOI{\tempurl}


\bibitem[\protect\citeauthoryear{Fagerholm and Pagels}{Fagerholm and
  Pagels}{2014}]%
        {fagerholm2014}
\bibfield{author}{\bibinfo{person}{Fabian Fagerholm} {and} \bibinfo{person}{Max
  Pagels}.} \bibinfo{year}{2014}\natexlab{}.
\newblock \showarticletitle{Examining the Structure of Lean and Agile Values
  among Software Developers}.
\newblock In \bibinfo{booktitle}{\emph{Lecture Notes in Business Information
  Processing: Agile Processes in Software Engineering and Extreme
  Programming}}. \bibinfo{publisher}{Springer International Publishing},
  \bibinfo{address}{Cham}, \bibinfo{pages}{218--233}.
\newblock


\bibitem[\protect\citeauthoryear{Feldt and Magazinius}{Feldt and
  Magazinius}{2010}]%
        {feldt2010}
\bibfield{author}{\bibinfo{person}{Robert Feldt} {and} \bibinfo{person}{Ana
  Magazinius}.} \bibinfo{year}{2010}\natexlab{}.
\newblock \showarticletitle{Validity Threats in Empirical Software Engineering
  Research - An Initial Survey}. In \bibinfo{booktitle}{\emph{Proceedings of
  the 22nd International Conference on Software Engineering {\&} Knowledge
  Engineering (SEKE'2010), Redwood City, San Francisco Bay, CA, USA, July 1 -
  July 3, 2010}}. \bibinfo{publisher}{NA}, \bibinfo{address}{NA},
  \bibinfo{pages}{374--379}.
\newblock
\newblock
\shownote{Available:
  \url{http://www.cse.chalmers.se/~feldt/publications/feldt_2010_validity_threats_in_ese_initial_survey.pdf}.}


\bibitem[\protect\citeauthoryear{Feldt, Torkar, Angelis, and Samuelsson}{Feldt
  et~al\mbox{.}}{2008}]%
        {feldt2008}
\bibfield{author}{\bibinfo{person}{Robert Feldt}, \bibinfo{person}{Richard
  Torkar}, \bibinfo{person}{Lefteris Angelis}, {and} \bibinfo{person}{Maria
  Samuelsson}.} \bibinfo{year}{2008}\natexlab{}.
\newblock \showarticletitle{Towards individualized software engineering:
  empirical studies should collect psychometrics.},
  Vol.~\bibinfo{volume}{Proceedings of the 2008 International Workshop on
  Cooperative and Human Aspects of Software Engineering}.
  \bibinfo{publisher}{ACM Press}, \bibinfo{address}{New York, New York, USA},
  \bibinfo{pages}{49--52}.
\newblock


\bibitem[\protect\citeauthoryear{Fernández, Graziotin, Wagner, and
  Seibold}{Fernández et~al\mbox{.}}{2020}]%
        {mendez2020}
\bibfield{author}{\bibinfo{person}{Daniel~Méndez Fernández},
  \bibinfo{person}{Daniel Graziotin}, \bibinfo{person}{Stefan Wagner}, {and}
  \bibinfo{person}{Heidi Seibold}.} \bibinfo{year}{2020}\natexlab{}.
\newblock \showarticletitle{Open science in software engineering}.
\newblock In \bibinfo{booktitle}{\emph{Contemporary Empirical Methods in
  Software Engineering}}, \bibfield{editor}{\bibinfo{person}{Michael Felderer}
  {and} \bibinfo{person}{Guilherme~Horta Travassos}} (Eds.).
  \bibinfo{publisher}{Springer International Publishing},
  \bibinfo{address}{Cham, Switzerland}, \bibinfo{pages}{479--504}.
\newblock
\urldef\tempurl%
\url{https://doi.org/10.1007/978-3-030-32489-6_17}
\showDOI{\tempurl}


\bibitem[\protect\citeauthoryear{Flora and Curran}{Flora and Curran}{2004}]%
        {flora2004}
\bibfield{author}{\bibinfo{person}{David~B Flora} {and}
  \bibinfo{person}{Patrick~J Curran}.} \bibinfo{year}{2004}\natexlab{}.
\newblock \showarticletitle{An empirical evaluation of alternative methods of
  estimation for confirmatory factor analysis with ordinal data.}
\newblock \bibinfo{journal}{\emph{Psychological methods}} \bibinfo{volume}{9},
  \bibinfo{number}{4} (\bibinfo{year}{2004}), \bibinfo{pages}{466}.
\newblock


\bibitem[\protect\citeauthoryear{Fox}{Fox}{2017}]%
        {fox2017}
\bibfield{author}{\bibinfo{person}{John Fox}.} \bibinfo{year}{2017}\natexlab{}.
\newblock \bibinfo{booktitle}{\emph{sem: Structural Equation Models}}.
\newblock \bibinfo{type}{{T}echnical {R}eport}. \bibinfo{institution}{The
  Comprehensive R Archive Network}. \bibinfo{pages}{1–79} pages.
\newblock


\bibitem[\protect\citeauthoryear{Franca, da~Silva, and Sharp}{Franca
  et~al\mbox{.}}{2020}]%
        {franca2018}
\bibfield{author}{\bibinfo{person}{Cesar Franca}, \bibinfo{person}{Fabio Q.~B.
  da Silva}, {and} \bibinfo{person}{Helen Sharp}.}
  \bibinfo{year}{2020}\natexlab{}.
\newblock \showarticletitle{Motivation and Satisfaction of Software Engineers}.
\newblock \bibinfo{journal}{\emph{IEEE Transactions on Software Engineering}}
  \bibinfo{volume}{46}, \bibinfo{number}{2} (\bibinfo{year}{2020}),
  \bibinfo{pages}{118–140}.
\newblock
\urldef\tempurl%
\url{https://doi.org/10.1109/tse.2018.2842201}
\showDOI{\tempurl}


\bibitem[\protect\citeauthoryear{Furia, Feldt, and Torkar}{Furia
  et~al\mbox{.}}{2019}]%
        {furia2019bayesian}
\bibfield{author}{\bibinfo{person}{Carlo~Alberto Furia},
  \bibinfo{person}{Robert Feldt}, {and} \bibinfo{person}{Richard Torkar}.}
  \bibinfo{year}{2019}\natexlab{}.
\newblock \showarticletitle{Bayesian data analysis in empirical software
  engineering research}.
\newblock \bibinfo{journal}{\emph{IEEE Transactions on Software Engineering}}
  \bibinfo{volume}{1}, \bibinfo{number}{1} (\bibinfo{year}{2019}),
  \bibinfo{pages}{1--26}.
\newblock
\urldef\tempurl%
\url{https://doi.org/10.1109/TSE.2019.2935974}
\showDOI{\tempurl}


\bibitem[\protect\citeauthoryear{Ginty}{Ginty}{2013}]%
        {ginty2013}
\bibfield{author}{\bibinfo{person}{Annie~T. Ginty}.}
  \bibinfo{year}{2013}\natexlab{}.
\newblock \showarticletitle{Psychometric Properties}.
\newblock In \bibinfo{booktitle}{\emph{Encyclopedia of Behavioral Medicine}},
  \bibfield{editor}{\bibinfo{person}{Marc~D. Gellman} {and}
  \bibinfo{person}{J.~Rick Turner}} (Eds.). \bibinfo{publisher}{Springer New
  York}, \bibinfo{address}{New York, NY}, \bibinfo{pages}{1563--1564}.
\newblock
\showISBNx{978-1-4419-1005-9}
\urldef\tempurl%
\url{https://doi.org/10.1007/978-1-4419-1005-9_480}
\showDOI{\tempurl}


\bibitem[\protect\citeauthoryear{Glaser}{Glaser}{1963}]%
        {glaser1963}
\bibfield{author}{\bibinfo{person}{Robert Glaser}.}
  \bibinfo{year}{1963}\natexlab{}.
\newblock \showarticletitle{Instructional technology and the measurement of
  learing outcomes: Some questions.}
\newblock \bibinfo{journal}{\emph{American Psychologist}} \bibinfo{volume}{18},
  \bibinfo{number}{8} (\bibinfo{year}{1963}), \bibinfo{pages}{519--521}.
\newblock
\urldef\tempurl%
\url{https://doi.org/10.1037/h0049294}
\showDOI{\tempurl}


\bibitem[\protect\citeauthoryear{Graziotin, Fagerholm, Wang, and
  Abrahamsson}{Graziotin et~al\mbox{.}}{2017}]%
        {graziotin2017unhappy}
\bibfield{author}{\bibinfo{person}{Daniel Graziotin}, \bibinfo{person}{Fabian
  Fagerholm}, \bibinfo{person}{Xiaofeng Wang}, {and} \bibinfo{person}{Pekka
  Abrahamsson}.} \bibinfo{year}{2017}\natexlab{}.
\newblock \showarticletitle{On the Unhappiness of Software Developers},
  \bibfield{editor}{\bibinfo{person}{Emilia Mendes}, \bibinfo{person}{Steve
  Counsell}, {and} \bibinfo{person}{Kai Petersen}} (Eds.),
  Vol.~\bibinfo{volume}{21st International Conference on Evaluation and
  Assessment in Software Engineering}. \bibinfo{publisher}{ACM Press},
  \bibinfo{address}{New York, New York, USA}, \bibinfo{pages}{324--333}.
\newblock


\bibitem[\protect\citeauthoryear{Graziotin, Lenberg, Feldt, and
  Wagner}{Graziotin et~al\mbox{.}}{2021}]%
        {graziotin2020}
\bibfield{author}{\bibinfo{person}{Daniel Graziotin}, \bibinfo{person}{Per
  Lenberg}, \bibinfo{person}{Robert Feldt}, {and} \bibinfo{person}{Stefan
  Wagner}.} \bibinfo{year}{2021}\natexlab{}.
\newblock \bibinfo{title}{{Behavioral Software Engineering - Example of
  psychometric evaluation with R}}.
\newblock
\newblock
\urldef\tempurl%
\url{https://doi.org/10.5281/zenodo.3799603}
\showDOI{\tempurl}


\bibitem[\protect\citeauthoryear{Graziotin, Wang, and Abrahamsson}{Graziotin
  et~al\mbox{.}}{2015a}]%
        {graziotin2015affect}
\bibfield{author}{\bibinfo{person}{Daniel Graziotin}, \bibinfo{person}{Xiaofeng
  Wang}, {and} \bibinfo{person}{Pekka Abrahamsson}.}
  \bibinfo{year}{2015}\natexlab{a}.
\newblock \showarticletitle{The Affect of Software Developers: Common
  Misconceptions and Measurements}. In \bibinfo{booktitle}{\emph{2015 IEEE/ACM
  8th International Workshop on Cooperative and Human Aspects of Software
  Engineering (CHASE)}}. \bibinfo{publisher}{IEEE},
  \bibinfo{address}{Piscataway, NJ}, \bibinfo{pages}{123--124}.
\newblock


\bibitem[\protect\citeauthoryear{Graziotin, Wang, and Abrahamsson}{Graziotin
  et~al\mbox{.}}{2015b}]%
        {graziotin2015you}
\bibfield{author}{\bibinfo{person}{Daniel Graziotin}, \bibinfo{person}{Xiaofeng
  Wang}, {and} \bibinfo{person}{Pekka Abrahamsson}.}
  \bibinfo{year}{2015}\natexlab{b}.
\newblock \showarticletitle{Do feelings matter? On the correlation of affects
  and the self-assessed productivity in software engineering}.
\newblock \bibinfo{journal}{\emph{Journal of Software: Evolution and Process}}
  \bibinfo{volume}{27}, \bibinfo{number}{7} (\bibinfo{year}{2015}),
  \bibinfo{pages}{467–487}.
\newblock


\bibitem[\protect\citeauthoryear{Graziotin, Wang, and Abrahamsson}{Graziotin
  et~al\mbox{.}}{2015c}]%
        {graziotin2015}
\bibfield{author}{\bibinfo{person}{Daniel Graziotin}, \bibinfo{person}{Xiaofeng
  Wang}, {and} \bibinfo{person}{Pekka Abrahamsson}.}
  \bibinfo{year}{2015}\natexlab{c}.
\newblock \showarticletitle{Understanding the Affect of Developers: Theoretical
  Background and Guidelines for Psychoempirical Software Engineering}. In
  \bibinfo{booktitle}{\emph{Proceedings of the 7th International Workshop on
  Social Software Engineering}} (Bergamo, Italy) \emph{(\bibinfo{series}{SSE
  2015})}. \bibinfo{publisher}{ACM}, \bibinfo{address}{New York, NY, USA},
  \bibinfo{pages}{25--32}.
\newblock
\showISBNx{978-1-4503-3818-9}
\urldef\tempurl%
\url{https://doi.org/10.1145/2804381.2804386}
\showDOI{\tempurl}


\bibitem[\protect\citeauthoryear{Green}{Green}{2009}]%
        {green2009}
\bibfield{author}{\bibinfo{person}{Christopher~D. Green}.}
  \bibinfo{year}{2009}\natexlab{}.
\newblock \showarticletitle{Darwinian theory, functionalism, and the first
  American psychological revolution.}
\newblock \bibinfo{journal}{\emph{American Psychologist}} \bibinfo{volume}{64},
  \bibinfo{number}{2} (\bibinfo{year}{2009}), \bibinfo{pages}{75--83}.
\newblock
\urldef\tempurl%
\url{https://doi.org/10.1037/a0013338}
\showDOI{\tempurl}


\bibitem[\protect\citeauthoryear{Gren}{Gren}{2018}]%
        {gren2018}
\bibfield{author}{\bibinfo{person}{Lucas Gren}.}
  \bibinfo{year}{2018}\natexlab{}.
\newblock \showarticletitle{Standards of validity and the validity of standards
  in behavioral software engineering research}. \bibinfo{publisher}{ACM Press},
  \bibinfo{address}{New York, New York, USA}.
\newblock
\urldef\tempurl%
\url{https://doi.org/10.1145/3239235.3267437}
\showDOI{\tempurl}


\bibitem[\protect\citeauthoryear{Gren and Goldman}{Gren and Goldman}{2016}]%
        {gren2016}
\bibfield{author}{\bibinfo{person}{Lucas Gren} {and} \bibinfo{person}{Alfredo
  Goldman}.} \bibinfo{year}{2016}\natexlab{}.
\newblock \showarticletitle{Useful Statistical Methods for Human Factors
  Research in Software Engineering: A Discussion on Validation with
  Quantitative Data}. In \bibinfo{booktitle}{\emph{Proceedings of the 9th
  International Workshop on Cooperative and Human Aspects of Software
  Engineering}} (Austin, Texas) \emph{(\bibinfo{series}{CHASE ’16})}.
  \bibinfo{publisher}{Association for Computing Machinery},
  \bibinfo{address}{New York, NY, USA}, \bibinfo{pages}{121–124}.
\newblock
\showISBNx{9781450341554}
\urldef\tempurl%
\url{https://doi.org/10.1145/2897586.2897588}
\showDOI{\tempurl}


\bibitem[\protect\citeauthoryear{Guilford}{Guilford}{1954a}]%
        {guilford1938}
\bibfield{author}{\bibinfo{person}{Joy~Paul Guilford}.}
  \bibinfo{year}{1954}\natexlab{a}.
\newblock \bibinfo{booktitle}{\emph{Psychometric methods} (\bibinfo{edition}{1}
  ed.)}.
\newblock \bibinfo{publisher}{McGraw-Hill Book Co.}, \bibinfo{address}{New
  York, London}.
\newblock


\bibitem[\protect\citeauthoryear{Guilford}{Guilford}{1954b}]%
        {guilford1954}
\bibfield{author}{\bibinfo{person}{Joy~Paul Guilford}.}
  \bibinfo{year}{1954}\natexlab{b}.
\newblock \bibinfo{booktitle}{\emph{Psychometric methods} (\bibinfo{edition}{2}
  ed.)}.
\newblock \bibinfo{publisher}{McGraw-Hill Book Co.}, \bibinfo{address}{New
  York, London}.
\newblock


\bibitem[\protect\citeauthoryear{Hambleton and Rodgers}{Hambleton and
  Rodgers}{1995}]%
        {hambleton1995item}
\bibfield{author}{\bibinfo{person}{Ronald~K Hambleton} {and}
  \bibinfo{person}{Jane Rodgers}.} \bibinfo{year}{1995}\natexlab{}.
\newblock \bibinfo{booktitle}{\emph{Item bias review}}.
\newblock \bibinfo{type}{{T}echnical {R}eport}. \bibinfo{institution}{ERIC
  Clearinghouse on Assessment and Evaluation, the Catholic University of
  America, Department of Education}.
\newblock


\bibitem[\protect\citeauthoryear{Hambleton, Swaminathan, and Rogers}{Hambleton
  et~al\mbox{.}}{1991}]%
        {hambleton1991}
\bibfield{author}{\bibinfo{person}{Ronald~K Hambleton},
  \bibinfo{person}{Hariharan Swaminathan}, {and} \bibinfo{person}{H~Jane
  Rogers}.} \bibinfo{year}{1991}\natexlab{}.
\newblock \bibinfo{booktitle}{\emph{Fundamentals of item response theory}}.
\newblock \bibinfo{publisher}{Sage Publications}, \bibinfo{address}{Newbury
  Park, Calif}.
\newblock
\showISBNx{978-0803936478}


\bibitem[\protect\citeauthoryear{Hilgard}{Hilgard}{1980}]%
        {hilgard1980}
\bibfield{author}{\bibinfo{person}{Ernest~R Hilgard}.}
  \bibinfo{year}{1980}\natexlab{}.
\newblock \showarticletitle{The trilogy of mind: Cognition, affection, and
  conation}.
\newblock \bibinfo{journal}{\emph{Journal of the History of the Behavioral
  Sciences}} \bibinfo{volume}{16}, \bibinfo{number}{2} (\bibinfo{year}{1980}),
  \bibinfo{pages}{107--117}.
\newblock


\bibitem[\protect\citeauthoryear{Hilton}{Hilton}{2017}]%
        {hilton2017}
\bibfield{author}{\bibinfo{person}{Charlotte~Emma Hilton}.}
  \bibinfo{year}{2017}\natexlab{}.
\newblock \showarticletitle{The importance of pretesting questionnaires: a
  field research example of cognitive pretesting the Exercise referral Quality
  of Life Scale (ER-QLS)}.
\newblock \bibinfo{journal}{\emph{International Journal of Social Research
  Methodology}} \bibinfo{volume}{20}, \bibinfo{number}{1}
  (\bibinfo{year}{2017}), \bibinfo{pages}{21–34}.
\newblock
\urldef\tempurl%
\url{https://doi.org/10.1080/13645579.2015.1091640}
\showDOI{\tempurl}


\bibitem[\protect\citeauthoryear{Hogan}{Hogan}{2007}]%
        {hogan2017}
\bibfield{author}{\bibinfo{person}{Robert Hogan}.}
  \bibinfo{year}{2007}\natexlab{}.
\newblock \bibinfo{booktitle}{\emph{Personality and the fate of
  organizations}}.
\newblock \bibinfo{publisher}{Erlbaum}, \bibinfo{address}{Mahwah, N.J}.
\newblock
\showISBNx{978-0805841428}


\bibitem[\protect\citeauthoryear{Horn}{Horn}{1965}]%
        {horn1965}
\bibfield{author}{\bibinfo{person}{John~L Horn}.}
  \bibinfo{year}{1965}\natexlab{}.
\newblock \showarticletitle{A Rationale And Test For The Number Of Factors In
  Factor Analysis.}
\newblock \bibinfo{journal}{\emph{Psychometrika}}  \bibinfo{volume}{30}
  (\bibinfo{year}{1965}), \bibinfo{pages}{179–185}.
\newblock


\bibitem[\protect\citeauthoryear{Ji, Li, Conradi, Liu, Ma, and Chen}{Ji
  et~al\mbox{.}}{2008}]%
        {ji2008}
\bibfield{author}{\bibinfo{person}{Junzhong Ji}, \bibinfo{person}{Jingyue Li},
  \bibinfo{person}{Reidar Conradi}, \bibinfo{person}{Chunnian Liu},
  \bibinfo{person}{Jianqiang Ma}, {and} \bibinfo{person}{Weibing Chen}.}
  \bibinfo{year}{2008}\natexlab{}.
\newblock \showarticletitle{Some lessons learned in conducting software
  engineering surveys in china}, Vol.~\bibinfo{volume}{Proceedings of the
  Second ACM-IEEE international symposium on Empirical software engineering and
  measurement}. \bibinfo{publisher}{ACM}, \bibinfo{address}{Piscataway, NJ},
  \bibinfo{pages}{168--177}.
\newblock


\bibitem[\protect\citeauthoryear{Johanson and Brooks}{Johanson and
  Brooks}{2010}]%
        {johanson2010}
\bibfield{author}{\bibinfo{person}{George~A. Johanson} {and}
  \bibinfo{person}{Gordon~P. Brooks}.} \bibinfo{year}{2010}\natexlab{}.
\newblock \showarticletitle{Initial Scale Development: Sample Size for Pilot
  Studies}.
\newblock \bibinfo{journal}{\emph{Educational and Psychological Measurement}}
  \bibinfo{volume}{70}, \bibinfo{number}{3} (\bibinfo{year}{2010}),
  \bibinfo{pages}{394–400}.
\newblock
\urldef\tempurl%
\url{https://doi.org/10.1177/0013164409355692}
\showDOI{\tempurl}


\bibitem[\protect\citeauthoryear{Jones and Thissen}{Jones and Thissen}{2006}]%
        {jones2006}
\bibfield{author}{\bibinfo{person}{Lyle~V. Jones} {and} \bibinfo{person}{David
  Thissen}.} \bibinfo{year}{2006}\natexlab{}.
\newblock \showarticletitle{A History and Overview of Psychometrics}.
\newblock In \bibinfo{booktitle}{\emph{Handbook of Statistics: Psychometrics}}.
  \bibinfo{publisher}{Elsevier}, \bibinfo{address}{Amsterdam Boston},
  \bibinfo{pages}{1--27}.
\newblock
\urldef\tempurl%
\url{https://doi.org/10.1016/s0169-7161(06)26001-2}
\showDOI{\tempurl}


\bibitem[\protect\citeauthoryear{Kaiser}{Kaiser}{1958}]%
        {kaiser1958}
\bibfield{author}{\bibinfo{person}{Henry~F. Kaiser}.}
  \bibinfo{year}{1958}\natexlab{}.
\newblock \showarticletitle{The varimax criterion for analytic rotation in
  factor analysis}.
\newblock \bibinfo{journal}{\emph{Psychometrika}} \bibinfo{volume}{23},
  \bibinfo{number}{3} (\bibinfo{year}{1958}), \bibinfo{pages}{187–200}.
\newblock


\bibitem[\protect\citeauthoryear{Kaiser}{Kaiser}{1960}]%
        {kaiser1960}
\bibfield{author}{\bibinfo{person}{Henry~F. Kaiser}.}
  \bibinfo{year}{1960}\natexlab{}.
\newblock \showarticletitle{The Application of Electronic Computers to Factor
  Analysis}.
\newblock \bibinfo{journal}{\emph{Educational and Psychological Measurement}}
  \bibinfo{volume}{20}, \bibinfo{number}{1} (\bibinfo{year}{1960}),
  \bibinfo{pages}{141–151}.
\newblock


\bibitem[\protect\citeauthoryear{Kalton and Schuman}{Kalton and
  Schuman}{1982}]%
        {kalton1982}
\bibfield{author}{\bibinfo{person}{Graham Kalton} {and} \bibinfo{person}{Howard
  Schuman}.} \bibinfo{year}{1982}\natexlab{}.
\newblock \showarticletitle{The effect of the question on survey responses: A
  review}.
\newblock \bibinfo{journal}{\emph{Journal of the Royal Statistical Society:
  Series A (General)}} \bibinfo{volume}{145}, \bibinfo{number}{1}
  (\bibinfo{year}{1982}), \bibinfo{pages}{42–57}.
\newblock


\bibitem[\protect\citeauthoryear{Kaplan}{Kaplan}{2008}]%
        {kaplan2008structural}
\bibfield{author}{\bibinfo{person}{David Kaplan}.}
  \bibinfo{year}{2008}\natexlab{}.
\newblock \bibinfo{booktitle}{\emph{Structural equation modeling: Foundations
  and extensions}}. Vol.~\bibinfo{volume}{10}.
\newblock \bibinfo{publisher}{SAGE}, \bibinfo{address}{Los Angeles}.
\newblock


\bibitem[\protect\citeauthoryear{Kitchenham}{Kitchenham}{2007}]%
        {kitchenham2007}
\bibfield{author}{\bibinfo{person}{B~A Kitchenham}.}
  \bibinfo{year}{2007}\natexlab{}.
\newblock \bibinfo{booktitle}{\emph{Guidelines for performing systematic
  literature reviews in software engineering}}.
\newblock \bibinfo{type}{{T}echnical {R}eport}. \bibinfo{institution}{Keele
  University and University of Durham Keele and Durham, UK}.
  \bibinfo{pages}{1--65} pages.
\newblock


\bibitem[\protect\citeauthoryear{Kitchenham and Pfleeger}{Kitchenham and
  Pfleeger}{2008}]%
        {kitchenham2008}
\bibfield{author}{\bibinfo{person}{Barbara~A. Kitchenham} {and}
  \bibinfo{person}{Shari~L. Pfleeger}.} \bibinfo{year}{2008}\natexlab{}.
\newblock \showarticletitle{Personal Opinion Surveys}.
\newblock In \bibinfo{booktitle}{\emph{Guide to Advanced Empirical Software
  Engineering}}, \bibfield{editor}{\bibinfo{person}{Forrest Shull},
  \bibinfo{person}{Janice Singer}, {and} \bibinfo{person}{Dag I.~K. Sjøberg}}
  (Eds.). \bibinfo{publisher}{Springer London}, \bibinfo{address}{London},
  \bibinfo{pages}{63--92}.
\newblock
\urldef\tempurl%
\url{https://doi.org/10.1007/978-1-84800-044-5_3}
\showDOI{\tempurl}


\bibitem[\protect\citeauthoryear{Kline}{Kline}{2015}]%
        {kline2015handbook}
\bibfield{author}{\bibinfo{person}{Paul Kline}.}
  \bibinfo{year}{2015}\natexlab{}.
\newblock \bibinfo{booktitle}{\emph{A handbook of test construction (psychology
  revivals): introduction to psychometric design}}.
\newblock \bibinfo{publisher}{Routledge}, \bibinfo{address}{London New York}.
\newblock


\bibitem[\protect\citeauthoryear{Koltko-Rivera}{Koltko-Rivera}{2004}]%
        {koltkorivera2004}
\bibfield{author}{\bibinfo{person}{Mark~E Koltko-Rivera}.}
  \bibinfo{year}{2004}\natexlab{}.
\newblock \showarticletitle{The psychology of worldviews}.
\newblock \bibinfo{journal}{\emph{Review of general psychology}}
  \bibinfo{volume}{8}, \bibinfo{number}{1} (\bibinfo{year}{2004}),
  \bibinfo{pages}{3–58}.
\newblock
\urldef\tempurl%
\url{https://doi.org/10.1021/bi00696a014}
\showDOI{\tempurl}


\bibitem[\protect\citeauthoryear{Kootstra}{Kootstra}{2006}]%
        {kootstra2006}
\bibfield{author}{\bibinfo{person}{GJ Kootstra}.}
  \bibinfo{year}{2006}\natexlab{}.
\newblock \bibinfo{booktitle}{\emph{Exploratory Factor Analysis: Theory and
  Application}}.
\newblock \bibinfo{type}{{T}echnical {R}eport}.
  \bibinfo{institution}{University of Groningen}. \bibinfo{pages}{1–15}
  pages.
\newblock


\bibitem[\protect\citeauthoryear{Lee}{Lee}{1981}]%
        {lee1981}
\bibfield{author}{\bibinfo{person}{Sik-Yum Lee}.}
  \bibinfo{year}{1981}\natexlab{}.
\newblock \showarticletitle{A bayesian approach to confirmatory factor
  analysis}.
\newblock \bibinfo{journal}{\emph{Psychometrika}} \bibinfo{volume}{46},
  \bibinfo{number}{2} (\bibinfo{year}{1981}), \bibinfo{pages}{153–160}.
\newblock


\bibitem[\protect\citeauthoryear{Lenberg, Feldt, Tengberg, Tidefors, and
  Graziotin}{Lenberg et~al\mbox{.}}{2017a}]%
        {lenberg2017a}
\bibfield{author}{\bibinfo{person}{Per Lenberg}, \bibinfo{person}{Robert
  Feldt}, \bibinfo{person}{Lars Göran~Wallgren Tengberg},
  \bibinfo{person}{Inga Tidefors}, {and} \bibinfo{person}{Daniel Graziotin}.}
  \bibinfo{year}{2017}\natexlab{a}.
\newblock \bibinfo{title}{Behavioral software engineering - guidelines for
  qualitative studies}.
\newblock
\newblock
\showeprint[arxiv]{1712.08341}~[cs.SE]
\newblock
\shownote{Available https://arxiv.org/abs/1712.08341.}


\bibitem[\protect\citeauthoryear{Lenberg, Feldt, and Wallgren}{Lenberg
  et~al\mbox{.}}{2015}]%
        {lenberg2015}
\bibfield{author}{\bibinfo{person}{Per Lenberg}, \bibinfo{person}{Robert
  Feldt}, {and} \bibinfo{person}{Lars~G{\"o}ran Wallgren}.}
  \bibinfo{year}{2015}\natexlab{}.
\newblock \showarticletitle{Behavioral software engineering: A definition and
  systematic literature review}.
\newblock \bibinfo{journal}{\emph{Journal of Systems and Software}}
  \bibinfo{volume}{107} (\bibinfo{year}{2015}), \bibinfo{pages}{15--37}.
\newblock
\urldef\tempurl%
\url{https://doi.org/10.1016/j.jss.2015.04.084}
\showDOI{\tempurl}


\bibitem[\protect\citeauthoryear{Lenberg, Wallgren~Tengberg, and Feldt}{Lenberg
  et~al\mbox{.}}{2017b}]%
        {lenberg2017b}
\bibfield{author}{\bibinfo{person}{Per Lenberg}, \bibinfo{person}{Lars~Göran
  Wallgren~Tengberg}, {and} \bibinfo{person}{Robert Feldt}.}
  \bibinfo{year}{2017}\natexlab{b}.
\newblock \showarticletitle{An initial analysis of software engineers’
  attitudes towards organizational change}.
\newblock \bibinfo{journal}{\emph{Empirical Software Engineering}}
  \bibinfo{volume}{22}, \bibinfo{number}{4} (\bibinfo{year}{2017}),
  \bibinfo{pages}{2179–2205}.
\newblock


\bibitem[\protect\citeauthoryear{Levy and Mislevy}{Levy and Mislevy}{2016}]%
        {levy2016bayesian}
\bibfield{author}{\bibinfo{person}{Roy Levy} {and} \bibinfo{person}{Robert~J
  Mislevy}.} \bibinfo{year}{2016}\natexlab{}.
\newblock \bibinfo{booktitle}{\emph{Bayesian psychometric modeling}}.
\newblock \bibinfo{publisher}{CRC Press}, \bibinfo{address}{Boca Raton,
  Florida}.
\newblock


\bibitem[\protect\citeauthoryear{Lewis}{Lewis}{2020}]%
        {lewis2020personality}
\bibfield{author}{\bibinfo{person}{Clayton Lewis}.}
  \bibinfo{year}{2020}\natexlab{}.
\newblock \showarticletitle{On personality testing and software engineering}.
  In \bibinfo{booktitle}{\emph{31st Annual Workshop of the Psychology of
  Programming Interest Group (PPIG 2020)}}. \bibinfo{publisher}{PPIG},
  \bibinfo{address}{PPIG}, \bibinfo{pages}{39--41}.
\newblock


\bibitem[\protect\citeauthoryear{Likert}{Likert}{1932}]%
        {likert1932}
\bibfield{author}{\bibinfo{person}{Rensis Likert}.}
  \bibinfo{year}{1932}\natexlab{}.
\newblock \showarticletitle{A technique for the measurement of attitudes.}
\newblock \bibinfo{journal}{\emph{Archives of psychology}}
  \bibinfo{volume}{22}, \bibinfo{number}{40} (\bibinfo{year}{1932}),
  \bibinfo{pages}{1--55}.
\newblock


\bibitem[\protect\citeauthoryear{Loevinger}{Loevinger}{1957}]%
        {loevinger1957}
\bibfield{author}{\bibinfo{person}{Jane Loevinger}.}
  \bibinfo{year}{1957}\natexlab{}.
\newblock \showarticletitle{Objective Tests as Instruments of Psychological
  Theory}.
\newblock \bibinfo{journal}{\emph{Psychological Reports}}  \bibinfo{volume}{3}
  (\bibinfo{year}{1957}), \bibinfo{pages}{635–694}.
\newblock
\urldef\tempurl%
\url{https://doi.org/10.2466/pr0.1957.3.3.635}
\showDOI{\tempurl}


\bibitem[\protect\citeauthoryear{Lu, Chow, and Loken}{Lu et~al\mbox{.}}{2016}]%
        {lu2016}
\bibfield{author}{\bibinfo{person}{ZH Lu}, \bibinfo{person}{SM Chow}, {and}
  \bibinfo{person}{E Loken}.} \bibinfo{year}{2016}\natexlab{}.
\newblock \showarticletitle{Bayesian Factor Analysis as a Variable-Selection
  Problem: Alternative Priors and Consequences.}
\newblock \bibinfo{journal}{\emph{Multivariate Behav Res}}
  \bibinfo{volume}{51}, \bibinfo{number}{4} (\bibinfo{year}{2016}),
  \bibinfo{pages}{519–539}.
\newblock


\bibitem[\protect\citeauthoryear{MacCallum and Austin}{MacCallum and
  Austin}{2000}]%
        {maccallum2000}
\bibfield{author}{\bibinfo{person}{RC MacCallum} {and} \bibinfo{person}{JT
  Austin}.} \bibinfo{year}{2000}\natexlab{}.
\newblock \showarticletitle{Applications of structural equation modeling in
  psychological research.}
\newblock \bibinfo{journal}{\emph{Annu Rev Psychol}}  \bibinfo{volume}{51}
  (\bibinfo{year}{2000}), \bibinfo{pages}{201–226}.
\newblock


\bibitem[\protect\citeauthoryear{MacCallum, Widaman, Zhang, and Hong}{MacCallum
  et~al\mbox{.}}{1999}]%
        {maccallum1999}
\bibfield{author}{\bibinfo{person}{Robert~C. MacCallum},
  \bibinfo{person}{Keith~F. Widaman}, \bibinfo{person}{Shaobo Zhang}, {and}
  \bibinfo{person}{Sehee Hong}.} \bibinfo{year}{1999}\natexlab{}.
\newblock \showarticletitle{Sample size in factor analysis.}
\newblock \bibinfo{journal}{\emph{Psychological Methods}} \bibinfo{volume}{4},
  \bibinfo{number}{1} (\bibinfo{year}{1999}), \bibinfo{pages}{84–99}.
\newblock


\bibitem[\protect\citeauthoryear{Masters}{Masters}{1988}]%
        {masters1988}
\bibfield{author}{\bibinfo{person}{Geofferey~N. Masters}.}
  \bibinfo{year}{1988}\natexlab{}.
\newblock \showarticletitle{Item Discrimination: When More Is Worse}.
\newblock \bibinfo{journal}{\emph{Journal of Educational Measurement}}
  \bibinfo{volume}{25}, \bibinfo{number}{1} (\bibinfo{year}{1988}),
  \bibinfo{pages}{15–29}.
\newblock
\urldef\tempurl%
\url{https://doi.org/10.1111/j.1745-3984.1988.tb00288.x}
\showDOI{\tempurl}


\bibitem[\protect\citeauthoryear{McDonald and Edwards}{McDonald and
  Edwards}{2007}]%
        {mcdonald2007}
\bibfield{author}{\bibinfo{person}{Sharon McDonald} {and}
  \bibinfo{person}{Helen~M. Edwards}.} \bibinfo{year}{2007}\natexlab{}.
\newblock \showarticletitle{Who should test whom}.
\newblock \bibinfo{journal}{\emph{Commun. ACM}} \bibinfo{volume}{50},
  \bibinfo{number}{1} (\bibinfo{year}{2007}), \bibinfo{pages}{66–71}.
\newblock


\bibitem[\protect\citeauthoryear{Michell}{Michell}{2000}]%
        {michell2000}
\bibfield{author}{\bibinfo{person}{Joel Michell}.}
  \bibinfo{year}{2000}\natexlab{}.
\newblock \showarticletitle{Normal science, pathological science and
  psychometrics}.
\newblock \bibinfo{journal}{\emph{Theory \& Psychology}} \bibinfo{volume}{10},
  \bibinfo{number}{5} (\bibinfo{year}{2000}), \bibinfo{pages}{639–667}.
\newblock


\bibitem[\protect\citeauthoryear{Miller, Chepp, Willson, and Padilla}{Miller
  et~al\mbox{.}}{2014}]%
        {miller2014}
\bibfield{author}{\bibinfo{person}{Kristen Miller}, \bibinfo{person}{Valerie
  Chepp}, \bibinfo{person}{Stephanie Willson}, {and} \bibinfo{person}{Jose-Luis
  Padilla}.} \bibinfo{year}{2014}\natexlab{}.
\newblock \bibinfo{booktitle}{\emph{Cognitive interviewing methodology}}.
\newblock \bibinfo{publisher}{John Wiley \& Sons}, \bibinfo{address}{Hoboken,
  New Jersey}.
\newblock


\bibitem[\protect\citeauthoryear{Molléri, Petersen, and Mendes}{Molléri
  et~al\mbox{.}}{2016}]%
        {molleri2016}
\bibfield{author}{\bibinfo{person}{Jefferson~Seide Molléri},
  \bibinfo{person}{Kai Petersen}, {and} \bibinfo{person}{Emilia Mendes}.}
  \bibinfo{year}{2016}\natexlab{}.
\newblock \showarticletitle{Survey guidelines in software engineering: An
  annotated review}, Vol.~\bibinfo{volume}{Proceedings of the 10th ACM/IEEE
  International Symposium on Empirical Software Engineering and Measurement}.
  \bibinfo{publisher}{IEEE}, \bibinfo{address}{Piscataway, NJ},
  \bibinfo{pages}{58}.
\newblock


\bibitem[\protect\citeauthoryear{Morgan}{Morgan}{1980}]%
        {morgan1980}
\bibfield{author}{\bibinfo{person}{WP Morgan}.}
  \bibinfo{year}{1980}\natexlab{}.
\newblock \showarticletitle{The trait psychology controversy.}
\newblock \bibinfo{journal}{\emph{Res Q Exerc Sport}} \bibinfo{volume}{51},
  \bibinfo{number}{1} (\bibinfo{year}{1980}), \bibinfo{pages}{50–76}.
\newblock
\urldef\tempurl%
\url{https://doi.org/10.1080/02701367.1980.10609275}
\showDOI{\tempurl}


\bibitem[\protect\citeauthoryear{Muthén and Asparouhov}{Muthén and
  Asparouhov}{2012}]%
        {muthen2012}
\bibfield{author}{\bibinfo{person}{Bengt Muthén} {and}
  \bibinfo{person}{Tihomir Asparouhov}.} \bibinfo{year}{2012}\natexlab{}.
\newblock \showarticletitle{Bayesian structural equation modeling: A more
  flexible representation of substantive theory.}
\newblock \bibinfo{journal}{\emph{Psychological Methods}} \bibinfo{volume}{17},
  \bibinfo{number}{3} (\bibinfo{year}{2012}), \bibinfo{pages}{313–335}.
\newblock


\bibitem[\protect\citeauthoryear{Norman}{Norman}{2002}]%
        {norman2002}
\bibfield{author}{\bibinfo{person}{Geoff Norman}.}
  \bibinfo{year}{2002}\natexlab{}.
\newblock \showarticletitle{Research in medical education: three decades of
  progress}.
\newblock \bibinfo{journal}{\emph{Bmj}} \bibinfo{volume}{324},
  \bibinfo{number}{7353} (\bibinfo{year}{2002}), \bibinfo{pages}{1560–1562}.
\newblock
\urldef\tempurl%
\url{https://doi.org/10.1136/bmj.324.7353.1560}
\showDOI{\tempurl}


\bibitem[\protect\citeauthoryear{Norris and Lecavalier}{Norris and
  Lecavalier}{2010}]%
        {Norris2010}
\bibfield{author}{\bibinfo{person}{Megan Norris} {and} \bibinfo{person}{Luc
  Lecavalier}.} \bibinfo{year}{2010}\natexlab{}.
\newblock \showarticletitle{Evaluating the Use of Exploratory Factor Analysis
  in Developmental Disability Psychological Research}.
\newblock \bibinfo{journal}{\emph{Journal of Autism and Developmental
  Disorders}} \bibinfo{volume}{40}, \bibinfo{number}{1} (\bibinfo{date}{01 Jan}
  \bibinfo{year}{2010}), \bibinfo{pages}{8--20}.
\newblock
\showISSN{1573-3432}
\urldef\tempurl%
\url{https://doi.org/10.1007/s10803-009-0816-2}
\showDOI{\tempurl}


\bibitem[\protect\citeauthoryear{Nunnally}{Nunnally}{1994}]%
        {nunnally1994psychometric}
\bibfield{author}{\bibinfo{person}{Jum Nunnally}.}
  \bibinfo{year}{1994}\natexlab{}.
\newblock \bibinfo{booktitle}{\emph{Psychometric theory}}.
\newblock \bibinfo{publisher}{McGraw-Hill}, \bibinfo{address}{New York}.
\newblock
\showISBNx{978-0070478497}


\bibitem[\protect\citeauthoryear{Nunnally}{Nunnally}{1978}]%
        {nunnally1978}
\bibfield{author}{\bibinfo{person}{Jum~C. Nunnally}.}
  \bibinfo{year}{1978}\natexlab{}.
\newblock \showarticletitle{An Overview of Psychological Measurement}.
\newblock In \bibinfo{booktitle}{\emph{Clinical Diagnosis of Mental
  Disorders}}. \bibinfo{publisher}{Springer {US}}, \bibinfo{pages}{97--146}.
\newblock
\urldef\tempurl%
\url{https://doi.org/10.1007/978-1-4684-2490-4_4}
\showDOI{\tempurl}


\bibitem[\protect\citeauthoryear{Oppenheim}{Oppenheim}{1992}]%
        {oppenheim1992}
\bibfield{author}{\bibinfo{person}{A.N. Oppenheim}.}
  \bibinfo{year}{1992}\natexlab{}.
\newblock \bibinfo{booktitle}{\emph{Questionnaire Design, Interviewing and
  Attitude Measurement}}.
\newblock \bibinfo{publisher}{Pinter Pub Ltd}, \bibinfo{address}{London, UK}.
\newblock


\bibitem[\protect\citeauthoryear{Ostberg, Graziotin, Wagner, and
  Derntl}{Ostberg et~al\mbox{.}}{2017}]%
        {ostberg2017towards}
\bibfield{author}{\bibinfo{person}{Jan-Peter Ostberg}, \bibinfo{person}{Daniel
  Graziotin}, \bibinfo{person}{Stefan Wagner}, {and} \bibinfo{person}{Birgit
  Derntl}.} \bibinfo{year}{2017}\natexlab{}.
\newblock \showarticletitle{Towards the Assessment of Stress and Emotional
  Responses of a Salutogenesis-Enhanced Software Tool Using Psychophysiological
  Measurements}. In \bibinfo{booktitle}{\emph{Towards the Assessment of Stress
  and Emotional Responses of a Salutogenesis-Enhanced Software Tool Using
  Psychophysiological Measurements}}. \bibinfo{publisher}{IEEE},
  \bibinfo{address}{Piscataway, NJ}, \bibinfo{pages}{22--25}.
\newblock
\urldef\tempurl%
\url{https://doi.org/10.1109/semotion.2017.4}
\showDOI{\tempurl}


\bibitem[\protect\citeauthoryear{Pearson}{Pearson}{1901}]%
        {pearson1901}
\bibfield{author}{\bibinfo{person}{Karl Pearson}.}
  \bibinfo{year}{1901}\natexlab{}.
\newblock \showarticletitle{{LIII}. On lines and planes of closest fit to
  systems of points in space}.
\newblock \bibinfo{journal}{\emph{The London, Edinburgh, and Dublin
  Philosophical Magazine and Journal of Science}} \bibinfo{volume}{2},
  \bibinfo{number}{11} (\bibinfo{date}{Nov.} \bibinfo{year}{1901}),
  \bibinfo{pages}{559--572}.
\newblock
\urldef\tempurl%
\url{https://doi.org/10.1080/14786440109462720}
\showDOI{\tempurl}


\bibitem[\protect\citeauthoryear{Perrone}{Perrone}{2006}]%
        {perrone2006}
\bibfield{author}{\bibinfo{person}{Michael Perrone}.}
  \bibinfo{year}{2006}\natexlab{}.
\newblock \showarticletitle{Differential item functioning and item bias:
  Critical considerations in test fairness}.
\newblock \bibinfo{journal}{\emph{Teachers College, Columbia University Working
  Papers in TESOL and Applied Linguistics}}  \bibinfo{volume}{6}
  (\bibinfo{year}{2006}), \bibinfo{pages}{1–3}.
\newblock


\bibitem[\protect\citeauthoryear{Petersen and Gencel}{Petersen and
  Gencel}{2013}]%
        {petersen2013}
\bibfield{author}{\bibinfo{person}{Kai Petersen} {and} \bibinfo{person}{Cigdem
  Gencel}.} \bibinfo{year}{2013}\natexlab{}.
\newblock \showarticletitle{Worldviews, Research Methods, and their
  Relationship to Validity in Empirical Software Engineering Research}. In
  \bibinfo{booktitle}{\emph{Worldviews, Research Methods, and their
  Relationship to Validity in Empirical Software Engineering Research}},
  Vol.~\bibinfo{volume}{2013 Joint Conference of the 23nd International
  Workshop on Software Measurement and the 8th International Conference on
  Software Process and Product Measurement (IWSM-MENSURA)}.
  \bibinfo{publisher}{IEEE}, \bibinfo{address}{Piscataway, NJ},
  \bibinfo{pages}{81--89}.
\newblock
\urldef\tempurl%
\url{https://doi.org/10.1109/iwsm-mensura.2013.22}
\showDOI{\tempurl}


\bibitem[\protect\citeauthoryear{Petrillo, Cano, McLeod, and Coon}{Petrillo
  et~al\mbox{.}}{2015}]%
        {petrillo2015}
\bibfield{author}{\bibinfo{person}{Jennifer Petrillo},
  \bibinfo{person}{Stefan~J. Cano}, \bibinfo{person}{Lori~D. McLeod}, {and}
  \bibinfo{person}{Cheryl~D. Coon}.} \bibinfo{year}{2015}\natexlab{}.
\newblock \showarticletitle{Using Classical Test Theory, Item Response Theory,
  and Rasch Measurement Theory to Evaluate Patient-Reported Outcome Measures: A
  Comparison of Worked Examples}.
\newblock \bibinfo{journal}{\emph{Value in Health}} \bibinfo{volume}{18},
  \bibinfo{number}{1} (\bibinfo{year}{2015}), \bibinfo{pages}{25–34}.
\newblock
\urldef\tempurl%
\url{https://doi.org/10.1016/j.jval.2014.10.005}
\showDOI{\tempurl}


\bibitem[\protect\citeauthoryear{Pittenger}{Pittenger}{1993}]%
        {pittenger1993}
\bibfield{author}{\bibinfo{person}{David~J Pittenger}.}
  \bibinfo{year}{1993}\natexlab{}.
\newblock \showarticletitle{Measuring the MBTI... and coming up short}.
\newblock \bibinfo{journal}{\emph{Journal of Career Planning and Employment}}
  \bibinfo{volume}{54}, \bibinfo{number}{1} (\bibinfo{year}{1993}),
  \bibinfo{pages}{48--52}.
\newblock


\bibitem[\protect\citeauthoryear{Pukelsheim}{Pukelsheim}{1994}]%
        {pukelsheim1994}
\bibfield{author}{\bibinfo{person}{Friedrich Pukelsheim}.}
  \bibinfo{year}{1994}\natexlab{}.
\newblock \showarticletitle{The Three Sigma Rule}.
\newblock \bibinfo{journal}{\emph{The American Statistician}}
  \bibinfo{volume}{48}, \bibinfo{number}{2} (\bibinfo{year}{1994}),
  \bibinfo{pages}{88–91}.
\newblock


\bibitem[\protect\citeauthoryear{Ralph, Baltes, Adisaputri, Torkar, Kovalenko,
  Kalinowski, Novielli, Yoo, Devroey, Tan, Zhou, Turhan, Hoda, Hata, Robles,
  Fard, and Alkadhi}{Ralph et~al\mbox{.}}{2020}]%
        {ralph2020}
\bibfield{author}{\bibinfo{person}{Paul Ralph}, \bibinfo{person}{Sebastian
  Baltes}, \bibinfo{person}{Gianisa Adisaputri}, \bibinfo{person}{Richard
  Torkar}, \bibinfo{person}{Vladimir Kovalenko}, \bibinfo{person}{Marcos
  Kalinowski}, \bibinfo{person}{Nicole Novielli}, \bibinfo{person}{Shin Yoo},
  \bibinfo{person}{Xavier Devroey}, \bibinfo{person}{Xin Tan},
  \bibinfo{person}{Minghui Zhou}, \bibinfo{person}{Burak Turhan},
  \bibinfo{person}{Rashina Hoda}, \bibinfo{person}{Hideaki Hata},
  \bibinfo{person}{Gregorio Robles}, \bibinfo{person}{Amin~Milani Fard}, {and}
  \bibinfo{person}{Rana Alkadhi}.} \bibinfo{year}{2020}\natexlab{}.
\newblock \showarticletitle{Pandemic programming}.
\newblock \bibinfo{journal}{\emph{Empirical Software Engineering}}
  \bibinfo{volume}{25}, \bibinfo{number}{6} (\bibinfo{date}{Sept.}
  \bibinfo{year}{2020}), \bibinfo{pages}{4927--4961}.
\newblock
\urldef\tempurl%
\url{https://doi.org/10.1007/s10664-020-09875-y}
\showDOI{\tempurl}


\bibitem[\protect\citeauthoryear{Ralph and Tempero}{Ralph and Tempero}{2018}]%
        {ralph2018}
\bibfield{author}{\bibinfo{person}{Paul Ralph} {and} \bibinfo{person}{Ewan
  Tempero}.} \bibinfo{year}{2018}\natexlab{}.
\newblock \showarticletitle{Construct Validity in Software Engineering Research
  and Software Metrics}. In \bibinfo{booktitle}{\emph{Proceedings of the 22nd
  International Conference on Evaluation and Assessment in Software Engineering
  2018}} (Christchurch, New Zealand) \emph{(\bibinfo{series}{EASE'18})}.
  \bibinfo{publisher}{Association for Computing Machinery},
  \bibinfo{address}{New York, NY, USA}, \bibinfo{pages}{13–23}.
\newblock
\showISBNx{9781450364034}
\urldef\tempurl%
\url{https://doi.org/10.1145/3210459.3210461}
\showDOI{\tempurl}


\bibitem[\protect\citeauthoryear{Rattray and Jones}{Rattray and Jones}{2007}]%
        {rattray2007}
\bibfield{author}{\bibinfo{person}{J Rattray} {and} \bibinfo{person}{MC
  Jones}.} \bibinfo{year}{2007}\natexlab{}.
\newblock \showarticletitle{Essential elements of questionnaire design and
  development.}
\newblock \bibinfo{journal}{\emph{J Clin Nurs}} \bibinfo{volume}{16},
  \bibinfo{number}{2} (\bibinfo{year}{2007}), \bibinfo{pages}{234–243}.
\newblock
\urldef\tempurl%
\url{https://doi.org/10.1111/j.1365-2702.2006.01573.x}
\showDOI{\tempurl}


\bibitem[\protect\citeauthoryear{Regehr, Bogo, Regehr, and Power}{Regehr
  et~al\mbox{.}}{2007}]%
        {regehr2007}
\bibfield{author}{\bibinfo{person}{Glenn Regehr}, \bibinfo{person}{Marion
  Bogo}, \bibinfo{person}{Cheryl Regehr}, {and} \bibinfo{person}{Roxanne
  Power}.} \bibinfo{year}{2007}\natexlab{}.
\newblock \showarticletitle{Can we build a better mousetrap? Improving the
  measures of practice performance in the field practicum}.
\newblock \bibinfo{journal}{\emph{Journal of Social Work Education}}
  \bibinfo{volume}{43}, \bibinfo{number}{2} (\bibinfo{year}{2007}),
  \bibinfo{pages}{327–344}.
\newblock
\urldef\tempurl%
\url{https://doi.org/10.5175/jswe.2007.200600607}
\showDOI{\tempurl}


\bibitem[\protect\citeauthoryear{Revelle}{Revelle}{2009}]%
        {revelle2009}
\bibfield{author}{\bibinfo{person}{William Revelle}.}
  \bibinfo{year}{2009}\natexlab{}.
\newblock \bibinfo{booktitle}{\emph{An introduction to psychometric theory with
  applications in R}}.
\newblock \bibinfo{publisher}{Available at personality-project.org},
  \bibinfo{address}{Online}.
\newblock


\bibitem[\protect\citeauthoryear{Revelle}{Revelle}{2018a}]%
        {revelle2018a}
\bibfield{author}{\bibinfo{person}{William Revelle}.}
  \bibinfo{year}{2018}\natexlab{a}.
\newblock \bibinfo{booktitle}{\emph{An introduction to the psych package: Part
  II Scale construction and psychometrics}}.
\newblock \bibinfo{type}{{T}echnical {R}eport}. \bibinfo{institution}{The
  Comprehensive R Archive Network}. \bibinfo{pages}{1–97} pages.
\newblock


\bibitem[\protect\citeauthoryear{Revelle}{Revelle}{2018b}]%
        {revelle2018}
\bibfield{author}{\bibinfo{person}{William Revelle}.}
  \bibinfo{year}{2018}\natexlab{b}.
\newblock \bibinfo{booktitle}{\emph{Using the psych package to generate and
  test structural models}}.
\newblock \bibinfo{type}{{T}echnical {R}eport}. \bibinfo{institution}{The
  Comprehensive R Archive Network}. \bibinfo{pages}{1–52} pages.
\newblock


\bibitem[\protect\citeauthoryear{Revelle}{Revelle}{2019}]%
        {revelle2017}
\bibfield{author}{\bibinfo{person}{William Revelle}.}
  \bibinfo{year}{2019}\natexlab{}.
\newblock \bibinfo{booktitle}{\emph{psych: Procedures for Psychological,
  Psychometric, and Personality Research}}.
\newblock Northwestern University, Evanston, Illinois.
\newblock
\urldef\tempurl%
\url{https://cran.r-project.org/package=psych}
\showURL{%
\tempurl}
\newblock
\shownote{R package version 1.9.12.}


\bibitem[\protect\citeauthoryear{Revelle}{Revelle}{2020}]%
        {revelle2020}
\bibfield{author}{\bibinfo{person}{William Revelle}.}
  \bibinfo{year}{2020}\natexlab{}.
\newblock \bibinfo{booktitle}{\emph{How To: Use the psych package for Factor
  Analysis and data reduction}}.
\newblock \bibinfo{type}{{T}echnical {R}eport}. \bibinfo{institution}{The
  Comprehensive R Archive Network}. \bibinfo{pages}{1--96} pages.
\newblock


\bibitem[\protect\citeauthoryear{Revelle and Rocklin}{Revelle and
  Rocklin}{1979}]%
        {revelle1979}
\bibfield{author}{\bibinfo{person}{William Revelle} {and}
  \bibinfo{person}{Thomas Rocklin}.} \bibinfo{year}{1979}\natexlab{}.
\newblock \showarticletitle{Very Simple Structure: An Alternative Procedure For
  Estimating The Optimal Number Of Interpretable Factors}.
\newblock \bibinfo{journal}{\emph{Multivariate Behavioral Research}}
  \bibinfo{volume}{14}, \bibinfo{number}{4} (\bibinfo{date}{Oct.}
  \bibinfo{year}{1979}), \bibinfo{pages}{403--414}.
\newblock
\urldef\tempurl%
\url{https://doi.org/10.1207/s15327906mbr1404_2}
\showDOI{\tempurl}


\bibitem[\protect\citeauthoryear{Runnels}{Runnels}{2013}]%
        {runnels2013}
\bibfield{author}{\bibinfo{person}{Judith Runnels}.}
  \bibinfo{year}{2013}\natexlab{}.
\newblock \showarticletitle{Measuring differential item and test functioning
  across academic disciplines}.
\newblock \bibinfo{journal}{\emph{Language Testing in Asia}}
  \bibinfo{volume}{3}, \bibinfo{number}{1} (\bibinfo{year}{2013}),
  \bibinfo{pages}{9}.
\newblock


\bibitem[\protect\citeauthoryear{Rus, Lindvall, and Sinha}{Rus
  et~al\mbox{.}}{2002}]%
        {rus2002}
\bibfield{author}{\bibinfo{person}{Ioana Rus}, \bibinfo{person}{Mikael
  Lindvall}, {and} \bibinfo{person}{S Sinha}.} \bibinfo{year}{2002}\natexlab{}.
\newblock \showarticletitle{Knowledge management in software engineering}.
\newblock \bibinfo{journal}{\emph{IEEE software}} \bibinfo{volume}{19},
  \bibinfo{number}{3} (\bibinfo{year}{2002}), \bibinfo{pages}{26--38}.
\newblock


\bibitem[\protect\citeauthoryear{Russell}{Russell}{2016}]%
        {russell2016}
\bibfield{author}{\bibinfo{person}{Daniel~W. Russell}.}
  \bibinfo{year}{2016}\natexlab{}.
\newblock \showarticletitle{In Search of Underlying Dimensions: The Use (and
  Abuse) of Factor Analysis in Personality and Social Psychology Bulletin}.
\newblock \bibinfo{journal}{\emph{Personality and Social Psychology Bulletin}}
  \bibinfo{volume}{28}, \bibinfo{number}{12} (\bibinfo{year}{2016}),
  \bibinfo{pages}{1629–1646}.
\newblock
\urldef\tempurl%
\url{https://doi.org/10.1177/014616702237645}
\showDOI{\tempurl}


\bibitem[\protect\citeauthoryear{Rust}{Rust}{2009}]%
        {rust2009}
\bibfield{author}{\bibinfo{person}{John Rust}.}
  \bibinfo{year}{2009}\natexlab{}.
\newblock \bibinfo{booktitle}{\emph{Modern psychometrics : the science of
  psychological assessment}}.
\newblock \bibinfo{publisher}{Routledge}, \bibinfo{address}{Hove, East Sussex
  New York}.
\newblock
\showISBNx{978-0415442152}


\bibitem[\protect\citeauthoryear{Schad, Betancourt, and Vasishth}{Schad
  et~al\mbox{.}}{2021}]%
        {schad2019toward}
\bibfield{author}{\bibinfo{person}{Daniel~J. Schad}, \bibinfo{person}{Michael
  Betancourt}, {and} \bibinfo{person}{Shravan Vasishth}.}
  \bibinfo{year}{2021}\natexlab{}.
\newblock \showarticletitle{Toward a principled Bayesian workflow in cognitive
  science.}
\newblock \bibinfo{journal}{\emph{Psychological Methods}} \bibinfo{volume}{26},
  \bibinfo{number}{1} (\bibinfo{date}{Feb.} \bibinfo{year}{2021}),
  \bibinfo{pages}{103--126}.
\newblock
\urldef\tempurl%
\url{https://doi.org/10.1037/met0000275}
\showDOI{\tempurl}


\bibitem[\protect\citeauthoryear{Schmidt}{Schmidt}{1992}]%
        {schmidt1992}
\bibfield{author}{\bibinfo{person}{Frank~L. Schmidt}.}
  \bibinfo{year}{1992}\natexlab{}.
\newblock \showarticletitle{What do data really mean? Research findings,
  meta-analysis, and cumulative knowledge in psychology.}
\newblock \bibinfo{journal}{\emph{American Psychologist}} \bibinfo{volume}{47},
  \bibinfo{number}{10} (\bibinfo{year}{1992}), \bibinfo{pages}{1173–1181}.
\newblock
\urldef\tempurl%
\url{https://doi.org/10.1037/0003-066x.47.10.1173}
\showDOI{\tempurl}


\bibitem[\protect\citeauthoryear{Schoenherr and Hamstra}{Schoenherr and
  Hamstra}{2016}]%
        {schoenherr2016}
\bibfield{author}{\bibinfo{person}{JR Schoenherr} {and} \bibinfo{person}{SJ
  Hamstra}.} \bibinfo{year}{2016}\natexlab{}.
\newblock \showarticletitle{Psychometrics and its discontents: an historical
  perspective on the discourse of the measurement tradition.}
\newblock \bibinfo{journal}{\emph{Adv Health Sci Educ Theory Pract}}
  \bibinfo{volume}{21}, \bibinfo{number}{3} (\bibinfo{year}{2016}),
  \bibinfo{pages}{719–729}.
\newblock
\urldef\tempurl%
\url{https://doi.org/10.1007/s10459-015-9623-z}
\showDOI{\tempurl}


\bibitem[\protect\citeauthoryear{Schwarz and Oyserman}{Schwarz and
  Oyserman}{2016}]%
        {schwarz2016}
\bibfield{author}{\bibinfo{person}{Norbert Schwarz} {and}
  \bibinfo{person}{Daphna Oyserman}.} \bibinfo{year}{2016}\natexlab{}.
\newblock \showarticletitle{Asking Questions About Behavior: Cognition,
  Communication, and Questionnaire Construction}.
\newblock \bibinfo{journal}{\emph{American Journal of Evaluation}}
  \bibinfo{volume}{22}, \bibinfo{number}{2} (\bibinfo{year}{2016}),
  \bibinfo{pages}{127–160}.
\newblock
\urldef\tempurl%
\url{https://doi.org/10.1177/109821400102200202}
\showDOI{\tempurl}


\bibitem[\protect\citeauthoryear{Shea, Jacobs, Esserman, Bruce, and
  Weiner}{Shea et~al\mbox{.}}{2014}]%
        {shea2014}
\bibfield{author}{\bibinfo{person}{CM Shea}, \bibinfo{person}{SR Jacobs},
  \bibinfo{person}{DA Esserman}, \bibinfo{person}{K Bruce}, {and}
  \bibinfo{person}{BJ Weiner}.} \bibinfo{year}{2014}\natexlab{}.
\newblock \showarticletitle{Organizational readiness for implementing change: a
  psychometric assessment of a new measure.}
\newblock \bibinfo{journal}{\emph{Implement Sci}}  \bibinfo{volume}{9}
  (\bibinfo{year}{2014}), \bibinfo{pages}{7}.
\newblock


\bibitem[\protect\citeauthoryear{Siegmund, Siegmund, and Apel}{Siegmund
  et~al\mbox{.}}{2015}]%
        {siegmund2015}
\bibfield{author}{\bibinfo{person}{Janet Siegmund}, \bibinfo{person}{Norbert
  Siegmund}, {and} \bibinfo{person}{Sven Apel}.}
  \bibinfo{year}{2015}\natexlab{}.
\newblock \showarticletitle{Views on Internal and External Validity in
  Empirical Software Engineering}, Vol.~\bibinfo{volume}{2015 IEEE/ACM 37th
  IEEE International Conference on Software Engineering (ICSE)}.
  \bibinfo{publisher}{IEEE}, \bibinfo{address}{Piscataway, NJ},
  \bibinfo{pages}{9--19}.
\newblock
\urldef\tempurl%
\url{https://doi.org/10.1109/icse.2015.24}
\showDOI{\tempurl}


\bibitem[\protect\citeauthoryear{Sigelman}{Sigelman}{1981}]%
        {sigelman1981}
\bibfield{author}{\bibinfo{person}{Lee Sigelman}.}
  \bibinfo{year}{1981}\natexlab{}.
\newblock \showarticletitle{Question-Order Effects on Presidential Popularity}.
\newblock \bibinfo{journal}{\emph{Public Opinion Quarterly}}
  \bibinfo{volume}{45}, \bibinfo{number}{2} (\bibinfo{year}{1981}),
  \bibinfo{pages}{199}.
\newblock
\urldef\tempurl%
\url{https://doi.org/10.1086/268650}
\showDOI{\tempurl}


\bibitem[\protect\citeauthoryear{Singh, Junnarkar, Kaur, Singh, Junnarkar, and
  Kaur}{Singh et~al\mbox{.}}{2016}]%
        {singh2016}
\bibfield{author}{\bibinfo{person}{Kamlesh Singh}, \bibinfo{person}{Mohita
  Junnarkar}, \bibinfo{person}{Jasleen Kaur}, \bibinfo{person}{Kamlesh Singh},
  \bibinfo{person}{Mohita Junnarkar}, {and} \bibinfo{person}{Jasleen Kaur}.}
  \bibinfo{year}{2016}\natexlab{}.
\newblock \showarticletitle{Norms for Test Construction}.
\newblock In \bibinfo{booktitle}{\emph{Measures of Positive Psychology}}.
  \bibinfo{publisher}{Springer India}, \bibinfo{address}{New Delhi},
  \bibinfo{pages}{17--34}.
\newblock


\bibitem[\protect\citeauthoryear{Swart and Kinnie}{Swart and Kinnie}{2003}]%
        {swart2003}
\bibfield{author}{\bibinfo{person}{Juani Swart} {and} \bibinfo{person}{Nicholas
  Kinnie}.} \bibinfo{year}{2003}\natexlab{}.
\newblock \showarticletitle{Sharing knowledge in knowledge-intensive firms}.
\newblock \bibinfo{journal}{\emph{Human resource management journal}}
  \bibinfo{volume}{13}, \bibinfo{number}{2} (\bibinfo{year}{2003}),
  \bibinfo{pages}{60--75}.
\newblock


\bibitem[\protect\citeauthoryear{Tabachnick, Fidell, and Ullman}{Tabachnick
  et~al\mbox{.}}{2007}]%
        {tabachnick2007}
\bibfield{author}{\bibinfo{person}{Barbara~G Tabachnick},
  \bibinfo{person}{Linda~S Fidell}, {and} \bibinfo{person}{Jodie~B Ullman}.}
  \bibinfo{year}{2007}\natexlab{}.
\newblock \bibinfo{booktitle}{\emph{Using multivariate statistics}}.
  Vol.~\bibinfo{volume}{5}.
\newblock \bibinfo{publisher}{Pearson Boston, MA}, \bibinfo{address}{Boston}.
\newblock


\bibitem[\protect\citeauthoryear{Thabane, Ma, Chu, Cheng, Ismaila, Rios,
  Robson, Thabane, Giangregorio, and Goldsmith}{Thabane et~al\mbox{.}}{2010}]%
        {thabane2010}
\bibfield{author}{\bibinfo{person}{Lehana Thabane}, \bibinfo{person}{Jinhui
  Ma}, \bibinfo{person}{Rong Chu}, \bibinfo{person}{Ji Cheng},
  \bibinfo{person}{Afisi Ismaila}, \bibinfo{person}{Lorena~P Rios},
  \bibinfo{person}{Reid Robson}, \bibinfo{person}{Marroon Thabane},
  \bibinfo{person}{Lora Giangregorio}, {and} \bibinfo{person}{Charles~H
  Goldsmith}.} \bibinfo{year}{2010}\natexlab{}.
\newblock \showarticletitle{A tutorial on pilot studies: the what, why and
  how}.
\newblock \bibinfo{journal}{\emph{BMC Medical Research Methodology}}
  \bibinfo{volume}{10}, \bibinfo{number}{1} (\bibinfo{year}{2010}).
\newblock
\urldef\tempurl%
\url{https://doi.org/10.1186/1471-2288-10-1}
\showDOI{\tempurl}


\bibitem[\protect\citeauthoryear{Thurstone}{Thurstone}{1937}]%
        {thurstone1937}
\bibfield{author}{\bibinfo{person}{Louis~Leon Thurstone}.}
  \bibinfo{year}{1937}\natexlab{}.
\newblock \showarticletitle{Psychology as a quantitative rational science}.
\newblock \bibinfo{journal}{\emph{Science}} \bibinfo{volume}{85},
  \bibinfo{number}{2201} (\bibinfo{year}{1937}), \bibinfo{pages}{227–232}.
\newblock
\urldef\tempurl%
\url{https://doi.org/10.2307/1662685}
\showDOI{\tempurl}


\bibitem[\protect\citeauthoryear{Tinsley and Tinsley}{Tinsley and
  Tinsley}{1987}]%
        {tinsley1987}
\bibfield{author}{\bibinfo{person}{Howard~E. Tinsley} {and}
  \bibinfo{person}{Diane~J. Tinsley}.} \bibinfo{year}{1987}\natexlab{}.
\newblock \showarticletitle{Uses of factor analysis in counseling psychology
  research.}
\newblock \bibinfo{journal}{\emph{Journal of Counseling Psychology}}
  \bibinfo{volume}{34}, \bibinfo{number}{4} (\bibinfo{year}{1987}),
  \bibinfo{pages}{414–424}.
\newblock


\bibitem[\protect\citeauthoryear{Traub}{Traub}{2005}]%
        {traub2005}
\bibfield{author}{\bibinfo{person}{Ross~E. Traub}.}
  \bibinfo{year}{2005}\natexlab{}.
\newblock \showarticletitle{Classical Test Theory in Historical Perspective}.
\newblock \bibinfo{journal}{\emph{Educational Measurement: Issues and
  Practice}} \bibinfo{volume}{16}, \bibinfo{number}{4} (\bibinfo{date}{Oct.}
  \bibinfo{year}{2005}), \bibinfo{pages}{8--14}.
\newblock
\urldef\tempurl%
\url{https://doi.org/10.1111/j.1745-3992.1997.tb00603.x}
\showDOI{\tempurl}


\bibitem[\protect\citeauthoryear{Tripp, Riemenschneider, and Thatcher}{Tripp
  et~al\mbox{.}}{2016}]%
        {tripp2016}
\bibfield{author}{\bibinfo{person}{John~F Tripp}, \bibinfo{person}{Cindy
  Riemenschneider}, {and} \bibinfo{person}{Jason~B Thatcher}.}
  \bibinfo{year}{2016}\natexlab{}.
\newblock \showarticletitle{Job satisfaction in agile development teams: Agile
  development as work redesign}.
\newblock \bibinfo{journal}{\emph{Journal of the Association for Information
  Systems}} \bibinfo{volume}{17}, \bibinfo{number}{4} (\bibinfo{year}{2016}),
  \bibinfo{pages}{267}.
\newblock


\bibitem[\protect\citeauthoryear{Uher}{Uher}{2018}]%
        {uher2018}
\bibfield{author}{\bibinfo{person}{J Uher}.} \bibinfo{year}{2018}\natexlab{}.
\newblock \showarticletitle{Quantitative Data From Rating Scales: An
  Epistemological and Methodological Enquiry.}
\newblock \bibinfo{journal}{\emph{Front Psychol}}  \bibinfo{volume}{9}
  (\bibinfo{year}{2018}), \bibinfo{pages}{2599}.
\newblock
\urldef\tempurl%
\url{https://doi.org/10.3389/fpsyg.2018.02599}
\showDOI{\tempurl}


\bibitem[\protect\citeauthoryear{Uher}{Uher}{2021}]%
        {uher2021}
\bibfield{author}{\bibinfo{person}{Jana Uher}.}
  \bibinfo{year}{2021}\natexlab{}.
\newblock \showarticletitle{Quantitative psychology under scrutiny: Measurement
  requires not result-dependent but traceable data generation}.
\newblock \bibinfo{journal}{\emph{Personality and Individual Differences}}
  \bibinfo{volume}{170} (\bibinfo{year}{2021}), \bibinfo{pages}{110205}.
\newblock
\urldef\tempurl%
\url{https://doi.org/10.1016/j.paid.2020.110205}
\showDOI{\tempurl}


\bibitem[\protect\citeauthoryear{Wagenmakers, Marsman, Jamil, Ly, Verhagen,
  Love, Selker, Gronau, {\v{S}}m{\'\i}ra, Epskamp, et~al\mbox{.}}{Wagenmakers
  et~al\mbox{.}}{2018}]%
        {wagenmakers2018bayesian}
\bibfield{author}{\bibinfo{person}{Eric-Jan Wagenmakers},
  \bibinfo{person}{Maarten Marsman}, \bibinfo{person}{Tahira Jamil},
  \bibinfo{person}{Alexander Ly}, \bibinfo{person}{Josine Verhagen},
  \bibinfo{person}{Jonathon Love}, \bibinfo{person}{Ravi Selker},
  \bibinfo{person}{Quentin~F Gronau}, \bibinfo{person}{Martin
  {\v{S}}m{\'\i}ra}, \bibinfo{person}{Sacha Epskamp}, {et~al\mbox{.}}}
  \bibinfo{year}{2018}\natexlab{}.
\newblock \showarticletitle{Bayesian inference for psychology. Part I:
  Theoretical advantages and practical ramifications}.
\newblock \bibinfo{journal}{\emph{Psychonomic bulletin \& review}}
  \bibinfo{volume}{25}, \bibinfo{number}{1} (\bibinfo{year}{2018}),
  \bibinfo{pages}{35--57}.
\newblock


\bibitem[\protect\citeauthoryear{Wagner, Mendez, Felderer, Graziotin, and
  Kalinowski}{Wagner et~al\mbox{.}}{2020}]%
        {wagner2020}
\bibfield{author}{\bibinfo{person}{Stefan Wagner}, \bibinfo{person}{Daniel
  Mendez}, \bibinfo{person}{Michael Felderer}, \bibinfo{person}{Daniel
  Graziotin}, {and} \bibinfo{person}{Marcos Kalinowski}.}
  \bibinfo{year}{2020}\natexlab{}.
\newblock \showarticletitle{Challenges in Survey Research}.
\newblock In \bibinfo{booktitle}{\emph{Contemporary Empirical Methods in
  Software Engineering}}, \bibfield{editor}{\bibinfo{person}{Michael Felderer},
  \bibinfo{person}{}, {and} \bibinfo{person}{Guilherme~Horta Travassos}}
  (Eds.). \bibinfo{publisher}{Springer International Publishing},
  \bibinfo{address}{Cham}, \bibinfo{pages}{93--125}.
\newblock
\urldef\tempurl%
\url{https://doi.org/10.1007/978-3-030-32489-6_4}
\showDOI{\tempurl}
\newblock
\shownote{Available https://arxiv.org/abs/1908.05899.}


\bibitem[\protect\citeauthoryear{Wang and Zhang}{Wang and Zhang}{2020}]%
        {wang2020}
\bibfield{author}{\bibinfo{person}{Yi Wang} {and} \bibinfo{person}{Min Zhang}.}
  \bibinfo{year}{2020}\natexlab{}.
\newblock \showarticletitle{Reducing implicit gender biases in software
  development: does intergroup contact theory work},
  Vol.~\bibinfo{volume}{ESEC/FSE '20: 28th ACM Joint European Software
  Engineering Conference and Symposium on the Foundations of Software
  Engineering}. \bibinfo{publisher}{ACM}, \bibinfo{address}{New York, NY, USA}.
\newblock
\urldef\tempurl%
\url{https://doi.org/10.1145/3368089.3409762}
\showDOI{\tempurl}


\bibitem[\protect\citeauthoryear{Wasserstein, Schirm, and Lazar}{Wasserstein
  et~al\mbox{.}}{2019}]%
        {wasserstein2019moving}
\bibfield{author}{\bibinfo{person}{Ronald~L Wasserstein},
  \bibinfo{person}{Allen~L Schirm}, {and} \bibinfo{person}{Nicole~A Lazar}.}
  \bibinfo{year}{2019}\natexlab{}.
\newblock \bibinfo{title}{Moving to a world beyond “p< 0.05”}.
\newblock
\newblock


\bibitem[\protect\citeauthoryear{Weinberg}{Weinberg}{1971}]%
        {weinberg1971}
\bibfield{author}{\bibinfo{person}{Gerald~M Weinberg}.}
  \bibinfo{year}{1971}\natexlab{}.
\newblock \bibinfo{booktitle}{\emph{The psychology of computer programming}}.
\newblock \bibinfo{publisher}{Dorset House Pub}, \bibinfo{address}{New York}.
\newblock


\bibitem[\protect\citeauthoryear{Weiss}{Weiss}{2002}]%
        {weiss2002}
\bibfield{author}{\bibinfo{person}{Howard~M Weiss}.}
  \bibinfo{year}{2002}\natexlab{}.
\newblock \showarticletitle{Deconstructing job satisfaction: Separating
  evaluations, beliefs and affective experiences}.
\newblock \bibinfo{journal}{\emph{Human resource management review}}
  \bibinfo{volume}{12}, \bibinfo{number}{2} (\bibinfo{year}{2002}),
  \bibinfo{pages}{173–194}.
\newblock


\bibitem[\protect\citeauthoryear{Widaman}{Widaman}{1993}]%
        {widaman1993}
\bibfield{author}{\bibinfo{person}{Keith~F. Widaman}.}
  \bibinfo{year}{1993}\natexlab{}.
\newblock \showarticletitle{Common Factor Analysis Versus Principal Component
  Analysis: Differential Bias in Representing Model Parameters}.
\newblock \bibinfo{journal}{\emph{Multivariate Behavioral Research}}
  \bibinfo{volume}{28}, \bibinfo{number}{3} (\bibinfo{year}{1993}),
  \bibinfo{pages}{263–311}.
\newblock
\urldef\tempurl%
\url{https://doi.org/10.1207/s15327906mbr2803_1}
\showDOI{\tempurl}


\bibitem[\protect\citeauthoryear{Willis}{Willis}{2004}]%
        {willis2004}
\bibfield{author}{\bibinfo{person}{Gordon~B. Willis}.}
  \bibinfo{year}{2004}\natexlab{}.
\newblock \showarticletitle{Cognitive Interviewing Revisited: A Useful
  Technique, in Theory}.
\newblock In \bibinfo{booktitle}{\emph{Methods for Testing and Evaluating
  Survey Questionnaires: Wiley Series in Survey Methodology}}.
  \bibinfo{publisher}{John Wiley \& Sons, Inc.}, \bibinfo{address}{Hoboken, NJ,
  USA}, \bibinfo{pages}{23--43}.
\newblock
\urldef\tempurl%
\url{https://doi.org/10.1002/0471654728.ch2}
\showDOI{\tempurl}


\bibitem[\protect\citeauthoryear{Wohlin, Runeson, Höst, Ohlsson, Regnell, and
  Wesslén}{Wohlin et~al\mbox{.}}{2012}]%
        {wohlin2012}
\bibfield{author}{\bibinfo{person}{Claes Wohlin}, \bibinfo{person}{Per
  Runeson}, \bibinfo{person}{Martin Höst}, \bibinfo{person}{Magnus~C.
  Ohlsson}, \bibinfo{person}{Björn Regnell}, {and} \bibinfo{person}{Anders
  Wesslén}.} \bibinfo{year}{2012}\natexlab{}.
\newblock \bibinfo{booktitle}{\emph{Experimentation in Software Engineering}}.
\newblock \bibinfo{publisher}{Springer Berlin Heidelberg},
  \bibinfo{address}{Berlin, Heidelberg}.
\newblock


\bibitem[\protect\citeauthoryear{Wood, Gardner, and Harms}{Wood
  et~al\mbox{.}}{2015}]%
        {wood2015}
\bibfield{author}{\bibinfo{person}{D Wood}, \bibinfo{person}{MH Gardner}, {and}
  \bibinfo{person}{PD Harms}.} \bibinfo{year}{2015}\natexlab{}.
\newblock \showarticletitle{How functionalist and process approaches to
  behavior can explain trait covariation.}
\newblock \bibinfo{journal}{\emph{Psychol Rev}} \bibinfo{volume}{122},
  \bibinfo{number}{1} (\bibinfo{year}{2015}), \bibinfo{pages}{84–111}.
\newblock
\urldef\tempurl%
\url{https://doi.org/10.1037/a0038423}
\showDOI{\tempurl}


\bibitem[\protect\citeauthoryear{Wynd, Schmidt, and Schaefer}{Wynd
  et~al\mbox{.}}{2003}]%
        {wynd2003}
\bibfield{author}{\bibinfo{person}{CA Wynd}, \bibinfo{person}{B Schmidt}, {and}
  \bibinfo{person}{MA Schaefer}.} \bibinfo{year}{2003}\natexlab{}.
\newblock \showarticletitle{Two quantitative approaches for estimating content
  validity.}
\newblock \bibinfo{journal}{\emph{West J Nurs Res}} \bibinfo{volume}{25},
  \bibinfo{number}{5} (\bibinfo{year}{2003}), \bibinfo{pages}{508–518}.
\newblock
\urldef\tempurl%
\url{https://doi.org/10.1177/0193945903252998}
\showDOI{\tempurl}


\bibitem[\protect\citeauthoryear{Wyrich, Graziotin, and Wagner}{Wyrich
  et~al\mbox{.}}{2019}]%
        {wyrich2019}
\bibfield{author}{\bibinfo{person}{Marvin Wyrich}, \bibinfo{person}{Daniel
  Graziotin}, {and} \bibinfo{person}{Stefan Wagner}.}
  \bibinfo{year}{2019}\natexlab{}.
\newblock \showarticletitle{A theory on individual characteristics of
  successful coding challenge solvers}.
\newblock \bibinfo{journal}{\emph{PeerJ Computer Science}}  \bibinfo{volume}{5}
  (\bibinfo{year}{2019}), \bibinfo{pages}{e173}.
\newblock


\bibitem[\protect\citeauthoryear{Wyrich, Preikschat, Graziotin, and
  Wagner}{Wyrich et~al\mbox{.}}{2021}]%
        {wyrich2020}
\bibfield{author}{\bibinfo{person}{Marvin Wyrich}, \bibinfo{person}{Andreas
  Preikschat}, \bibinfo{person}{Daniel Graziotin}, {and}
  \bibinfo{person}{Stefan Wagner}.} \bibinfo{year}{2021}\natexlab{}.
\newblock \showarticletitle{The Mind Is a Powerful Place: How Showing Code
  Comprehensibility Metrics Influences Code Understanding}. In
  \bibinfo{booktitle}{\emph{2021 {IEEE}/{ACM} 43rd International Conference on
  Software Engineering ({ICSE})}}. \bibinfo{publisher}{{IEEE}},
  \bibinfo{address}{Piscataway, NJ}, \bibinfo{pages}{512--523}.
\newblock
\urldef\tempurl%
\url{https://doi.org/10.1109/icse43902.2021.00055}
\showDOI{\tempurl}


\bibitem[\protect\citeauthoryear{Yong and Pearce}{Yong and Pearce}{2013}]%
        {yong2013}
\bibfield{author}{\bibinfo{person}{An~Gie Yong} {and} \bibinfo{person}{Sean
  Pearce}.} \bibinfo{year}{2013}\natexlab{}.
\newblock \showarticletitle{A Beginner’s Guide to Factor Analysis: Focusing
  on Exploratory Factor Analysis}.
\newblock \bibinfo{journal}{\emph{Tutorials in Quantitative Methods for
  Psychology}} \bibinfo{volume}{9}, \bibinfo{number}{2} (\bibinfo{year}{2013}),
  \bibinfo{pages}{79–94}.
\newblock


\bibitem[\protect\citeauthoryear{Zumbo}{Zumbo}{2007}]%
        {zumbo2007}
\bibfield{author}{\bibinfo{person}{Bruno~D Zumbo}.}
  \bibinfo{year}{2007}\natexlab{}.
\newblock \showarticletitle{Three generations of DIF analyses: Considering
  where it has been, where it is now, and where it is going}.
\newblock \bibinfo{journal}{\emph{Language assessment quarterly}}
  \bibinfo{volume}{4}, \bibinfo{number}{2} (\bibinfo{year}{2007}),
  \bibinfo{pages}{223–233}.
\newblock


\end{thebibliography}

\appendix
\section{Appendix\label{sec:appendix}}
The following appendix (explained in section~\ref{sec:example}) is available as an always updated replication package~\citep{graziotin2020}.


\section*{Supplementary material}

# Behavioral Software Engineering - Example of psychometric evaluation with R

Daniel Graziotin, Per Lenberg, Robert Feldt, Stefan Wagner

2021-06-08

# Contents

# Introduction

The present document provides an executable hands-on introductory psychometric validation example written for a behavioral software engineering audience.

The present document is part of a paper:

> Graziotin, D., Lenberg, P., Feldt, R., & Wagner, S. (2021). Psychometrics in Behavioral Software Engineering: A Methodological Introduction with Guidelines. ACM Transactions on Software Engineering and Methodology, 1, 1-56. In press. Available: https://arxiv.org/abs/2005.09959

The present document should be available as online appendix of the paper. If not, we invite you to retrieve the latest version of this document at:

> Daniel Graziotin, Per Lenberg, Robert Feldt, & Stefan Wagner. (2021). Behavioral Software Engineering - Example of psychometric evaluation with R [Data set]. Zenodo. https://doi.org/10.5281/zenodo.3799603

Even though this document is as self-contained as possible, we recommend reading the document only after reading the paper. The present document is also an R Markdown file. Its text version is interactive and can be executed directly in R Studio, making it completely reproducible.

# Context

The present document contains an example, with simulated data, of the psychometric validation phases of a norm-based measurement instrument for assessing an overall construct with five sub-constructs.

Our fictitious construct is the "individual perception styles of source code" that, through a literature review (or, perhaps, after a grounded theory study), we believe is mainly composed of, or highly related, to the following five constructs: code curiosity, programming paradigm flexibility, learning disposition, collaboration propensity, and comfort in novelty.

We develop a measurement instrument with 31 items, all represented by Likert items ranging from 1 to 5, which is self-assessed by the individual software engineer. The Likert items would ideally form five Likert scales, that would actually emerge after an exploratory factors analysis. However, when we develop scales we have a rough idea of the related constructs anyway. We represent here what we expect the exploratory factor analysis to show, with all items grouped by what we might see as potential factors.

- F1, "Code curiosity"
  - 7 items, F1_1 to F1_7
- F2 "Programming paradigm flexibility"
  - 4 items, F2_8 to F2_11
- F3 "Learning disposition"
  - 9 items, F3_12 to F3_20
- F4 "Collaboration propensity"
  - 4 items, F4_21 to F4_24
- F5 "Comfort in novelty"
  - 7 items, F5_25 to F5_31

What the items actually are is not interesting for the purposes of the present document. Our team proceeds to psychometrically evaluate the measurement instrument, in particular, to reduce the number of items. Then, to validate the items belonging to the factors and, possibly, reducing the number of items agin. Finally, to offer a reasoning on statistical properties, reliability, and validity.

### Requirements and options

The following are requirements that are typically found in a basic psychometric validation, with the exclusion of `fabricatr`, which is for fabricating data and we use for developing this example only.

```
if (!require("fabricatr")){
  install.packages("fabricatr", dep = T)
}
```

```
## Loading required package: fabricatr
```

```
require("fabricatr")

if (!require("psych")){
  install.packages("psych", dep = T)
}
```

```
## Loading required package: psych
```

```
require("psych")

if (!require(corrplot)) {
  install.packages("corrplot", dep = T)
}
```

```
## Loading required package: corrplot
```

```
## Warning: package 'corrplot' was built under R version 4.0.2
```

```
## corrplot 0.84 loaded
```

```
require(corrplot)

if (!require("knitr")){
  install.packages("knitr", dep = T)
}
```

```
## Loading required package: knitr
require("knitr")

knitr::opts_chunk$set(echo = TRUE, cache = FALSE, width = 50)
```

We offer a dataset with pre-populated data that allows repetition of the entire document. For discarding our provided data and simulating new data (which should conform, to a fair extent, to what we expect in the various sections), set the `safeguard` variable to FALSE or delete the `graziotin_et_al-bse_psychometrics_example.csv` file. The new dataset will behave very similarly to the provided one–after all, we used the very same code to generate it. This option will allow reproducibility only.

For full repeatability, but mostly for better clarity, we recommend to start with our provided dataset. The reason is that the rest of this tutorial will provide reasoning around certain values for items that will slightly change in case a new dataset is generated.

The following are options that are useful for repetition and replication of our example.

```
# set to FALSE to regenerate the dataset
safeguard <- TRUE && file.exists('graziotin_et_al-bse_psychometrics_example.csv')
overall.sample.size = 142
pilot.test.sample.size = 23
reliability.sample.size = 48
```

## Data simulation

This section explains how we constructed the simulated dataset `graziotin_et_al-bse_psychometrics_example.csv`.

**This section can be safely skipped as it does not pertain to psychometric evaluation**. We provide the data simulation part for full reproducibility.

```
if (!safeguard) {
  # 31 items belonging to 5 clusters
  items.initialset <- c(rep(1, 7), # cluster 1
                        rep(2, 4), # cluster 2
                        rep(3, 9), # cluster 3
                        rep(4, 4), # cluster 4
                        rep(5, 7)) # cluster 5

  Likert.brakes <- c(-Inf,-1.5,-0.5, 0.5, 1.5, Inf)
  Likert.values <- c("1", "2", "3", "4", "5")

  counter <- 0
  var.names <-
    unlist(lapply(items.initialset, function(x) {
      paste0("F", x, "_", counter <<- counter + 1)
    }))
  normalized.response <-
    draw_normal_icc(clusters = items.initialset,
                    ICC = sample(seq(0.4, 0.7, by = 0.05), 1),
                    sd = sample(seq(0.5, 0.9, by = 0.1), 1))
  ordered.response <- draw_ordered(x = normalized.response,
                                   breaks = Likert.brakes,
                                   break_labels = Likert.values)
  full.df <- rbind(ordered.response)
  colnames(full.df) <- var.names
```

```r
  for (i in seq(1, overall.sample.size - 1)) {
    normalized.response <-
      draw_normal_icc(clusters = items.initialset,
                      ICC = sample(seq(0.4, 0.7, by = 0.05), 1),
                      sd = sample(seq(0.5, 0.9, by = 0.1), 1))
    ordered.response <- draw_ordered(x = normalized.response,
                                     breaks = Likert.brakes,
                                     break_labels = Likert.values)
    full.df <- rbind(full.df, ordered.response)
  }
  # each row is a participant id, from 1 to overall.sample.size
  rownames(full.df) <- seq(1, overall.sample.size)
}
```

Up to this point we have generated a perfect dataset. All items cluster as we wish them to be. We now add noise to the data. For two items, `F1_4` and `F5_25`, we simulate oddly worded or uninteresting items that most users rate at the maximum and minimum values, respectively. This will allow us later on to drop these items during item facility analysis.

```r
if (!safeguard) {
  full.df[, 'F1_4'] <- draw_ordered(
    x = rnorm(n = overall.sample.size, mean = 2.7, sd = 0.8),
    breaks = Likert.brakes,
    break_labels = Likert.values
  )
  full.df[, 'F5_25'] <- draw_ordered(
    x = rnorm(n = overall.sample.size, mean = 1.2, sd = 0.8),
    breaks = Likert.brakes,
    break_labels = Likert.values
  )
}
```

For other three items, `F4_21` to `F4_23`, we get even more malicious. Factor 4 only had 4 candidates for a representative cluster. We generate values for these three items again such that they do not likely belong to the cluster 'F4' anymore, but they do not provide clear extreme values that would get caught during item analysis.

```r
if (!safeguard) {
  full.df[, 'F4_21'] <- draw_ordered(
    x = rnorm(n = overall.sample.size, mean = 1.5, sd = 2.3),
    breaks = Likert.brakes,
    break_labels = Likert.values
  )
  full.df[, 'F4_22'] <- draw_ordered(
    x = rnorm(n = overall.sample.size, mean = 3, sd = 2.1),
    breaks = Likert.brakes,
    break_labels = Likert.values
  )
  full.df[, 'F4_23'] <- draw_ordered(
    x = rnorm(n = overall.sample.size, mean = 4.5, sd = 1.8),
    breaks = Likert.brakes,
    break_labels = Likert.values
  )

  write.csv(x = as.data.frame(full.df),
```

```
          file = 'graziotin_et_al-bse_psychometrics_example.csv',
          row.names = F)
}
```

## Item Review and Item Analysis

For brevity, as item review is mostly a review and reasoning of items, we assume that the 31 items for five factors are already the result of item review. So we proceed to perform a pilot test for item analysis.

We simulate a pilot study with the first `pilot.test.sample.size` lines of the dataframe. This is `pilot.df`. We call psych function describe() and start our item analysis.

```
full.df <-
  read.csv(file = 'graziotin_et_al-bse_psychometrics_example.csv')
pilot.df <- head(full.df, n = pilot.test.sample.size)
desc <- describe(pilot.df)
desc

##        vars  n mean   sd median trimmed  mad min max range  skew kurtosis   se
## F1_1      1 23 2.91 1.08      3    2.89 1.48   1   5     4  0.16    -0.65 0.23
## F1_2      2 23 2.74 0.96      3    2.79 1.48   1   4     3 -0.37    -0.90 0.20
## F1_3      3 23 2.78 1.09      3    2.79 1.48   1   5     4  0.00    -0.88 0.23
## F1_4      4 23 4.96 0.21      5    5.00 0.00   4   5     1 -4.19    16.26 0.04
## F1_5      5 23 2.83 1.11      3    2.79 1.48   1   5     4  0.33    -0.79 0.23
## F1_6      6 23 2.87 1.18      3    2.84 1.48   1   5     4  0.08    -0.99 0.25
## F1_7      7 23 2.87 1.01      3    2.84 1.48   1   5     4  0.25    -1.03 0.21
## F2_8      8 23 3.04 0.98      3    3.00 1.48   1   5     4  0.20    -0.32 0.20
## F2_9      9 23 3.30 0.97      3    3.32 1.48   1   5     4 -0.32    -0.33 0.20
## F2_10    10 23 3.04 0.93      3    2.95 1.48   2   5     3  0.57    -0.60 0.19
## F2_11    11 23 3.17 0.65      3    3.16 0.00   2   5     3  0.79     1.08 0.14
## F3_12    12 23 2.96 1.15      3    2.95 1.48   1   5     4  0.25    -0.74 0.24
## F3_13    13 23 2.83 0.94      3    2.89 1.48   1   4     3 -0.31    -0.94 0.20
## F3_14    14 23 2.83 1.07      3    2.79 1.48   1   5     4  0.33    -0.51 0.22
## F3_15    15 23 2.91 1.04      3    2.95 1.48   1   5     4 -0.07    -0.84 0.22
## F3_16    16 23 2.65 1.07      3    2.63 1.48   1   5     4  0.26    -0.77 0.22
## F3_17    17 23 2.74 0.92      3    2.68 1.48   1   5     4  0.50    -0.18 0.19
## F3_18    18 23 2.83 1.15      3    2.79 1.48   1   5     4 -0.02    -0.64 0.24
## F3_19    19 23 2.78 0.95      3    2.84 1.48   1   4     3 -0.19    -1.09 0.20
## F3_20    20 23 3.04 0.71      3    3.00 0.00   2   5     3  0.69     0.84 0.15
## F4_21    21 23 3.70 1.43      4    3.84 1.48   1   5     4 -0.47    -1.29 0.30
## F4_22    22 23 4.57 1.04      5    4.84 0.00   1   5     4 -2.39     4.74 0.22
## F4_23    23 23 4.96 0.21      5    5.00 0.00   4   5     1 -4.19    16.26 0.04
## F4_24    24 23 2.78 0.85      3    2.84 0.00   1   4     3 -0.88     0.13 0.18
## F5_25    25 23 4.04 0.71      4    4.11 0.00   2   5     3 -0.80     1.21 0.15
## F5_26    26 23 2.91 1.04      3    2.84 1.48   1   5     4  0.39    -0.68 0.22
## F5_27    27 23 3.04 1.02      3    3.11 0.00   1   5     4 -0.57    -0.08 0.21
## F5_28    28 23 2.65 1.11      3    2.63 1.48   1   5     4  0.11    -0.88 0.23
## F5_29    29 23 2.91 1.08      3    2.89 1.48   1   5     4  0.16    -0.65 0.23
## F5_30    30 23 2.96 0.98      3    3.05 1.48   1   4     3 -0.48    -0.94 0.20
## F5_31    31 23 2.91 1.08      3    2.95 1.48   1   5     4 -0.25    -0.78 0.23
```

The function provides useful descriptive statistics for each item. Our first observation is that some items have mean and median that deviate notably from the (in our case) central value of 3. These are `F1_4`, `F4_22`, `F4_23`, with a mean value approaching 5, and `F5_25`, with a mean value of 4.04, which is high but not alarming yet. Items `F1_4` and `F4_23` are also those with the smallest standard deviation of 0.21, which results

in a variance of 0.04. The other two items show standard deviations in line with the rest of the dataset. Furthermore, `F1_4` and `F4_23` have unusual values for skew and kurtosis as well, indicating that the data is strongly biased on a value (indeed, their range is from 4 to 5). We conclude that items `F1_4` and `F4_23` have a high item facility (item difficulty), so they are candidate for deletion.

As for item discrimination, being our example one for trait measurement, we calculate how the two items correlate with those that are supposedly part of their factors.

```
cor.f1.discrimination <- cor(pilot.df[, grepl("F1_" , names(pilot.df))], method = c("kendall"))
corrplot(cor.f1.discrimination, method="number")
```

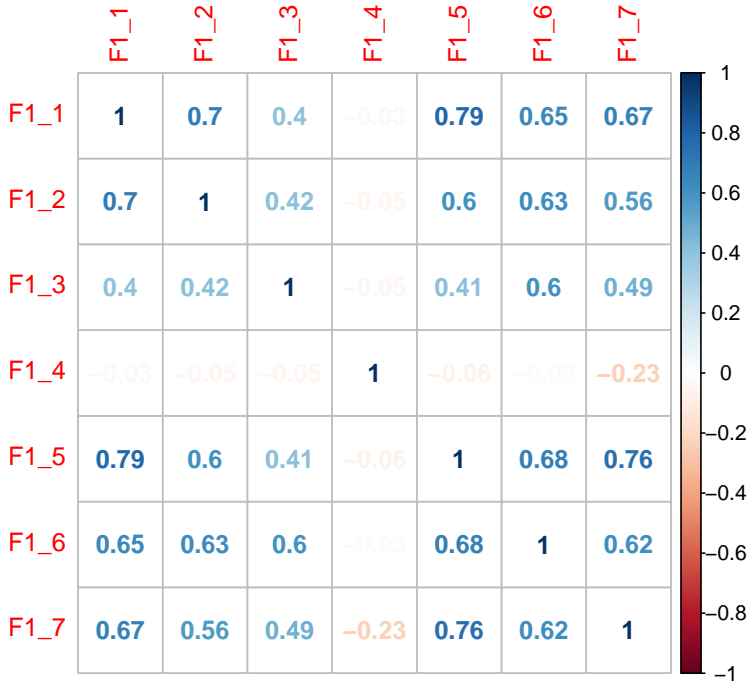

Item `F1_4` behaves differently to other items of its cluster as it shows near zero correlation to all items but a weak one to `F1_7`.

```
cor.f4.discrimination <- cor(pilot.df[, grepl("F4_" , names(pilot.df))], method = c("kendall"))
corrplot(cor.f4.discrimination, method="number")
```

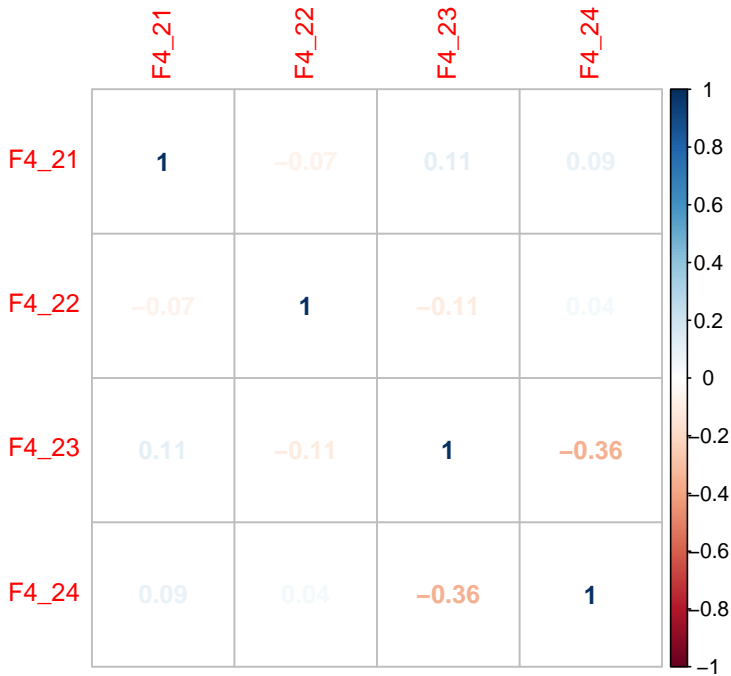

Item `F4_23` does not show particular differences in its cluster (we manipulated Factor 4 responses to result in a likely void factor).

We conclude that only item `F1_4` possesses a high discrimination and we therefore are confident in removing it from the measurement instrument. `F4_23` would show high discrimination as well, but not compared to the rest of its theorized cluster, yet it has high item facility. We decide to drop it as well. We decide to not drop `F4_22` yet, but the entire cluster for Factor 4 looks suspicious already.

```
pilot.afteritemanalysis.df <-
  pilot.df[,-which(names(pilot.df) %in% c("F1_4", "F4_23"))]
```

### Factor Analysis

Our psychometric evaluation study foresees an exploratory factor analysis (EFA) with an idea of items belonging to some possible constructs. While EFA does not require known factors at all (it is, after all, an exploration), we can still use it to cut down on items rather than on discovering how many factors we require. EFA *will* show us suggestions for factors, anyway.

We gather data with a new round of participants, this time aiming at a larger sample.

```
# we drop the pilot participants
study.df <-
  tail(full.df, n = NROW(full.df) - pilot.test.sample.size)
# we drop those columns that we identified after item analysis
study.beforeefa.df <-
  study.df[,-which(names(study.df) %in% c("F1_4", "F4_23"))]
```

Following the structure of our paper, we show here three possibilities for factor extraction, namely the scree plot analysis, the very simple structure analysis, and parallel analysis, and then we provide an entire session

of factor loading estimation, factor extraction, and factor rotation with principal axis factoring.

**Factor extraction: scree plot**

The scree test is based on eigenvalues. The `psych` package is nice enough to perform a principal component analysis (PCA) as well as of a principal axis factoring (PAF) for generating the plot.

We plot the scree test using our sample with 29 variables.

```
scree(study.beforeefa.df)
```

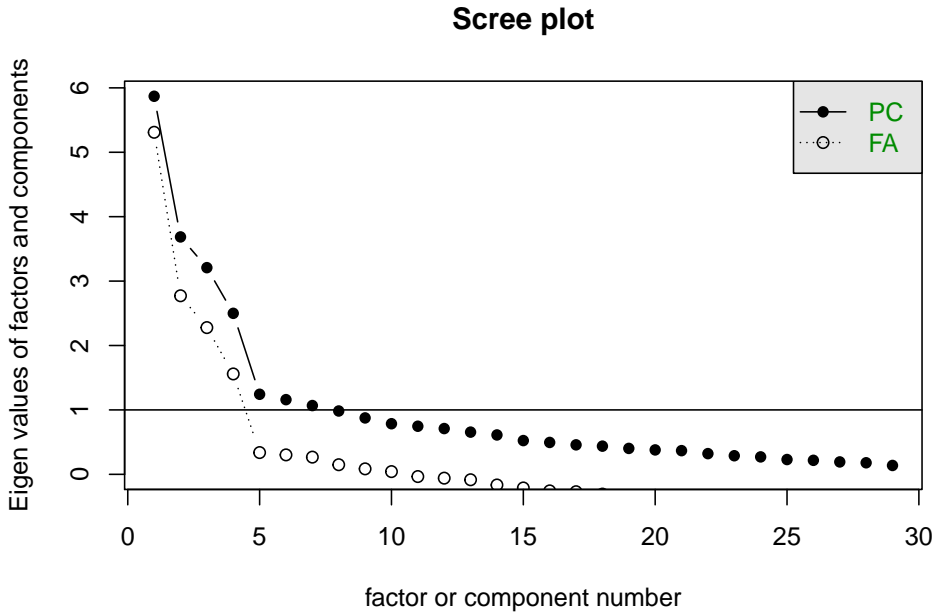

The breaking point starts with the fifth component for PCA and with the fifth factor for PAF. As written in the paper, "the strategy is to keep all factors before the breaking point", meaning that 4 is the number of factors to extract.

**Factor extraction: Very Simple Structure**

An alternative for factor extraction is the Very Simple Structure, or VSS, which can be used for extracting factors before their rotation.

We plot the structure using our initial five factor model with 29 items. As written in our paper, the VSS criterion requires a user specified number of factors, which should be above the target number of factors. We specify seven factors using the `n = 7` parameter.

```
vss.test.five <-
  vss(study.beforeefa.df,
      title = "Very Simple Structure of our fictitious five factor model", n = 7)
```

```
## Warning in fa.stats(r = r, f = f, phi = phi, n.obs = n.obs, np.obs = np.obs, :
## The estimated weights for the factor scores are probably incorrect. Try a
## different factor score estimation method.
```

8

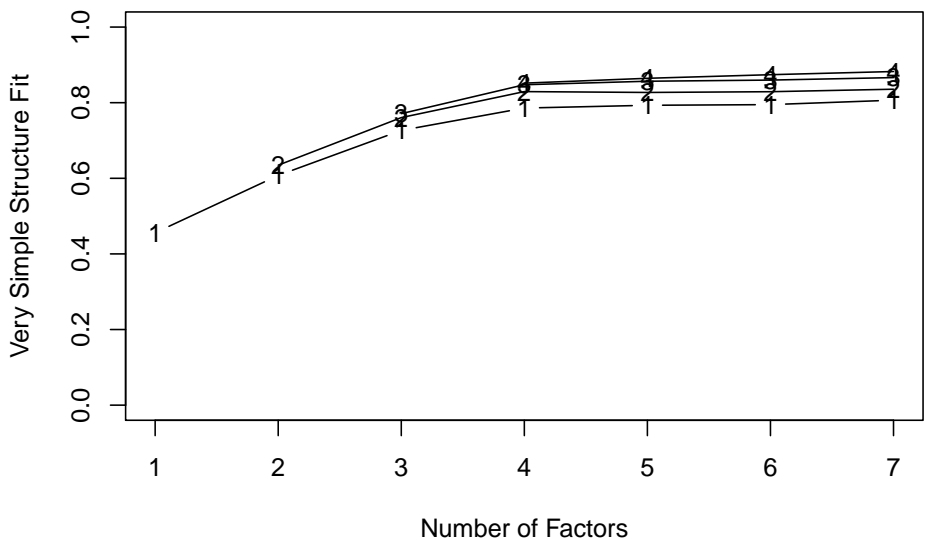

**Very Simple Structure of our fictitious five factor model**

```
vss.test.five
```

```
## 
## Very Simple Structure of  Very Simple Structure of our fictitious five factor model
## Call: vss(x = study.beforeefa.df, n = 7, title = "Very Simple Structure of our fictitious five facto
## VSS complexity 1 achieves a maximimum of 0.81  with  7  factors
## VSS complexity 2 achieves a maximimum of 0.84  with  7  factors
## 
## The Velicer MAP achieves a minimum of 0.02  with  4  factors
## BIC achieves a minimum of  -1071.41  with  4  factors
## Sample Size adjusted BIC achieves a minimum of  -135.64  with  4  factors
## 
## Statistics by number of factors 
##   vss1 vss2    map dof chisq     prob sqresid  fit RMSEA   BIC SABIC complex
## 1 0.46 0.00 0.036 377   995  3.3e-57    40.6 0.46 0.117  -807   385     1.0
## 2 0.61 0.63 0.027 349   714  3.0e-27    27.3 0.63 0.093  -954   149     1.2
## 3 0.73 0.76 0.018 322   487  7.6e-09    17.1 0.77 0.065 -1052   -34     1.2
## 4 0.79 0.83 0.015 296   343  3.1e-02    11.1 0.85 0.036 -1071  -136     1.2
## 5 0.79 0.83 0.018 271   310  5.1e-02    10.0 0.87 0.034  -985  -128     1.3
## 6 0.79 0.83 0.020 247   274  1.2e-01     9.0 0.88 0.029  -907  -126     1.3
## 7 0.81 0.84 0.023 224   247  1.4e-01     7.9 0.89 0.028  -823  -115     1.3
##   eChisq  SRMR eCRMS  eBIC
## 1   2359 0.156 0.162   558
## 2   1323 0.117 0.126  -345
## 3    575 0.077 0.087  -964
## 4    199 0.045 0.053 -1216
## 5    167 0.042 0.051 -1128
## 6    138 0.038 0.048 -1042
## 7    113 0.034 0.046  -958
```

Besides the warning that the estimated weights are probably incorrect, which would prompt us to use a different factor extraction method anyway, we can observe that the VSS criterion does achieve its maximum with 7 factors.

However, the increase in `vss1` and `vss2` score decreases rapidly after four factors. This can be noted graphically as well as in the console output. The `fit` behaves the same. The VSS still suggests that having four factors is optimal.

**Factor extraction: Parallel Analysis**

Finally, a robust method for factor extraction is Parallel Analysis (PA), which performs a constant comparison of our solution with randomly generated data.

```
pa.test.five <-
  fa.parallel(study.beforeefa.df)
```

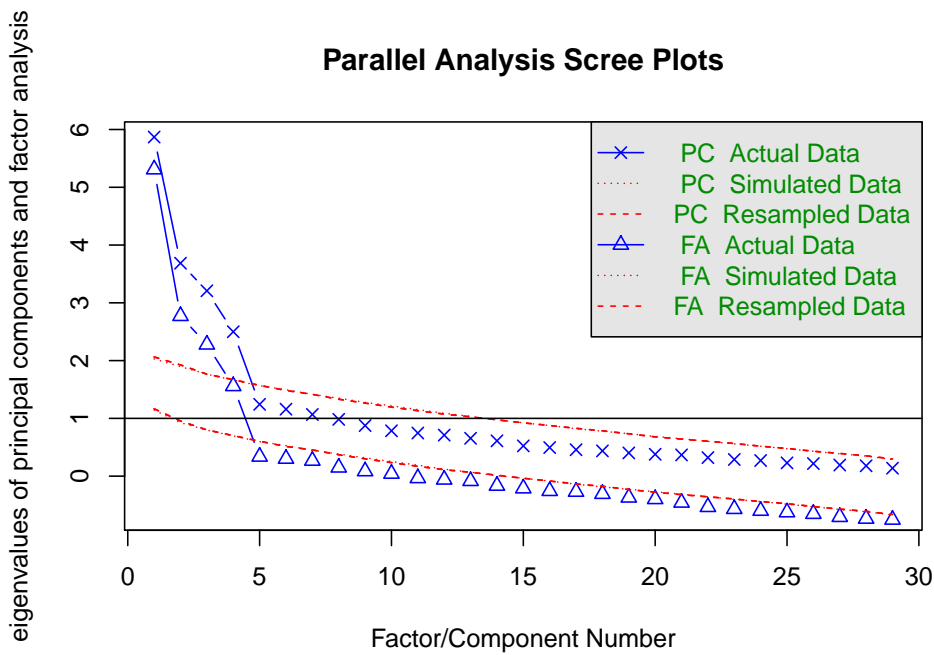

```
## Parallel analysis suggests that the number of factors =  4  and the number of components =  4
```

As with the Scree Plot function, we see that psych `fa.parallel` function performs both a PCA and a PAF with the data. The textual output suggests already that four factors should be extracted. The graph shows our data and with calculated components (crosses) and factors (triangles). They are placed the same as with a Scree Plot. The red lines (upper one for PCA, lower one for PAF), on the other hand, represent the mean generated eingenvalues. Factors are retained as long as they are greater than the mean eigenvalue generated from the random matrices.

For both PCA and PAF, the generated data crosses the lines of the actual eigenvalue between 4 and 5 components and factors. This suggests that 4 factors should be extracted.

**Factor loading, extraction, and rotation: principal axis factoring**

The R psych package makes it easy to perform loading estimation, factor extraction, and factor rotation with a single function call. We will perform here a principal axis factoring (PAF) to fully go from five factors (the initial assumption) to four factors. We show in this section how a reasoning can be conducted by observing values as returned by functions for loading, extration, and rotation.

Besides our reasoning on pure numbers, we will add elements of confirmatory factory analysis (CFA) to guide our retention of items and factors, that is, measures of fit. First, the change in chi square from factor n to factor n - 1 should go from significant (there is is a bad fit) to insignificant. To the chi square test, we add other two measure of fit, namely the root mean square of the residuals (RMSR) and the root mean square error of approximation (RMSEA) index. RMSR should be as close to zero as possible while RMSEA should be below 0.05.

It should be kept in mind that measures of fit are mere indicators in EFA, being them part of a confirmatory factor analysis (CFA) strategy mostly, and should be taken into consideration together with a sensitivity analysis.

As suggested in our paper, we opt for Principal Axis Factoring (PAF) with a hybrid orthogonal first, oblique rotation then, namely a promax rotation. This means that we have a further criterion for factor reduction, that is reducing factors while the sum of squared loadings is below 1.00.

```
five.factors <-
  fa(
    study.beforeefa.df ,
    fm = "pa",
    nfactors = 5,
    rotate = "promax"
  )
```

```
## Loading required namespace: GPArotation
```

```
five.factors
```

```
## Factor Analysis using method =  pa
## Call: fa(r = study.beforeefa.df, nfactors = 5, rotate = "promax", fm = "pa")
## Standardized loadings (pattern matrix) based upon correlation matrix
##         PA1   PA2   PA3   PA4   PA5   h2   u2  com
## F1_1  -0.06  0.71 -0.09  0.00 -0.04 0.550 0.45 1.1
## F1_2   0.00  0.74  0.04 -0.10 -0.08 0.558 0.44 1.1
## F1_3  -0.05  0.70  0.12  0.00 -0.04 0.486 0.51 1.1
## F1_5   0.00  0.70 -0.01  0.02  0.12 0.507 0.49 1.1
## F1_6  -0.08  0.76  0.05  0.05  0.19 0.607 0.39 1.2
## F1_7  -0.05  0.73  0.09 -0.06 -0.13 0.548 0.45 1.1
## F2_8  -0.02 -0.10  0.01  0.69 -0.15 0.484 0.52 1.1
## F2_9   0.04 -0.01  0.02  0.84  0.11 0.733 0.27 1.0
## F2_10  0.04 -0.04  0.15  0.64 -0.06 0.433 0.57 1.1
## F2_11  0.05  0.01  0.04  0.63  0.15 0.436 0.56 1.1
## F3_12  0.77  0.00  0.00 -0.03  0.08 0.585 0.42 1.0
## F3_13  0.64  0.04  0.00  0.04  0.07 0.419 0.58 1.0
## F3_14  0.72  0.00 -0.09  0.03 -0.07 0.515 0.48 1.1
## F3_15  0.82  0.05  0.01 -0.07  0.28 0.697 0.30 1.3
## F3_16  0.69  0.05  0.03  0.06  0.00 0.504 0.50 1.0
## F3_17  0.81 -0.02 -0.04  0.08 -0.11 0.700 0.30 1.1
## F3_18  0.65 -0.20  0.06  0.10  0.03 0.517 0.48 1.3
## F3_19  0.79  0.02 -0.08 -0.02  0.03 0.600 0.40 1.0
## F3_20  0.75 -0.10  0.01 -0.07 -0.18 0.639 0.36 1.2
## F4_21 -0.16 -0.01 -0.13  0.19  0.06 0.089 0.91 3.0
```

11

```
## F4_22  0.04  0.03  0.02  0.01  0.54 0.284 0.72 1.0
## F4_24 -0.06 -0.29  0.05  0.01 -0.05 0.094 0.91 1.2
## F5_25 -0.04 -0.09  0.02 -0.10  0.06 0.026 0.97 3.1
## F5_26  0.03 -0.04  0.72 -0.01  0.04 0.533 0.47 1.0
## F5_27 -0.13  0.02  0.77 -0.05 -0.06 0.579 0.42 1.1
## F5_28 -0.03 -0.03  0.68  0.16 -0.10 0.490 0.51 1.2
## F5_29 -0.06  0.01  0.66  0.07  0.03 0.424 0.58 1.0
## F5_30  0.05 -0.05  0.66  0.10  0.16 0.474 0.53 1.2
## F5_31  0.03  0.16  0.69 -0.07  0.00 0.488 0.51 1.1
##
##                        PA1  PA2  PA3  PA4  PA5
## SS loadings           5.02 3.31 2.95 2.12 0.60
## Proportion Var        0.17 0.11 0.10 0.07 0.02
## Cumulative Var        0.17 0.29 0.39 0.46 0.48
## Proportion Explained  0.36 0.24 0.21 0.15 0.04
## Cumulative Proportion 0.36 0.59 0.81 0.96 1.00
##
##  With factor correlations of
##       PA1   PA2   PA3   PA4   PA5
## PA1  1.00 -0.06  0.20  0.13 -0.09
## PA2 -0.06  1.00 -0.14  0.07 -0.05
## PA3  0.20 -0.14  1.00 -0.06 -0.07
## PA4  0.13  0.07 -0.06  1.00  0.05
## PA5 -0.09 -0.05 -0.07  0.05  1.00
##
## Mean item complexity =  1.2
## Test of the hypothesis that 5 factors are sufficient.
##
## The degrees of freedom for the null model are  406  and the objective function was  14.23 with Chi S
## The degrees of freedom for the model are 271  and the objective function was  2.98
##
## The root mean square of the residuals (RMSR) is  0.04
## The df corrected root mean square of the residuals is  0.05
##
## The harmonic number of observations is  119 with the empirical chi square  167.43  with prob <  1
## The total number of observations was  119  with Likelihood Chi Square =  310.07  with prob <  0.051
##
## Tucker Lewis Index of factoring reliability =  0.946
## RMSEA index =  0.034  and the 90 % confidence intervals are  0 0.052
## BIC =  -985.07
## Fit based upon off diagonal values = 0.97
## Measures of factor score adequacy
##                                                   PA1  PA2  PA3  PA4  PA5
## Correlation of (regression) scores with factors  0.96 0.94 0.93 0.92 0.74
## Multiple R square of scores with factors         0.93 0.88 0.86 0.84 0.55
## Minimum correlation of possible factor scores    0.85 0.75 0.72 0.68 0.10
```

Let us now break down the output of this function.

The function represents the candidate factors with labels `PA1` to `PA5`. The factor numbering does not necessarily correspond to the factors we developed, that is, `PA1` is not automatically assigned to items `F1_x`. Factors with lower numbering are those that provide the highest sum of squared loadings (`SS loadings` in R output) and also the variance explained.

Let us first inspect the computed loadings (which could also be singled out with the `loadings(five.factors)`

function.), in particular, the summary table that provides the sum of squared loadings (SS loadings) and the variance explained for each factor. A first observation is that `PA5` provides a sum of squared loadings below 1, which is an indication for removal. The factor accounts also for only 0.04 variance, and the cumulative variance moves from 0.46 to 0.48 from four factors to five factors.

Goodness of fit with Likelihood Chi Square is already non-significant but approaching significance levels (p = 0.052). RMSR and RMSEA are within desirable levels.

If we then move to inspect the loadings of all our single items, we see a table of our F items crossed with the candidate factors PA. Furthermore, the table provides, for each item, the communalities (h2), the unique variance (u2), and the complexity of the component loadings (com).

We observe that the item possessing the highest load to `PA5` is `F4_22` with 0.536. All other items have loadings near 0. Furthermore, most of our items for a factor F have a corresponding factor PA (e.g., `F1` items load mostly on factor `PA2`) while `PA5` seems to meet all Fx items.

Another observation is on items for `F4`. As stated above, `F4_22` is the only one with a meaningful loading on a factor, that is `PA5`, while `F4_21` does not seem to have a meaningful loading on anything. The only remaining item, `F4_24`, provides a loading on `PA2`, which seems associated with `F1`items mostly. Given our previous suspicious on `F4`items, after an internal review, we decide to drop the F4 items entirely and test for a four factors model.

**Factor reduction**

We decide to drop F4 (Collaboration propensity) and test for a four factors model.

```
# we drop those columns that we identified after item analysis
study.efa.df <-
  study.beforeefa.df[,-which(names(study.beforeefa.df) %in% c("F4_21", "F4_22", "F4_24"))]
four.factors <-
  fa(study.efa.df ,
     fm = "pa",
     nfactors = 4,
     rotate = "promax")
four.factors
```

```
## Factor Analysis using method =  pa
## Call: fa(r = study.efa.df, nfactors = 4, rotate = "promax", fm = "pa")
## Standardized loadings (pattern matrix) based upon correlation matrix
##         PA1   PA2   PA3   PA4   h2   u2 com
## F1_1  -0.03  0.73 -0.11  0.00 0.568 0.43 1.0
## F1_2   0.04  0.75  0.03 -0.12 0.556 0.44 1.1
## F1_3  -0.02  0.68  0.11 -0.02 0.462 0.54 1.1
## F1_5   0.01  0.69 -0.04  0.03 0.481 0.52 1.0
## F1_6  -0.07  0.74  0.02  0.07 0.576 0.42 1.0
## F1_7  -0.01  0.73  0.07 -0.07 0.530 0.47 1.0
## F2_8  -0.06 -0.09 -0.01  0.67 0.432 0.57 1.1
## F2_9  -0.03 -0.02 -0.02  0.86 0.720 0.28 1.0
## F2_10 -0.01 -0.04  0.13  0.65 0.438 0.56 1.1
## F2_11 -0.01 -0.01  0.00  0.65 0.416 0.58 1.0
## F3_12  0.77  0.00  0.01 -0.04 0.582 0.42 1.0
## F3_13  0.63  0.04  0.00  0.04 0.414 0.59 1.0
## F3_14  0.72  0.01 -0.07  0.00 0.512 0.49 1.0
## F3_15  0.80  0.04  0.02 -0.07 0.615 0.38 1.0
## F3_16  0.70  0.05  0.04  0.04 0.506 0.49 1.0
## F3_17  0.82 -0.01 -0.02  0.04 0.683 0.32 1.0
## F3_18  0.64 -0.19  0.08  0.08 0.519 0.48 1.2
```

```
## F3_19  0.79  0.02 -0.07 -0.03 0.596 0.40 1.0
## F3_20  0.76 -0.09  0.04 -0.11 0.589 0.41 1.1
## F5_25 -0.04 -0.09  0.03 -0.11 0.024 0.98 2.4
## F5_26  0.05 -0.03  0.71 -0.03 0.528 0.47 1.0
## F5_27 -0.09  0.04  0.76 -0.08 0.567 0.43 1.1
## F5_28 -0.01 -0.01  0.67  0.12 0.473 0.53 1.1
## F5_29 -0.05  0.00  0.66  0.06 0.427 0.57 1.0
## F5_30  0.05 -0.06  0.64  0.09 0.451 0.55 1.1
## F5_31  0.07  0.17  0.69 -0.11 0.495 0.51 1.2
##
##                        PA1  PA2  PA3  PA4
## SS loadings           4.94 3.21 2.92 2.09
## Proportion Var        0.19 0.12 0.11 0.08
## Cumulative Var        0.19 0.31 0.43 0.51
## Proportion Explained  0.38 0.24 0.22 0.16
## Cumulative Proportion 0.38 0.62 0.84 1.00
##
##  With factor correlations of
##       PA1   PA2   PA3  PA4
## PA1  1.00 -0.08  0.16 0.24
## PA2 -0.08  1.00 -0.12 0.06
## PA3  0.16 -0.12  1.00 0.03
## PA4  0.24  0.06  0.03 1.00
##
## Mean item complexity =  1.1
## Test of the hypothesis that 4 factors are sufficient.
##
## The degrees of freedom for the null model are  325  and the objective function was  13.29 with Chi S
## The degrees of freedom for the model are 227  and the objective function was  2.51
##
## The root mean square of the residuals (RMSR) is  0.04
## The df corrected root mean square of the residuals is  0.05
##
## The harmonic number of observations is  119 with the empirical chi square  132.9  with prob <  1
## The total number of observations was  119  with Likelihood Chi Square =  265.54  with prob <  0.04
##
## Tucker Lewis Index of factoring reliability =  0.949
## RMSEA index =  0.037  and the 90 % confidence intervals are  0.009 0.056
## BIC =  -819.32
## Fit based upon off diagonal values = 0.97
## Measures of factor score adequacy
##                                                  PA1  PA2  PA3  PA4
## Correlation of (regression) scores with factors  0.96 0.93 0.92 0.92
## Multiple R square of scores with factors         0.92 0.87 0.85 0.84
## Minimum correlation of possible factor scores    0.84 0.75 0.71 0.68
```

The four factor model seems an improvement over the previous one. The variance explained has not decreased (it actually increases to 0.51), and each factor explains at minimum 16% of the variance. Where we did not improve was the Likelihood Chi Square test, which is now significant (p = 0.04). RMSR and RMSEA have not changed significantly ands are still within desirable levels.

If we inspect the single items and their loadings on factors, however, the situation is improved significantly. All our developed items for a factor F load majorly on a single cluster PA found by the PAF analysis.

The only exception appears to be `F5_25`, which seems to not load on any factor whatsoever. After reviewing

the item, we decide to remove it and run the EFA again.

```r
# we drop those columns that we identified after item analysis
study.efa.bis.df <-
  study.efa.df[,-which(names(study.efa.df) %in% c("F5_25"))]
four.factors.bis <-
  fa(study.efa.bis.df ,
     fm = "pa",
     nfactors = 4,
     rotate = "promax")
four.factors.bis
```

```
## Factor Analysis using method =  pa
## Call: fa(r = study.efa.bis.df, nfactors = 4, rotate = "promax", fm = "pa")
## Standardized loadings (pattern matrix) based upon correlation matrix
##         PA1   PA2   PA3   PA4   h2   u2 com
## F1_1  -0.01  0.73 -0.12  0.02 0.57 0.43 1.1
## F1_2   0.05  0.74  0.02 -0.10 0.55 0.45 1.0
## F1_3  -0.01  0.68  0.10  0.00 0.46 0.54 1.0
## F1_5   0.03  0.69 -0.05  0.05 0.48 0.52 1.0
## F1_6  -0.05  0.75  0.00  0.10 0.58 0.42 1.0
## F1_7   0.01  0.73  0.06 -0.04 0.53 0.47 1.0
## F2_8  -0.04 -0.04 -0.03  0.66 0.43 0.57 1.0
## F2_9   0.00  0.04 -0.05  0.85 0.72 0.28 1.0
## F2_10  0.02  0.00  0.11  0.63 0.43 0.57 1.1
## F2_11  0.02  0.04 -0.02  0.65 0.43 0.57 1.0
## F3_12  0.77  0.00  0.01 -0.04 0.58 0.42 1.0
## F3_13  0.64  0.05 -0.01  0.05 0.41 0.59 1.0
## F3_14  0.73  0.02 -0.07  0.01 0.51 0.49 1.0
## F3_15  0.80  0.05  0.01 -0.06 0.62 0.38 1.0
## F3_16  0.70  0.07  0.03  0.05 0.51 0.49 1.0
## F3_17  0.82  0.00 -0.02  0.04 0.68 0.32 1.0
## F3_18  0.64 -0.18  0.07  0.08 0.52 0.48 1.2
## F3_19  0.79  0.03 -0.07 -0.02 0.60 0.40 1.0
## F3_20  0.76 -0.08  0.04 -0.10 0.59 0.41 1.1
## F5_26  0.04 -0.04  0.72 -0.05 0.53 0.47 1.0
## F5_27 -0.10  0.02  0.76 -0.09 0.57 0.43 1.1
## F5_28 -0.01 -0.01  0.67  0.12 0.47 0.53 1.1
## F5_29 -0.06 -0.01  0.66  0.05 0.43 0.57 1.0
## F5_30  0.04 -0.07  0.64  0.07 0.45 0.55 1.1
## F5_31  0.06  0.15  0.69 -0.11 0.49 0.51 1.2
##
##                        PA1  PA2  PA3  PA4
## SS loadings           4.94 3.20 2.92 2.07
## Proportion Var        0.20 0.13 0.12 0.08
## Cumulative Var        0.20 0.33 0.44 0.53
## Proportion Explained  0.38 0.24 0.22 0.16
## Cumulative Proportion 0.38 0.62 0.84 1.00
##
##  With factor correlations of
##       PA1   PA2   PA3   PA4
## PA1  1.00 -0.13  0.18  0.20
## PA2 -0.13  1.00 -0.09 -0.05
## PA3  0.18 -0.09  1.00  0.08
## PA4  0.20 -0.05  0.08  1.00
```

```
## 
## Mean item complexity =  1
## Test of the hypothesis that 4 factors are sufficient.
## 
## The degrees of freedom for the null model are  300  and the objective function was  12.97 with Chi S
## The degrees of freedom for the model are 206  and the objective function was  2.21
## 
## The root mean square of the residuals (RMSR) is  0.04
## The df corrected root mean square of the residuals is  0.05
## 
## The harmonic number of observations is  119 with the empirical chi square  108.99  with prob <  1
## The total number of observations was  119  with Likelihood Chi Square =  234.17  with prob <  0.087
## 
## Tucker Lewis Index of factoring reliability =  0.962
## RMSEA index =  0.033  and the 90 % confidence intervals are  0 0.054
## BIC =  -750.33
## Fit based upon off diagonal values = 0.98
## Measures of factor score adequacy
##                                                    PA1  PA2  PA3  PA4
## Correlation of (regression) scores with factors   0.96 0.93 0.92 0.91
## Multiple R square of scores with factors          0.92 0.87 0.86 0.84
## Minimum correlation of possible factor scores     0.84 0.75 0.71 0.67
```

With the revised four factor model, the total variance explained rises to 0.53, which is another improvement.

It holds that all our developed items for a factor F load majorly on a single cluster PA found by the PAF analysis. In particular, our factors sorted for their importance (variance explained) are F3 (Learning disposition, PA1 in the output), F1 (Code curiosity, PA2), F5 (Comfort in novelty, PA3), and F2 (Programming paradign flexibility, PA4). We are satisfied with the resulting measurement instrument so far.

We improved on Likelihood Chi Square test, which is now back to non-significant (p = 0.087). RMSR and RMSEA have not changed significantly are are still within desirable levels. We decide to stop our EFA at this point and use the present form of our measurement instrument, with 4 factors and 26 items.

**The resulting measurement instrument**

We started with 31 items, which in our fictitious example were already a reduction, after item review, of an original measurement instrument with five factors. Thanks to item analysis and exploratory factor analysis, we were able to reduce the measurement instrument to 26 items (a 16% reduction) and four factors. All this with an improved confidence in what we can conclude from interpreting the results of employing the measurement instrument.

```
study.after.efa.df <- study.efa.bis.df
```

### Item statistical properties

```
describe(study.after.efa.df)
```

```
##        vars   n mean   sd median trimmed  mad min max range  skew kurtosis   se
## F1_1      1 119 2.99 1.13      3    2.99 1.48   1   5     4  0.05    -0.68 0.10
## F1_2      2 119 2.87 0.97      3    2.89 1.48   1   5     4 -0.01    -0.37 0.09
## F1_3      3 119 2.87 1.11      3    2.87 1.48   1   5     4 -0.01    -0.60 0.10
## F1_5      4 119 2.97 1.05      3    2.96 1.48   1   5     4  0.09    -0.38 0.10
## F1_6      5 119 2.87 1.11      3    2.87 1.48   1   5     4  0.06    -0.70 0.10
## F1_7      6 119 2.94 1.06      3    2.93 1.48   1   5     4  0.12    -0.51 0.10
## F2_8      7 119 2.99 1.06      3    2.95 1.48   1   5     4  0.23    -0.60 0.10
```

```
## F2_9      8 119 3.06 1.01        3   3.03 1.48    1    5    4  0.08   -0.71 0.09
## F2_10     9 119 3.18 1.05        3   3.19 1.48    1    5    4 -0.14   -0.45 0.10
## F2_11    10 119 2.97 1.03        3   2.94 1.48    1    5    4  0.16   -0.81 0.09
## F3_12    11 119 3.13 1.06        3   3.15 1.48    1    5    4 -0.22   -0.38 0.10
## F3_13    12 119 3.20 1.07        3   3.21 1.48    1    5    4 -0.07   -0.57 0.10
## F3_14    13 119 3.13 1.05        3   3.11 1.48    1    5    4  0.01   -0.55 0.10
## F3_15    14 119 3.00 1.05        3   3.02 1.48    1    5    4 -0.09   -0.44 0.10
## F3_16    15 119 3.15 1.10        3   3.15 1.48    1    5    4  0.00   -0.64 0.10
## F3_17    16 119 3.13 1.17        3   3.15 1.48    1    5    4 -0.27   -0.76 0.11
## F3_18    17 119 3.18 1.10        3   3.18 1.48    1    5    4  0.03   -0.69 0.10
## F3_19    18 119 3.03 1.10        3   3.04 1.48    1    5    4 -0.07   -0.64 0.10
## F3_20    19 119 3.08 1.04        3   3.10 1.48    1    5    4 -0.19   -0.61 0.10
## F5_26    20 119 2.94 1.05        3   2.95 1.48    1    5    4  0.03   -0.63 0.10
## F5_27    21 119 3.01 1.04        3   3.02 1.48    1    5    4 -0.06   -0.60 0.10
## F5_28    22 119 3.05 1.10        3   3.05 1.48    1    5    4  0.09   -0.51 0.10
## F5_29    23 119 3.02 1.08        3   3.03 1.48    1    5    4 -0.07   -0.60 0.10
## F5_30    24 119 3.02 1.02        3   2.99 1.48    1    5    4  0.11   -0.51 0.09
## F5_31    25 119 2.96 0.96        3   2.93 1.48    1    5    4  0.20   -0.48 0.09
```

Possible minimum and maximum scores versus obtained minimum and maximum scores for F1, F2, F3, and F5, respectively.

```r
f1.df <-
  study.after.efa.df[, grepl("F1_", names(study.after.efa.df))]
f2.df <-
  study.after.efa.df[, grepl("F2_", names(study.after.efa.df))]
f3.df <-
  study.after.efa.df[, grepl("F3_", names(study.after.efa.df))]
f5.df <-
  study.after.efa.df[, grepl("F5_", names(study.after.efa.df))]
print("F1")
```

```
## [1] "F1"
```

```r
c(c(1, NCOL(f1.df) * 5), c(min(rowSums(f1.df)), max(rowSums(f1.df))))
```

```
## [1]  1 30  8 29
```

```r
print("F2")
```

```
## [1] "F2"
```

```r
c(c(1, NCOL(f2.df) * 5), c(min(rowSums(f2.df)), max(rowSums(f2.df))))
```

```
## [1]  1 20  5 20
```

```r
print("F3")
```

```
## [1] "F3"
```

```r
c(c(1, NCOL(f3.df) * 5), c(min(rowSums(f3.df)), max(rowSums(f3.df))))
```

```
## [1]  1 45  9 44
```

```r
print("F5")
```

```
## [1] "F5"
```

```r
c(c(1, NCOL(f5.df) * 5), c(min(rowSums(f5.df)), max(rowSums(f5.df))))
```

```
## [1]  1 30  6 29
```

```r
factors.after.efa.df <-
  data.frame(
    F1 = rowSums(f1.df),
    F2 = rowSums(f2.df),
    F3 = rowSums(f3.df),
    F5 = rowSums(f5.df)
  )
describe(factors.after.efa.df)

##    vars   n  mean   sd median trimmed  mad min max range  skew kurtosis   se
## F1    1 119 17.52 4.97     17   17.43 4.45   8  29    21  0.20    -0.50 0.46
## F2    2 119 12.19 3.26     12   12.16 2.97   5  20    15  0.08    -0.34 0.30
## F3    3 119 28.03 7.51     29   28.14 7.41   9  44    35 -0.17    -0.49 0.69
## F5    4 119 17.99 4.68     18   17.97 4.45   6  29    23  0.01    -0.01 0.43
```

```r
error.bars(
  factors.after.efa.df,
  ylab = "Score",
  xlab = "Factor",
  main = "Confidence intervals \n for F1, F2, F3, and F5",
  eyes = F,
  sd = F
)
```

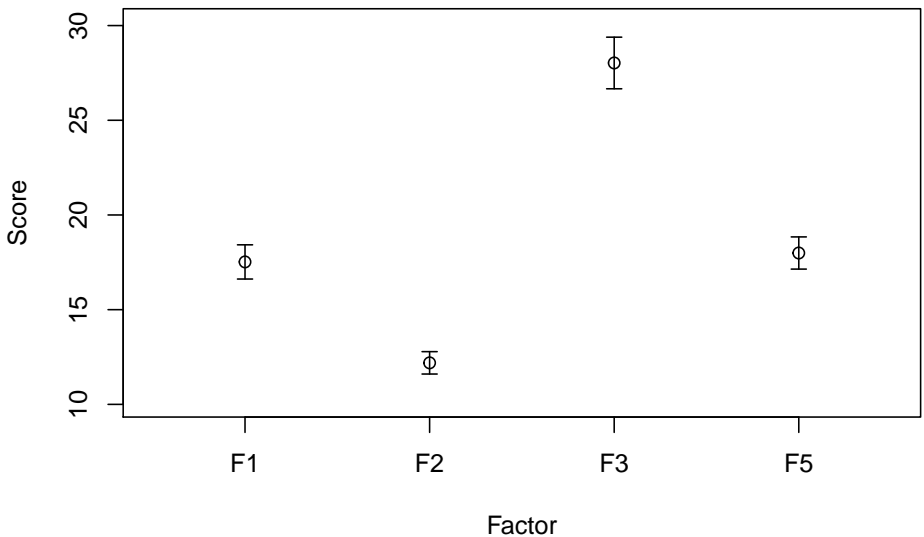

```r
print("Standardized scores available in factors.after.efa.standard.df but not shown here.")

## [1] "Standardized scores available in factors.after.efa.standard.df but not shown here."

factors.after.efa.standard.df <-
  data.frame(
```

```
    F1 = scale(rowSums(f1.df)),
    F2 = scale(rowSums(f2.df)),
    F3 = scale(rowSums(f3.df)),
    F5 = scale(rowSums(f5.df))
  )
```

## Reliability

Being our measurement instrument one for norm-referencing testing and not one for assessing skills, we perform a test-retest reliability assessment. Only a smaller number of participants (`reliability.sample.size`) accepted to take part to a second test.

```
reliablity.df.first <-
  tail(study.after.efa.df, reliability.sample.size)
reliablity.df.second <-
  tail(study.after.efa.df, reliability.sample.size)
f1.first <-
  rowSums(reliablity.df.first[, grepl("F1_" , names(reliablity.df.first))])
f2.first <-
  rowSums(reliablity.df.first[, grepl("F2_" , names(reliablity.df.first))])
f3.first <-
  rowSums(reliablity.df.first[, grepl("F3_" , names(reliablity.df.first))])
f5.first <-
  rowSums(reliablity.df.first[, grepl("F5_" , names(reliablity.df.first))])

f1.second <- round(jitter(unname(f1.first), amount = 3))
f2.second <- round(jitter(unname(f2.first), amount = 3))
f3.second <- round(jitter(unname(f3.first), amount = 3))
f5.second <- round(jitter(unname(f5.first), amount = 3))
```

Test-retest reliability for F1, F2, F3, and F5 are, respectively:

```
print("F1")
```

```
## [1] "F1"
```

```
cor(unname(f1.first), unname(f1.second))
```

```
## [1] 0.9187912
```

```
print("F2")
```

```
## [1] "F2"
```

```
cor(unname(f2.first), unname(f2.second))
```

```
## [1] 0.8393884
```

```
print("F3")
```

```
## [1] "F3"
```

```
cor(unname(f3.first), unname(f3.second))
```

```
## [1] 0.9708618
```

```
print("F5")
```

```
## [1] "F5"
```

```
cor(unname(f5.first), unname(f5.second))
```

```
## [1] 0.9372151
```

**Validity**

Our fictitious measurement instrument is novel and has no correspondence with existing measurement instruments. Therefore, any discussion on its validity depends on the instrument itself and the reasoning we can apply to it. Therefore, the present example can not elaborate on validity measurements.

**Conclusion**

The present document presented an introduction to psychometric evaluation with R for a behavioral software engineering audience. Through phases of item review and item analysis, exploratory factor analysis, item statistical properties, and reliability and validity evaluation, we show how we started with 31 items and 5 factors to represent our fictitious "individual perception styles of source code" to 26 items and four factors.

We could, therefore, offer the resulting measurement instrument in a paper describing its psychometric evaluation, for other researchers to use for further evaluation, and companies to use.



\end{document}